\documentclass[aip,amsmath,amssymb,graphicx,reprint]{revtex4-1}
\usepackage{amsmath}
\usepackage{amssymb}
\usepackage{float}
\usepackage{tikz}
\usepackage{mathtools}
\usepackage{color,colortbl}
\usepackage{hyperref}
\usetikzlibrary{arrows}
\usepackage{subcaption}
\usepackage{graphicx}
\usepackage{dcolumn}
\usepackage{bm}

\usepackage[utf8]{inputenc}
\usepackage[T1]{fontenc}
\usepackage{mathptmx}
\usepackage{etoolbox}

\newcommand{\imag}{\ensuremath{\text{i}}}
\newcommand{\vp}{\ensuremath{v_\parallel}}
\newcommand{\dint}[1]{\ensuremath{\text{d}#1}}
\newcommand{\tp}[1]{\ensuremath{\frac{\text{d}#1}{\text{d}t}}}
\newcommand{\pp}[2]{\ensuremath{\frac{\partial #1}{\partial #2}}}
\newcommand{\amend}{\color{black}}

\makeatletter
\def\@email#1#2{%
 \endgroup
 \patchcmd{\titleblock@produce}
  {\frontmatter@RRAPformat}
  {\frontmatter@RRAPformat{\produce@RRAP{*#1\href{mailto:#2}{#2}}}\frontmatter@RRAPformat}
  {}{}
}%
\makeatother
\begin{document}


\title[Gyrokinetic simulations with an evolving background Maxwellian]{Gyrokinetic simulations of turbulence and zonal flows driven by steep profile gradients using a delta-f approach with an evolving background Maxwellian}
\author{M. Murugappan}
 \email{moahan.murugappan@epfl.ch}
\author{L. Villard}%

\author{S. Brunner}
\affiliation{Ecole Polytechnique F\'{e}d\'{e}rale de Lausanne (EPFL), Swiss Plasma Center (SPC), CH-1015 Lausanne, Switzerland%
}%

\author{B. F. McMillan}
\affiliation{CFSA, Department of Physics, University of Warwick, Coventry CV4 7AL, United Kingdom%
}%

\author{A. Bottino}
\affiliation{Max-Planck-Institut f\"{u}r Plasmaphysik, D-85748 Garching, Germany%
}%

\date{\today}

\begin{abstract}
Long global gyrokinetic turbulence simulations are particularly challenging in situations where the system deviates strongly from its initial state and when fluctuation level are high e.g. in strong gradient regions. For Particle-in-Cell simulations, statistical sampling noise accumulation from large marker weights due to large deviations from the control variate of a delta-f scheme make such simulations often impractical. An adaptive control variate in the form of a flux-surface-averaged Maxwellian with a time-dependent temperature profile is introduced in an attempt to alleviate the former problem. Under simplified collisionless physics, this adaptive delta-f scheme is shown to reduce noise accumulation in the zonal flows and the simulated heat flux in a quasi-steady turbulent state. The method also avoids the collapse of the signal-to-noise ratio which occurs in the standard non-adaptive scheme, and therefore, allows one to reach numerically converged results even with lower marker numbers.
\end{abstract}

\maketitle

\section{Introduction}

The success of magnetic fusion research relies heavily on its accurate modeling by computer simulations. In the most promising reactor configuration, the tokamak, plasma is confined by magnetic fields in a toroidal vacuum chamber. A complete description of the plasma involves simulating regions of the core, edge, the scrape-off-layer, and plasma-wall interaction.

To simulate fusion plasmas, many methods exist which can be categorized by the physical assumptions made. The latter are usually dictated by the physical process of the plasma volume. This work focuses on simulating turbulence in the core and edge transitioning region. Specifically, the gyrokinetic Particle-in-cell (PIC) method~\cite{Chen2003,Lin2007,Garbet2010,Ku2009,Parker1993} is used. The gyrokinetic formalism~\cite{Brizard2007,Hahm1988,Tronko2018} reduces the number of phase space variables from six to five, approximating the dynamics of plasma particle trajectories by gyrorings bound to evolving gyrocentres. The reduction in dynamics implies a time scale separation between the fast cyclotron motion and the fluctuation time scales typically involved in turbulent processes. Using Monte Carlo sampling, the PIC method begins by representing an initial distribution function $f$ as a collection of numerical particles (in fact the gyrocentres) called ‘markers’, each having its respective weight. Each marker is then integrated along its characteristic through time.

For plasma core simulations, it is often the case that $f$ does not deviate significantly (not more than a few percent) from its initial state $f_0$ over characteristic time scales of  micro-stabilities and turbulent processes. This allows one to split the distribution function into a stationary analytic control variate~\cite{Parker1993,Aydemir1994,Hu1994} $f_0$ and a time-dependent deviation part $\delta f$, which is represented by numerical markers. This approach is referred to as the delta-f PIC method and is to be contrasted with the full-f PIC scheme which represents the whole $f$ in terms of markers. The gain in noise reduction of the delta-f scheme relies on the reduced variance of the marker weights, provided that the assumption $|\delta f|/|f_0|\ll1$ for some definition of the norm is met.

However, the plasma edge involves steep profile gradients and low density levels, which often leads to conditions with fluctuation levels as large as the background such that the assumption of the delta-f scheme will not be met. One could fall back to the full-f scheme, which entails using high marker numbers to achieve similar low noise levels as the delta-f scheme in the core. As marker numbers typically need to be at least at the order of $10$ per grid cell for an adequate simulation of the core of a medium-sized tokamak plasma, like that of the Tokamak a Configuration Variable (TCV) at EPFL~\cite{Villard2019}, larger marker numbers per grid cell may exceed the computational limits of most computers, even more so for particularly large plasma volumes like that of ITER. In order to still possibly retain some advantage of the delta-f scheme, one could also evolve $f_0$, albeit at a longer time-scale than that of the fluctuating $\delta f$. This approach has for example been suggested in references~\cite{Brunner1999,Allfrey2003,Ku2016}. This work explores the implications of a specific implementation of this approach. Namely, to have a time-evolving background by constraining $f_0$ to be a flux-surface-dependent Maxwellian which furthermore time-dependent via its evolving temperature profile. It will be shown that this adaptive control variate scheme is always effective in reducing the statistical sampling noise, especially in situations where turbulence saturation is controlled by the zonal flow shearing rate. \amend At time of writing, a similar work~\cite{Hager2022} is being done as a follow-up to Ref.~\onlinecite{Ku2016}. \color{black}

Another source of statistical sampling noise is due to `weight-spreading'~\cite{Brunner1999,Chen1997} as a result of the implementation of collision operators using a Langevin approach. However, this problem will not be addressed in this work as collisions are not considered. Though the collision-less limit is valid when simulating core plasmas which are weakly collisional, collisions are important when simulating plasma edge, the same region in which the adaptive delta-f scheme just mentioned could be useful.

This paper is organized as follows. A brief presentation of the slab-like model considered in the GKengine code used in this study is made in Section~\ref{sec:physical}. The numerical methods are then discussed in Section~\ref{sec:numerical}, in particular the adaptive delta-f PIC scheme, which is the main focus of this paper. This is followed in Section~\ref{sec:profiles} by a discussion of the profiles~\cite{Murugappan2021} and parameters used to demonstrate the feasibility and utility of the scheme. Various aspects of noise reduction involving the adaptive scheme are discussed in turn, with key diagnostics explained as they appear. This paper concludes with a discussion on the effectiveness of the adaptive scheme in high-flux and high-fluctuation level scenarios achieved by tuning the flux-surface-averaged (f.s.a.)~potential term $\langle\phi\rangle$ of the quasi-neutrality equation. As this work is a contribution towards the development of the tokamak edge/scrape-off-layer (SOL) code PICLS~\cite{Boesl2019}, this tuning emulates edge conditions when simulating slab-ITG-driven turbulence subject to strong zonal shear flow stabilisation.

\section{Physical model}\label{sec:physical}

\begin{figure}[H]
	\centering
	\includegraphics[width=0.6\linewidth]{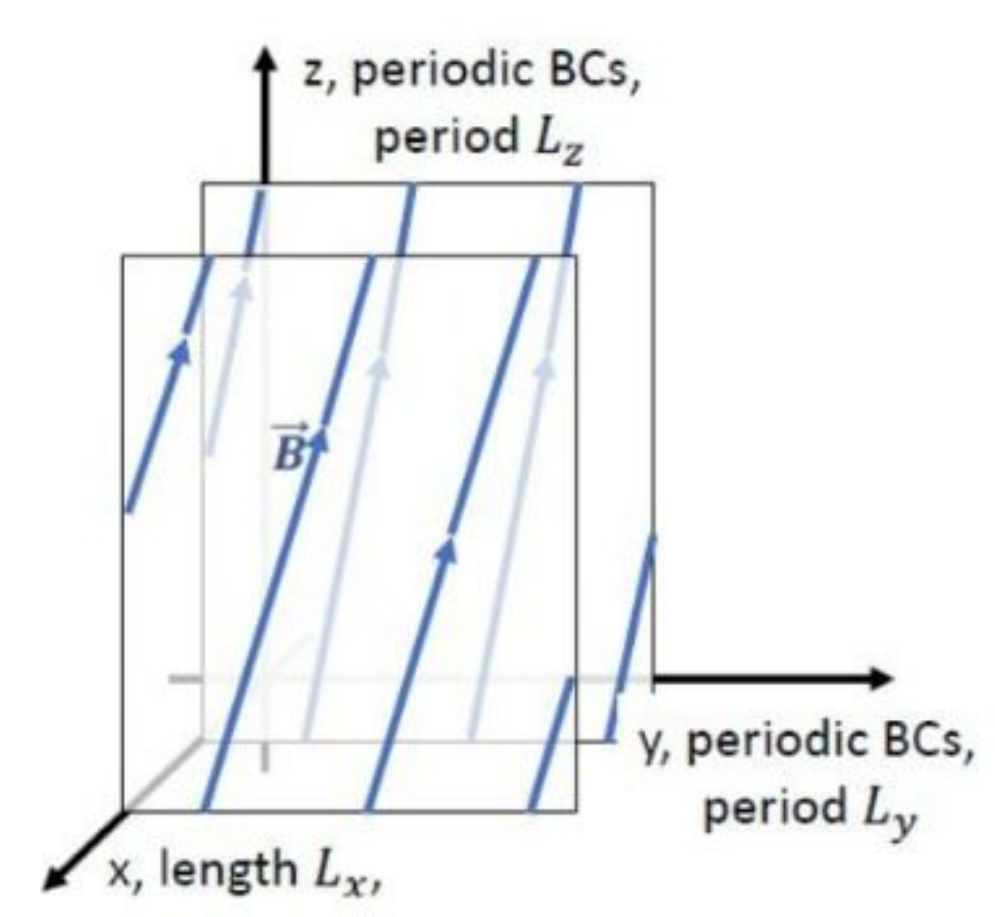}
	\caption{The sheared-slab magnetic geometry (see Eq.~(\ref{eq:B})) used in this work.}
	\label{fig:Bgeometry}
\end{figure}

All simulations carried out in the frame of this work are run with the GKengine code~\cite{Ohana2020,Ohana2016}, which solves for the distribution function $f$ of the single kinetic singly charged $Z=1$ ion species via the gyrokinetic equation
\begin{eqnarray}
\tp{f} &=& S \label{eq:vlasov},
\end{eqnarray}
where $S$ is a general source term. The electrons on the other hand are assumed to be adiabatic. A review of such a scenario is given for example in \cite{Bottino2015}. The left-hand side of Eq.~(\ref{eq:vlasov}) reads~\cite{Tronko2018}
\begin{eqnarray}
\tp{f} &=& \left\{\pp{}{t} + \tp{\vec{R}}\cdot\nabla_{\vec{R}} + \tp{\vp}\pp{}{\vp} + \tp{\mu}\pp{}{\mu}\right\}f \label{eq:dfdt_gk},
\end{eqnarray}
with $f$ function of the $5$D gyrocentre phase space coordinates $(\vec{R},\vp,\mu)$. Specifically, $\vec{R}=[X,Y,Z]$ is the gyrocentre real space position, $\vp$ its parallel velocity, and $\mu$ its magnetic moment, given by $\mu=m_iv_\perp^2/2B$, with $m_i$ is the ion mass, $v_\perp$ its gyrating velocity, and $B$ the strength of the local magnetic field $\vec{B}=B\hat{b}$, with $\hat{b}$ its direction. Two further assumptions are made. Firstly, the geometry we considered is that of sheared-slab in a spatial $3$D domain that spans $(x,y,z)\in[0,L_x]\times[0,L_y]\times[0,L_z]$, where the Cartesian coordinates $(x,y,z)$ can be related to the radial, poloidal and toroidal directions respectively in a tokamak system. Periodic boundary conditions are thus imposed in the $y$- and $z$- directions, which span $L_y=\pi a$ and $L_z=2\pi R_0$ respectively, with $a$ the minor radius and $R_0$ the major radius. Fig.~\ref{fig:Bgeometry} illustrates the sheared magnetic field, given by
\begin{eqnarray}
\vec{B}(x) &=& B_y(x)\textbf{e}_y + B_z\textbf{e}_z \nonumber \\
&=& B_z\left[\frac{L_y}{L_zq(x)}\textbf{e}_y + \textbf{e}_z\right] \label{eq:B},
\end{eqnarray}
where $q(x)$ is the safety factor profile.
The second assumption is that all fluctuations considered are electrostatic. Under these simplifications, the evolution of gyrocentre phase space coordinates is governed by~\cite{Ohana2016}
\begin{eqnarray}
\tp{\vec{R}} &=& \vp\hat{b} + \frac{\mu}{eB_\parallel^\star}\hat{b}\times\nabla B + \frac{1}{B_\parallel^\star}\hat{b}\times\nabla\tilde{\phi}; \nonumber \\
\tp{\vp} &=& -\frac{e}{m_i}\hat{b}\cdot\nabla\tilde{\phi}; \label{eq:coord_gk} \\
\tp{\mu} &=& 0. \nonumber
\end{eqnarray}
Here, $e$ is the ion charge, $B_\parallel^\star=B[1+m_iB_y'B_z\vp/(eB^3)]$, $\phi$ the electric potential, and $\tilde{\cdot}$ the gyro-averaging operator, given by
\begin{eqnarray}
\tilde{\phi}(\vec{R},\mu) &=& \frac{1}{2\pi}\int_0^{2\pi}\dint{\alpha}\,\phi(\vec{R}+\vec{\rho_L}(\mu,\alpha)), \nonumber
\end{eqnarray}
with $\vec{\rho}_L$ the Larmor vector with radius $\rho_L=\sqrt{2m_i\mu/B}$, and $\alpha$ the gyrophase. The set of Eqs.~(\ref{eq:coord_gk}) are nonlinear as these equations depend on the self-consistent electrostatic potential $\phi(x,y,z)$ satisfying the quasi-neutrality equation with $Z=1$,
\begin{eqnarray}
& & \frac{en_{e0}}{T_e}(\phi-\langle\phi\rangle)-\nabla_\perp\cdot\left(\frac{m_in_{i0}}{eB^2}\nabla_\perp\phi\right) \nonumber \\
&=& \int\dint{\alpha}\dint{^3R}\dint{\vp}\dint{\mu}\,\left\{\frac{B_\parallel^\star}{m_i}f(\vec{R},\vp,\mu,t)\delta[\vec{R}+\vec{\rho}_L(\mu,\alpha)-\vec{r}]\right\} - n_{e0}, \nonumber \\
& & \label{eq:qn}
\end{eqnarray}
where $n_{i0}$ is the background ion gyrocentre density, and $n_{e0}$ and $T_{e}$ are the background electron density and temperature profiles respectively, satisfying a local Maxwellian distribution due to the adiabatic assumption. Here, the $B_\parallel^\star/m_i$ term represents the Jacobian for the coordinate transformation from particle variables $(\vec{r},\vec{v})$ to gyrocentre variables $(\vec{R},\vp,\mu)$, and $\langle\cdot\rangle$ is the flux-surface-averaging operator defined by
\begin{eqnarray}
\langle\phi\rangle(x) &=& \frac{1}{L_yL_z}\int_0^{L_y}\dint{y}\int_0^{L_z}\dint{z}\,\phi(x,y,z). \nonumber
\end{eqnarray}

Looking at Eq.~(\ref{eq:qn}), the first term on the left-hand side represents the \amend linearized \color{black} adiabatic electron response and the second term represents the ion polarisation density in the long wavelength approximation $k_\perp\rho_{th}\ll1$, with $\rho_{th}$ the thermal ion Larmor radius, which together with the right-hand side representing the ion density fluctuation.

In order to simulate physics under strong profile gradients in quasi-steady state, heat sources are implemented to clamp ion temperature $T_i$ at profile edges to prevent relaxation below critical gradients. This Krook-like source $S_h$ with associated relaxation rate $\gamma_h(x)$ is stationary and radially dependent, maintaining over time the high and low ends of the $T_i$ profile of the initial background distribution function $f_0(t=0)$. Particle sources are not needed as there is no density profile relaxation. The assumed adiabatic electron response indeed enables no particle transport. Therefore, an additional correction term $S_{h,corr}$ to $S_h$ is included to conserve density, along with parallel momentum $\vp$.

Furthermore, it is shown~\cite{McMillan2008} that the inclusion of a Krook-like noise control operator $S_n$, with  uniform relaxation rate $\gamma_n$ (usually taken to be a few percent of the maximum linear growth rate) which relaxes to a (possibly time-dependent) reference distribution $f_n$, is important in PIC simulations to achieve a converged quasi-steady state at long simulation times. A corresponding correction term $S_{n,corr}$ is also included to conserve density, parallel momentum $\vp$ and energy $v^2$.

Taken together, considering Eqs.~(\ref{eq:dfdt_gk}) and (\ref{eq:coord_gk}), Eq.~(\ref{eq:vlasov}) is expanded to
\begin{eqnarray}
& & \left\{\pp{}{t} + \tp{\vec{R}}\cdot\nabla_{\vec{R}} + \tp{\vp}\pp{}{\vp}\right\}f \nonumber \\
&=& -\gamma_h(x)(f-f_0(t=0)) + S_{h,corr} -\gamma_n(f-f_n) + S_{n,corr} \label{eq:gk}.
\end{eqnarray}

\section{Numerical methods} \label{sec:numerical}

\subsection{Delta-f scheme with adaptive control variate} \label{sec:delta-f}

The solution method of the GKengine employs the delta-f PIC scheme, which splits the ion distribution $f$ into an unperturbed background part $f_0$, and a perturbed part $\delta f$~\cite{Parker1993,Hu1994}. Namely,
\begin{eqnarray}
f &=& f_0 + \delta f \label{eq:delta-f}.
\end{eqnarray}

$\delta f$ therefore represents the deviation component of $f$, including in particular fluctuations. The rationale for the splitting of $f$ is that $f_0$ acts as a control variate~\cite{Aydemir1994}. As long as $|\delta f|/|f|\ll 1$, i.e. as long as the system does not deviate too much from a known $f_0$, the scheme reduces sampling noise. However, processes involving large profile gradients and high fluctuation amplitudes will result in $f$ deviating far from its initial background $f_0$. Therefore, in order to reduce sampling noise in a delta-f scheme, we now allow the control variate $f_0$ to be a time-dependent local Maxwellian. Specifically,
\begin{eqnarray}
f_0 &=& f_M(\vec{R},\vp,\mu,t) \nonumber \\
&=& \frac{n_{i0}(X)}{[2\pi T_{i0}(X,t)/m_i]^{3/2}}\exp\left\{-\frac{m_i\vp^2/2+\mu B(X)}{T_{i0}(X,t)}\right\} \label{eq:f0},
\end{eqnarray}
with $T_{i0}$ the background ion temperature profile. Note that the profiles of $n_{i0}$ and $T_{i0}$ are spatially only functions of the `radial' variable. The time dependence of $f_0$ appears only through $T_{i0}$, which is governed by an ad-hoc relaxation equation~\cite{Brunner1999} of the form:
\begin{eqnarray}
\pp{}{t}\left(\frac{3}{2}n_{i0}(x)T_{i0}(x,t)\right) &=& \alpha_E\left\langle\int\dint{\vp}\dint{\mu}\frac{2\pi B_\parallel^\star}{m_i}\delta f\left(\frac{m_i\vp^2}{2}+\mu B\right)\right\rangle, \nonumber \\
& & \label{eq:adp_relax}
\end{eqnarray}
where $\alpha_E$ is the relaxation rate, which is a constant numerical parameter. The left-hand side of Eq.~(\ref{eq:adp_relax}) represents the variation in time of the background kinetic energy density $E_{kin0}(x,t)$ related to $f_0$:
\begin{eqnarray}
E_{kin0}(x,t) &=& \frac{3}{2}n_{i0}(x)T_{i0}(x,t), \label{eq:ekin0}
\end{eqnarray}
with
\begin{eqnarray*}
T_{i0}(x,t) &=& T_{i0}(x,0) + \delta T_{i0}(x,t)
\end{eqnarray*}
and
\begin{eqnarray*}
\delta E_{kin0}(x,t) &=& \frac{3}{2}n_{i0}(x)\delta T_{i0}(x,t).
\end{eqnarray*}
Here, $\delta E_{kin0}(x,t)$ and $\delta T_{i0}(x,t)$ are the deviations of the background ion kinetic energy density and temperature profiles from their initial states $E_{kin0}(x,0)$ and $T_{i0}(x,0)$, respectively.

Finally, let us explicate the contributions from $f_0$ and $\delta f$ to the gyrodensity on the right-hand side of the quasi-neutrality equation, Eq.(~\ref{eq:qn}). Assuming the gyrodensity associated to $f_0(t=0)$ verifies quasi-neutrality:
\begin{eqnarray*}
\int\dint{^3R}\dint{\alpha}\dint{\vp}\dint{\mu}\,\frac{B_\parallel^\star}{m_i}f_0(\vec{R},\vp,\mu,0)\delta[\vec{R}+\vec{\rho}_L(\mu,\alpha)-\vec{r}] &=& n_{e0},
\end{eqnarray*}
\amend
Eq.(~\ref{eq:qn}) becomes:
\begin{eqnarray}
& &
\frac{en_0}{T_e}\left(\phi-\langle\phi\rangle\right)-\nabla_\perp\cdot\left(\frac{m_in_0}{eB^2}\nabla_\perp\phi\right) \nonumber \\
&=&\int\dint{^3R}\dint{\alpha}\dint{\vp}\dint{\mu}\,\frac{B_\parallel^\star}{m_i}\delta[\vec{R}+\vec{\rho}_L(\mu,\alpha)-\vec{r}]\times \nonumber \\
& &[f_0(\vec{R},\vp,\mu,t)-f_0(\vec{R},\vp,\mu,0) + \delta f(\vec{R},\vp,\mu,t)] \label{eq:qn_df}.
\end{eqnarray}
Here, the electron density $n_{e0}$ has been approximated to the ion background density $n_{i0}$, both denoted by $n_0$. \color{black} The perpendicular gradient $\nabla_\perp\approx\nabla_{pol}=\textbf{e}_x\partial_x+\textbf{e}_y\partial_y$ has been approximated to the gradient in the $(x,y)$-plane (corresponding to the poloidal plane in a tokamak) due to the fact that micro-instabilities align along field-lines and assuming $B_y/B_z=L_y/(L_zq(x))\ll1$. Since $f_0$ has an analytic form, the integral involving $f_0$ terms are calculated using Gauss-Laguerre and Gauss-Chebyshev quadratures for the $\mu$ and $\alpha$ integrations respectively. Convergence of this scheme is detailed in \amend Appendix \color{black}\ref{sec:app1}. For this work, we use $30$ quadrature points for each of these dimensions. The $\vp$ integration on the other hand can be integrated analytically.

\subsection{Numerical discretization} \label{sec:discretization}

Under the PIC scheme, $\delta f$ is represented by markers, and is given by
\begin{eqnarray}
\delta f &=& \frac{1}{2\pi} \sum_{p}^{N_p}\frac{w_p(t)}{B_\parallel^\star/m_i}\delta[\vec{R}-\vec{R}_p(t)]\delta[\vp-v_{\parallel p}(t)]\delta[\mu-\mu_p(t)], \nonumber \\
& & \label{eq:df}
\end{eqnarray}
where the denominator represents the same Jacobian as in Eq.(~\ref{eq:qn}), $N_p$ is the total number of markers, and $w_p(t)$ and $(\vec{R}_p(t),v_{\parallel p}(t),\mu_p(t))$ are the weight and the phase space position in gyrocentre variables at time $t$ of the marker with index $p$, respectively. \amend All markers are initialized in phase space using Hammersley sequences~\cite{Wong1997}, and are time-integrated using a fourth order Runge-Kutta scheme~\cite{Ohana2016}. \color{black}

Under the delta-f scheme, with reference to the discussion leading to Eq.(~\ref{eq:gk}), the $f$ conservation term, the heat source $S_h$ and noise control $S_n$ operators with their corresponding corrections, are implemented consecutively. The control variate is the chosen reference function for $S_n$, $f_n=f_0$. Designating $f_{00}=f_0(t=0)$, Eq.(~\ref{eq:gk}) is split into three equations:
\begin{eqnarray}
\tp{\delta f^{(0)}} &=& -\tp{f_0} \label{eq:bg_term} \\
\tp{\delta f^{(1)}} &=& -\gamma_h(x)(f_0(t)-f_{00}+\delta f^{(0)}) + S_{h,corr} \label{eq:heat_term} \\
\tp{\delta f^{(2)}} &=& -\gamma_n\delta f^{(1)} + S_{n,corr}, \label{eq:noise_term}
\end{eqnarray}
with $\delta f^{(0)}(t) = \delta f(t)$ and $\delta f^{(2)}$ is assigned to $\delta f$ once these equations are solved. Eqs.~(\ref{eq:bg_term}), (\ref{eq:heat_term}) and (\ref{eq:noise_term}) are discretized by multiplying by the phase space volume $\Omega_p$ associated to the marker with index $p$, which remains constant along characteristics due to the collisionless physics under study, and are each integrated over the full time step $\Delta t$ considered for integrating the marker trajectories. Eq.~(\ref{eq:bg_term}) is first solved using the `direct-$\delta f$ approach'~\cite{Allfrey2003}.

Focusing on Eqs.~(\ref{eq:heat_term}) and (\ref{eq:noise_term}), let us denote $(1,\vp,v^2)$ to represent the density, parallel velocity and energy moments respectively. Then, the terms $S_{h,corr}$ and $S_{n,corr}$ are the corrections that are necessary to ensure that $\delta f$ conserves $(1,\vp)$ and $(1,\vp,v^2)$ with respect to $S_h$ and $S_n$ respectively. Indeed, these different conservations motivate the consecutive implementation of Eq.~(\ref{eq:gk}). The correction term $S_{corr}$ of either $S_{h,corr}$ or $S_{n,corr}$ is expressed as~\cite{McMillan2008}
\begin{eqnarray}
S_{corr} &=& \sum_c g_c(x)f_0(x,\vp,\mu,t)G_c, \label{eq:Scorr}
\end{eqnarray}
where $c\in\{0,1,2\}$, and $G_c\in\{1,\vp,v^2\}$, terms of which are included when their conservation is required. Here, $f_0$ is the control variate, which is the adaptive background local Maxwellian, Eq.~(\ref{eq:f0}).

Here, we demonstrate the implementation of the heat source. Since Eq.~(\ref{eq:Scorr}) is time-independent within a time step, Eq.~(\ref{eq:heat_term}) is first analytically integrated over $\Delta t$ to give the $p$-indexed marker weight change:
\begin{eqnarray*}
	\Delta w_p &=& (e^{-\gamma_{hj}\Delta t}-1)\times \\
	& & \left[w_p+\Omega_p(f_0-f_{00})_p-\frac{\Omega_pf_{0p}}{\gamma_{hj}}(g_{0j} + g_{1j}v_{\parallel p})\right],
\end{eqnarray*}
where $f_{0p}$ is a shorthand for $f_0$ evaluated at the phase point $(\vec{R},\vp,\mu)$ of the marker with index $p$, the subscript $j$ represents the profile evaluation at the centre of the $j$-th radial bin of width $\Delta x$, and having made use of the relation $\Omega_p\delta f_p=w_p$. Then, simultaneous conservation of $(1,\vp)$ in each $j$-th radial bin leads to the $2$-by-$2$ linear system with which the coefficients $g_{0j}$ and $g_{1j}$ are to be solved,
\begin{eqnarray*}
	& & \sum_{x_p\in[x_j,x_j+\Delta x]} \Omega_pf_{0p}
	\begin{bmatrix}
		1 & v_{\parallel p} \\
		v_{\parallel p} & v_{\parallel p}^2
	\end{bmatrix} \begin{bmatrix}
	g_{0j} \\
	g_{1j}
\end{bmatrix} \\
    &=& 
\sum_{x_q\in[x_j,x_j+\Delta x]} \gamma_{hj} \begin{bmatrix}
	\Omega_q(f_0-f_{00})_q + w_q\\
	\Omega_q(f_0-f_{00})_q v_{\parallel q} + w_q v_{\parallel q}
\end{bmatrix}.
\end{eqnarray*}
The implementation of the noise control operator, Eq.~(\ref{eq:noise_term}) is done analogously, which involves a $3$-by-$3$ linear system due to the simultaneous conservation of $(1,\vp,v^2)$.

Finally, under the adaptive scheme, the term on the right-hand side of Eq.~(\ref{eq:bg_term}) now includes the change in $f_0$ associated with the background ion temperature $T_{i0}(x,t)$ adaptation, which is derived via the background ion internal energy density deviation $\delta E_{kin0}(x,t)$ of Eq.~(\ref{eq:ekin0}). This field is represented with finite element cubic B-spline basis functions $\Lambda_i(x)$:
\begin{eqnarray}
\delta E_{kin0}(x,t) &=&  \frac{3}{2}n_{i0}(x)\delta T_{i0}(x,t) = \sum_k\xi_k(t)\Lambda_k(x) \label{eq:xi}.
\end{eqnarray}
The time-dependent coefficients $\xi_k(t)$ are obtained by first projecting Eq.~(\ref{eq:adp_relax}) on the same B-spline basis functions. Combining Eqs.~(\ref{eq:adp_relax}), (\ref{eq:ekin0}), (\ref{eq:df}) and (\ref{eq:xi}), the time derivatives $\dot{\xi}_k(t)$ are retrieved by back-solving
\begin{eqnarray}
\sum_k\dot{\xi}_k(t)M_{kj} &=& \alpha_E\sum_p^{N_p}w_p\Lambda_j(X_p)\left(\frac{m_iv_{\parallel p}^2}{2}+\mu_p B(X_p)\right)\label{eq:dxi_dt},
\end{eqnarray}
where $M_{kj} = \int_0^{L_x}\dint{x}\Lambda_k(x)\Lambda_j(x)$
are the mass matrix elements,
with $\xi_k(0)=0, \forall k$ as initial condition.

The terms in the parenthesis of Eq.~(\ref{eq:dxi_dt}) motivates the adaption of $E_{kin0}$ as it is the $v^2$-moment of the distribution function, contrary to $T_{i0}$ which is a derived measure, i.e. $2E_{kin0}/3n_{i0}$. Having solved Eq.~(\ref{eq:dxi_dt}) for $\dot{\xi}(t)$, these coefficients are then integrated in time by applying a first order accurate forward Euler scheme with fixed time step $N_\alpha\Delta t$, where $N_\alpha$ is a user-defined fixed integer. We have
\amend
\begin{eqnarray}
\alpha_EN_\alpha\Delta t &\leq& 2 \label{eq:Nalpha}
\end{eqnarray}
\color{black}
for all simulations performed in this paper. \amend Violation of Eq.~(\ref{eq:Nalpha}) has shown to lead to numerical instability. \color{black}

Obviously, marker weights $w_p$ need to be adapted if the background distribution $f_0$ is adapted as a result of applying Eqs.~(\ref{eq:xi}) and (\ref{eq:dxi_dt}). In the following lines of this section, let $\Delta$ represent the change before and after the adaptation of $f_0$. Then, invoking the invariance of the total distribution $f$, from Eq.~(\ref{eq:delta-f}) one obtains:
\begin{eqnarray}
0 = \Delta f_0 + \Delta\delta f \hspace{5mm}\Leftrightarrow\hspace{5mm} \Delta\delta f = -\Delta f_0, \nonumber
\end{eqnarray}
and after evaluating this relation at a marker position and multiplying by $\Omega_p$:
\begin{eqnarray}
\Delta w_p &=& -\Omega_p\Delta f_{0p}, \label{eq:dw_adp}
\end{eqnarray}
having made use of $\Delta\Omega_p=0$.

\section{Profiles and simulation parameters} \label{sec:profiles}

Let the normalized radial coordinate be $s=x/L_x\in[0,1]$. \amend Then, the profiles representing the initial unperturbed ion and electron background densities, $n_0$, \color{black} as well as their temperatures $T_{i0}$ and $T_{e0}$, respectively, are parameterized by three parameters given by the amplitude $A$, the normalized absolute maximum logarithmic gradient $\bar{\kappa}$, and the slope half-width $\bar{\Delta}$. For $s\in[0,0.5]$, such a profile $g(s)$ is given explicitly by
\begin{eqnarray}
g(s;\kappa,\Delta) &=& \begin{cases}
A\exp\left(\frac{2\bar{\kappa}\bar{\Delta}}{3}\right) & 0\le s<s_0-\bar{\Delta} \\
A\exp\left[-\bar{\kappa}(s-s_0)+\frac{\bar{\kappa}(s-s_0)^3}{3\bar{\Delta}^2}\right] & |s-s_0|\leq \bar{\Delta} \\
A\exp\left[-\frac{2\bar{\kappa}\bar{\Delta}}{3}\right] & s_0+\bar{\Delta}<s\le 0.5 \\
\end{cases} \nonumber \\
& &
\end{eqnarray}

Note that this definition ensures that $g(s)$ and $\frac{\dint{g}}{\dint{s}}$ are both continuous. At the reference radial position $s=s_0$, it has a value of $A$, and its parabolic normalized logarithmic gradient peaks at this same point with value $\bar{\kappa}$.

The heat source radial profile $\gamma_h$ is parameterized by the amplitude $A_h$, and the half-widths $\bar{\delta}_c$ and $\bar{\delta}_s$ of the clamp maximum and edge-slope regions respectively, with $\bar{\delta}_c\geq\bar{\delta}_s$. This profile for $s\in[0,0.5]$ is given explicitly by
\begin{widetext}
\begin{eqnarray}
\gamma_h(s;\bar{\delta}_c,\bar{\delta}_s) &=& \begin{cases}
A_h & 0\le s< \bar{\delta}_c-\bar{\delta}_s \\
\frac{A_h}{2}\left[1 - \frac{3}{2}\left(\frac{s-\bar{\delta}_c}{\bar{\delta}_s}\right) + \frac{1}{2}\left(\frac{s-\bar{\delta}_c}{\bar{\delta}_s}\right)^3\right] & |s-\bar{\delta}_c|\le\bar{\delta}_s \\
0 & \bar{\delta}_c+\bar{\delta}_s < s\leq \frac{1}{2}-(\bar{\delta}_c+\bar{\delta}_s) \\
\frac{A_h}{2}\left[1 + \frac{3}{2}\left(\frac{s-(1/2-\bar{\delta}_c)}{\bar{\delta}_s}\right) - \frac{1}{2}\left(\frac{s-(1/2-\bar{\delta}_c)}{\bar{\delta}_s}\right)^3\right] & \left|s-\left(\frac{1}{2}-\bar{\delta}_c\right)\right|\leq\bar{\delta}_s \\
A_h & \frac{1}{2}-(\bar{\delta}_c-\bar{\delta}_s)< s\le 0.5.
\end{cases} \nonumber \\
& & \label{eq:source}
\end{eqnarray}
\end{widetext}

The noise control profile is uniform and is parameterized by as single parameter $\gamma_n=A_n$.

\begin{figure}[H]
	\centering
	\includegraphics[width=0.8\linewidth]{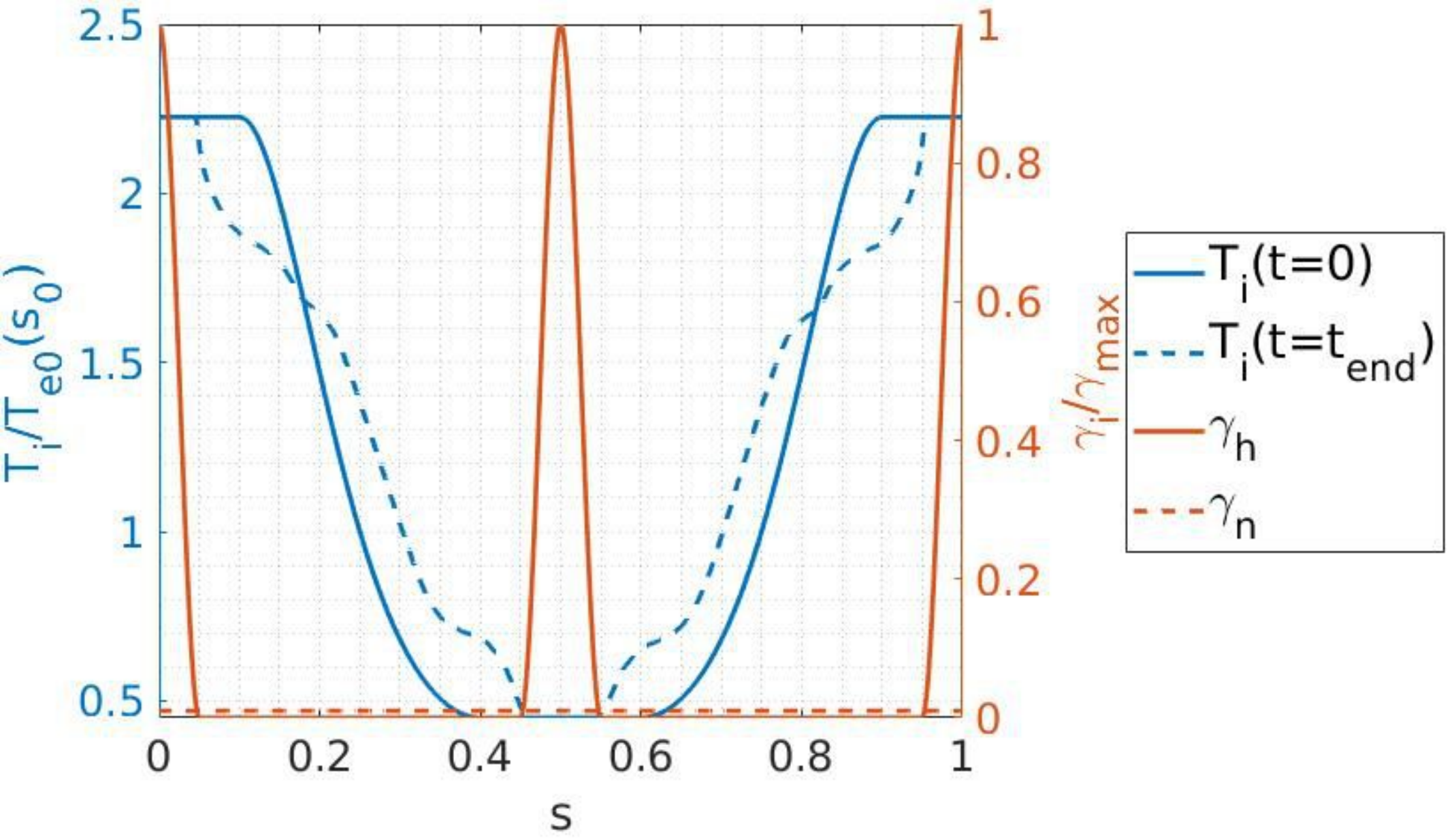}
	\caption{Symmetrized profiles used for this work. Blue: Initial (solid) and typical final (dashed) $T_i$. Red: Heat $\gamma_h$ (solid) and noise control $\gamma_n$ (dashed).}
	\label{fig:source}
\end{figure}

To avoid spurious marker build-up at the radial domain boundaries for long-time simulations with large $\kappa$ under Dirichlet boundary conditions $\phi(s=0)=\phi(s=1)=0$, all radial profiles are mirrored about $s=0.5$, and periodic boundary conditions are imposed. Examples of such profiles, along with heat source and noise control operator profiles, are shown in Fig.~\ref{fig:source}. Henceforth, only profiles from the left half will be shown, i.e. for $s\in[0,0.5]$.

\begin{figure}[H]
	\centering
	\includegraphics[width=0.7\linewidth]{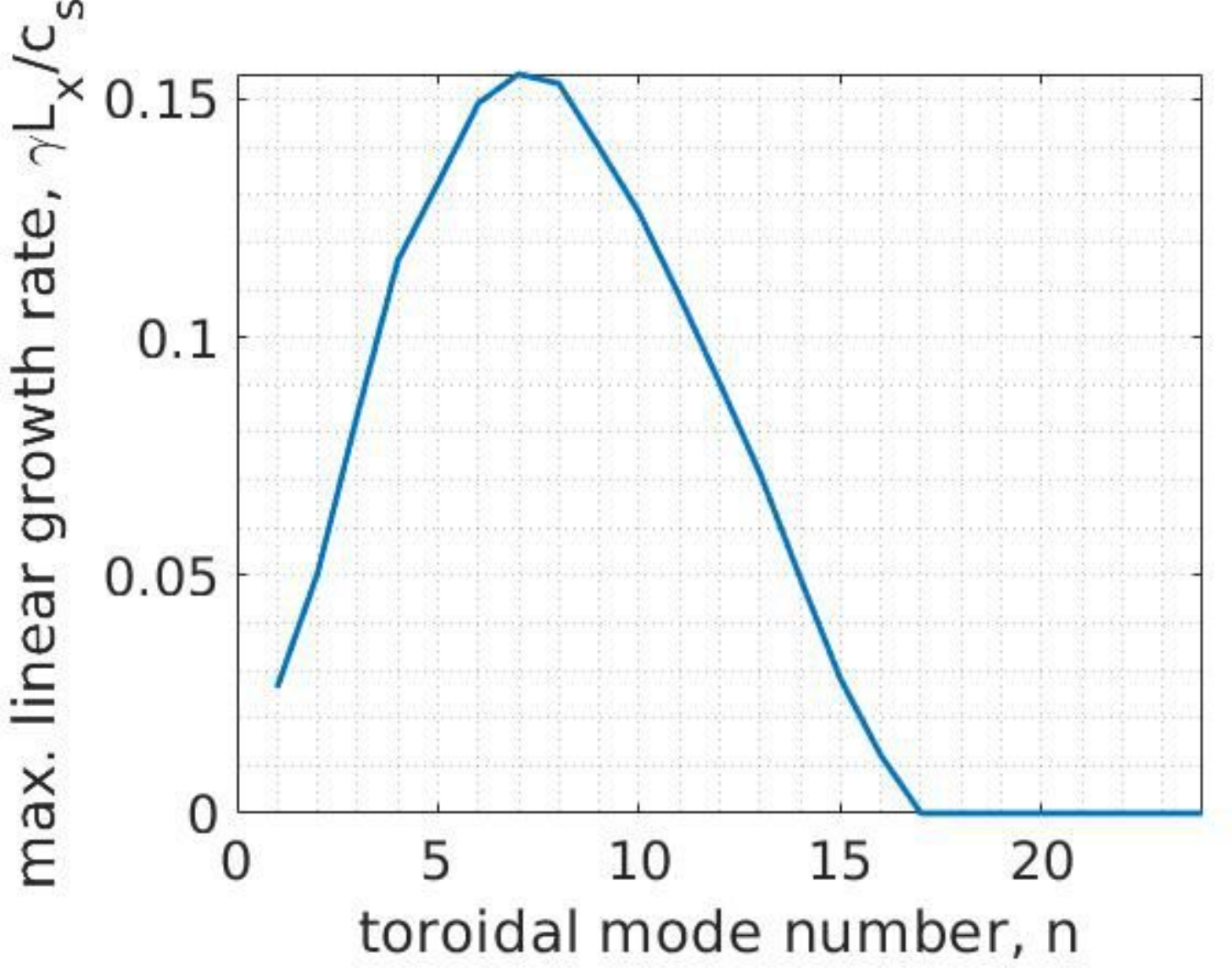}
	\caption{Linear studies: Maximum growth rate of linear modes as a function of each toroidal mode number $n$, for normalized ion temperature logarithmic gradient $a\left|\dint{\log T_i}/\dint{x}\right|=4.0$ and $(\dint{\log Ti}/\dint{x})/(\dint{\log n}/\dint{x})=10$. The poloidal mode number $m$ is radially dependent and is determined by the Fourier filter.}
	\label{fig:gamma_max}
\end{figure}

The GK-engine code works in units such that time and speeds are normalized to $\Omega_c^{-1}=m_i/eB(s_0)$ and $c_s=\sqrt{T_e(s_0)/m_i}$ ($Z=1$), representing the inverse ion cyclotron frequency and ion sound speed at $s_0=0.25$, respectively, which together gives the ion sound Larmor radius $\rho_s=c_s/\Omega_c$ for units of length. The magnetic and potential fields are normalized to $B_z$, and $T_{e0}(s_0)/e$ respectively. To simulate slab-ITG instabilities, we use the major and minor radii values of $R_0=243.5\rho_s$ and $a=66.4\rho_s$. The spatial domain is $L_x=2a$ for an $x$-periodic profile, $L_y=\pi a$ and $L_z=2\pi R_0$. The grid-cells number for $\phi$ is $(N_x,N_y,N_z)=(256,512,128)$. The time step used here is $\Delta t=20\Omega_c^{-1}=0.15L_x/c_s$. The safety factor is given by $q(s)=1.25 + 12s^2$ for the half-domain $s\in[0.0,0.5]$, and mirrored in the other half-domain. All normalized parameters describing profiles are converted to physical units via a multiplication/division by $L_x$. The initial profile gradients used in this paper are $\bar{\kappa}_n=0.8$ and $\bar{\kappa}_{Ti}=\bar{\kappa}_{Te}=8.0$, with $\bar{\kappa}_n=|\dint{\log n(s_0)}/\dint{s}|$, so that $\eta_i(s_0)=\bar{\kappa}_{Ti}/\bar{\kappa}_n=10$. \amend This value corresponds to a peak value of $T_i$ logarithmic gradient of $|\dint{\log T_i(s_0)}/\dint{x}|=4.0/a$, i.e. $R_0/L_T=14.6$. \color{black} The radial profile widths for $n_0$ and $T_{i0}$ are $\bar{\Delta}_n=0.3$ and $\bar{\Delta}_{Ti}=0.15$ respectively, and $T_{e0}(x)=T_{i0}(x,t=0)$. The maximum values $A_h$ and $A_n$ of the  profiles for heat source $\gamma_h$, and noise control $\gamma_n$, are at $100\%$ and $3\%$ of the maximum linear growth rate respectively, which is found to be $\gamma_{max}=1.169\times10^{-3}\Omega_c=0.155c_s/L_x$ for these values of gradient, as shown in Fig.~\ref{fig:gamma_max}. The radial parameters of $\gamma_h$ are $(\bar{\delta}_c,\bar{\delta}_s)=(0.025,0.025)$. The control variate at initial time is taken to be the initial background $f_0=f_M(t=0)$.

For the toroidal we chose to resolve modes in the range of $[n_{min},n_{max}]=[0,32]$, and poloidal modes $m$ determined by the field-aligned Fourier-filter $|nq(s)+m|\leq\Delta m$, with $\Delta m=5$~\cite{Jolliet2012}. All simulations are initialized with $f_{00}+\delta f(t=0)$. The initial background is taken to be $f_{00}=f_{M0}$, where $f_{M0}=f_M(\vec{R},\vp,\mu,0)$. $\delta f(t=0)$ represents a density perturbation of amplitude $10^{-4}$ and toroidal mode number $n=7$ corresponding to the strongest growing linear mode. The poloidal modes intialized are those within the field-aligned filter at $s=s_0$, i.e. $|12q(s_0)+m|\le5$.

Unless otherwise stated, all cases are run with $N_p=256$M and adaptive background cases consider the relaxation rate $\alpha_E=1.92\gamma_{max}$. The number of time steps after which the background temperature profile is adapted via Eq.~(\ref{eq:adp_relax}), is set to $N_\alpha=10$ for all cases.

\section{Results}

\subsection{Marker convergence} \label{sec:convergence}

\begin{figure}[H]
	\centering
	\begin{subfigure}[c]{1.0\columnwidth}
	    \caption{\label{fig:chi_np}}
	    \centering
		\includegraphics[width=0.7\linewidth]{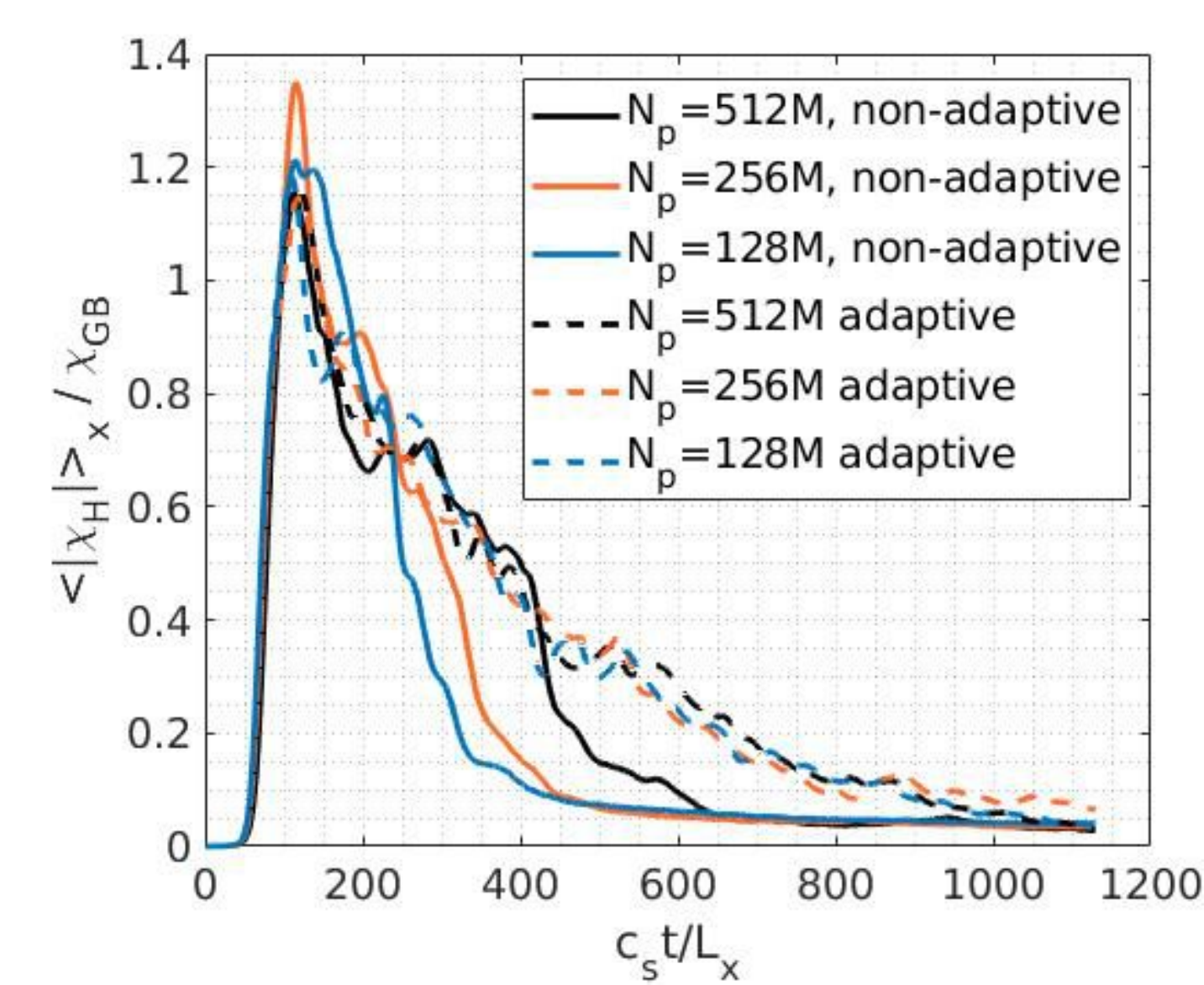}
	\end{subfigure}
	\begin{subfigure}[c]{1.0\columnwidth}
	    \caption{\label{fig:omega_np}}
	    \centering
		\includegraphics[width=0.7\linewidth]{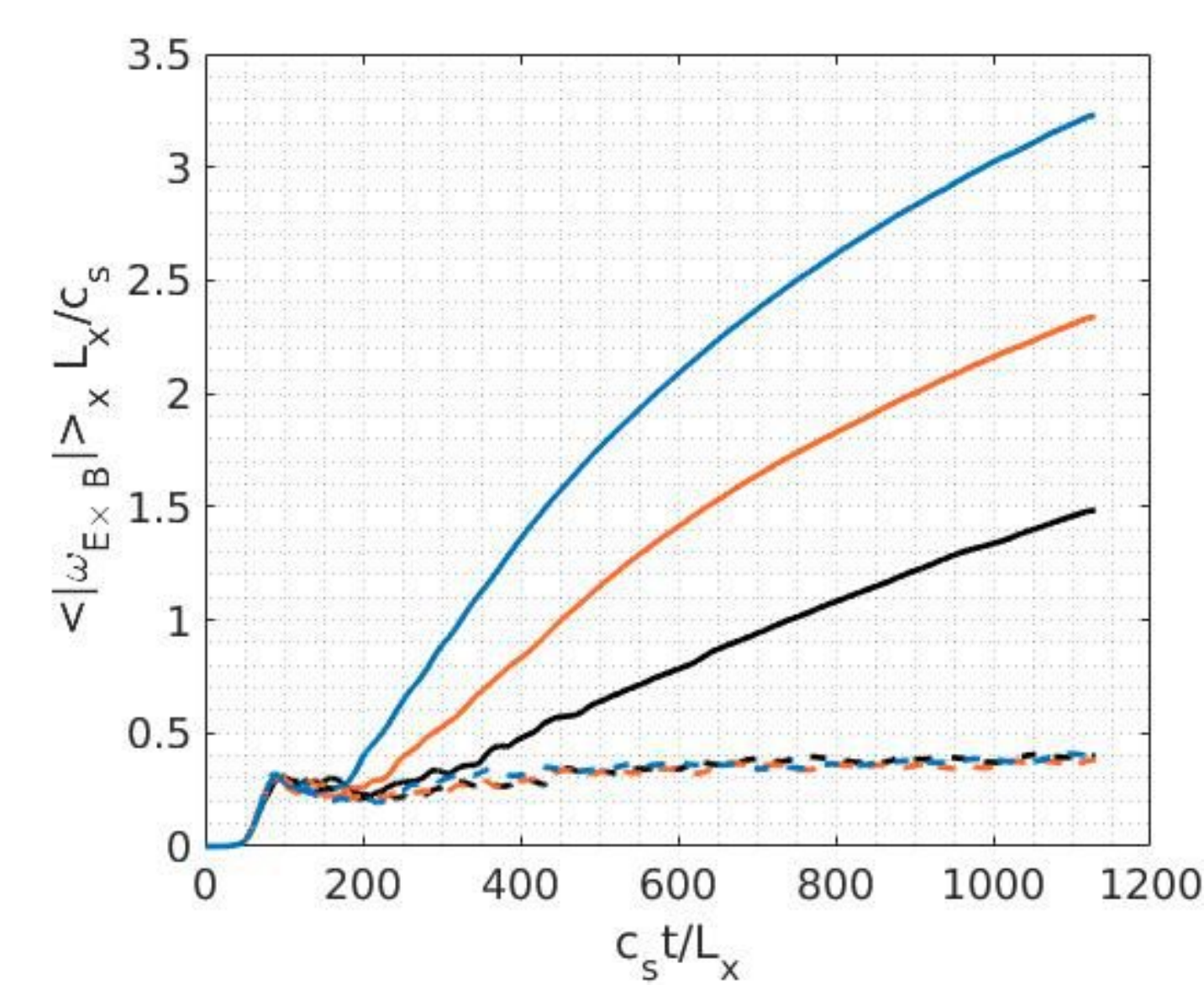}
	\end{subfigure}
	\caption{Time traces of the radially averaged absolute value of (a) heat diffusivity $\chi_H$ and (b) zonal flow shearing rate $\omega_{E\times B}$, for various marker numbers, considering the non-adaptive and adaptive cases. The adaptive rate is set to $\alpha_E=1.92\gamma_{max}$ where applicable. A moving time-averaging window of half-width $c_st/L_x=10$, which is equivalent to $\gamma_{max}t=1.6$, has been implemented.}
	\label{fig:chi_omega_np}
\end{figure}

\begin{figure*}
	\centering
	\begin{subfigure}[c]{0.24\textwidth}
	    \caption{\label{fig:reldevTi_np_adp}}
		\centering
		\includegraphics[width=1.0\linewidth]{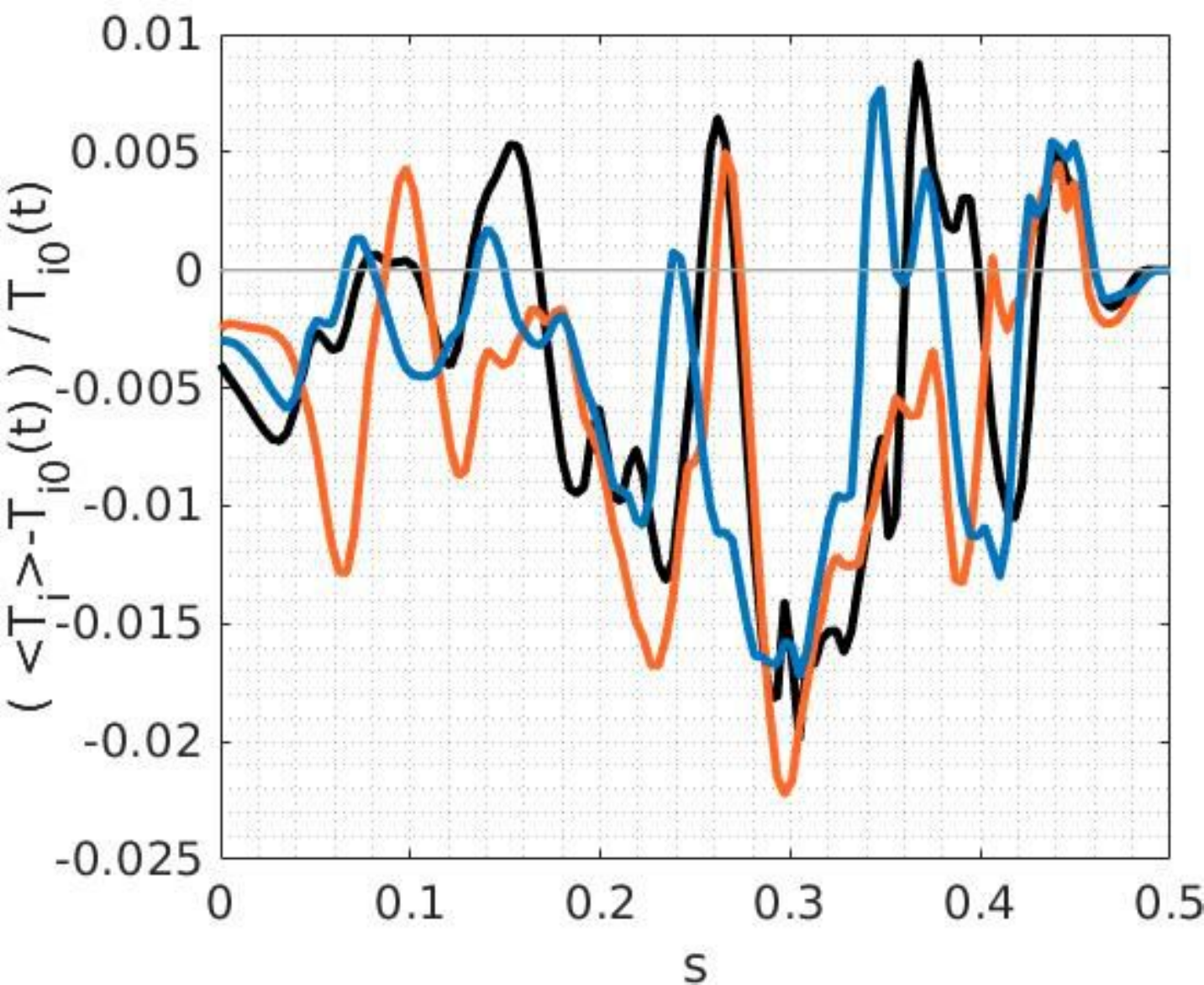}
	\end{subfigure}
	\begin{subfigure}[c]{0.24\textwidth}
	    \caption{\label{fig:reldev0Ti_np_adp}}
		\centering
		\includegraphics[width=1.0\linewidth]{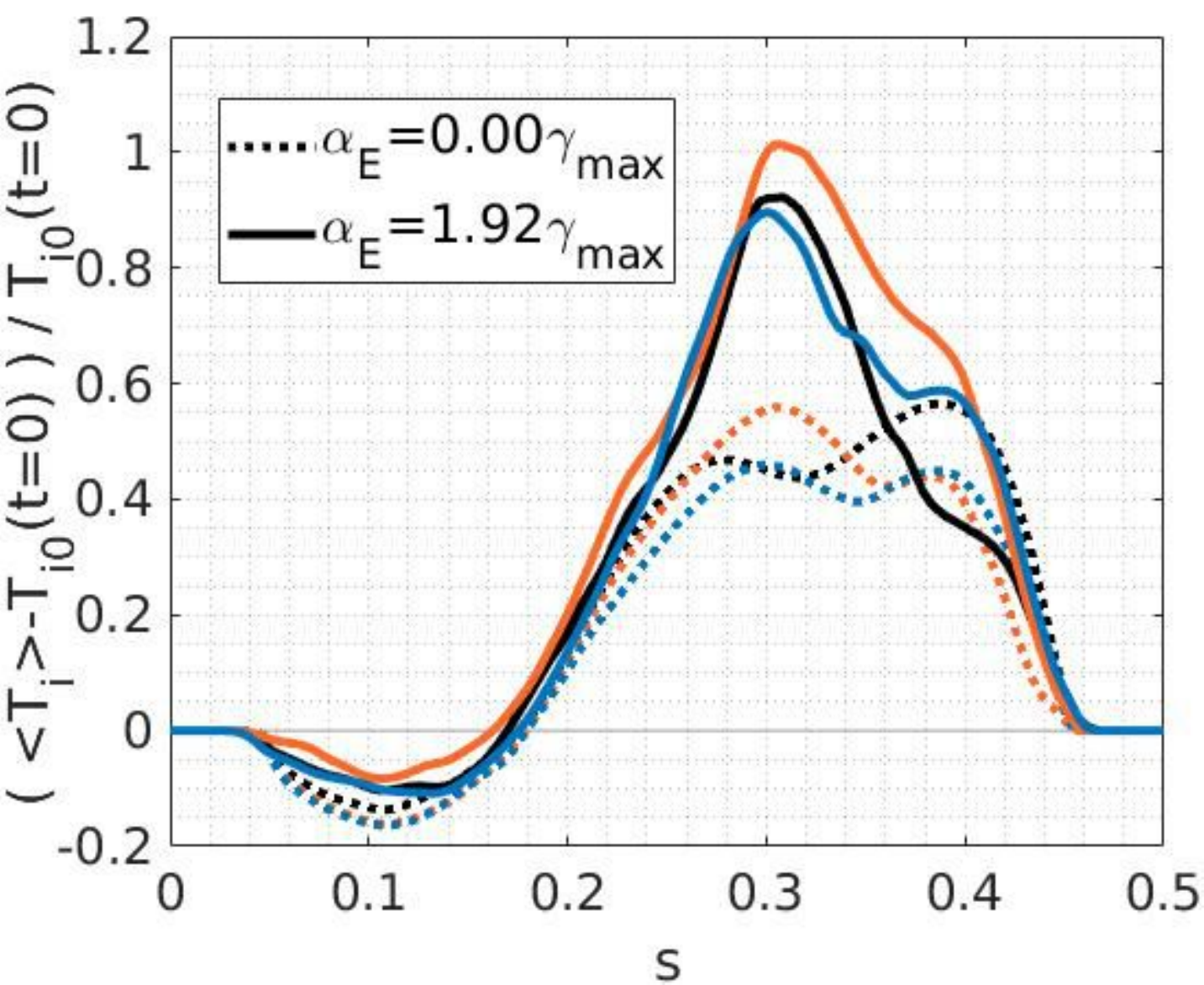}
	\end{subfigure}
	\begin{subfigure}[c]{0.24\textwidth}
	    \caption{\label{fig:kappa_np_adp}}
		\centering
		\includegraphics[width=1.0\linewidth]{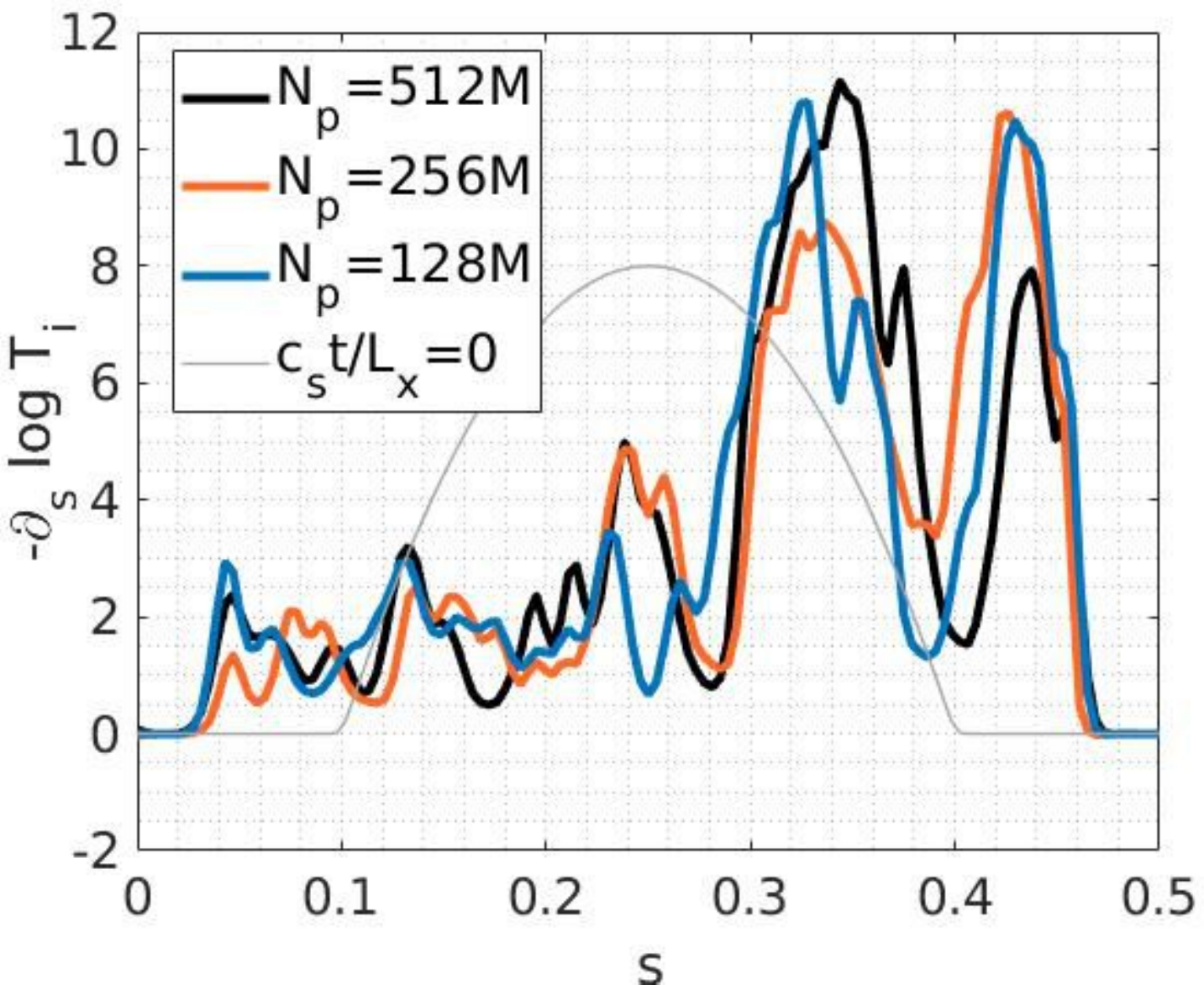}
	\end{subfigure}
	\begin{subfigure}[c]{0.24\textwidth}
	    \caption{\label{fig:shear_np_adp}}
		\centering
		\includegraphics[width=1.0\linewidth]{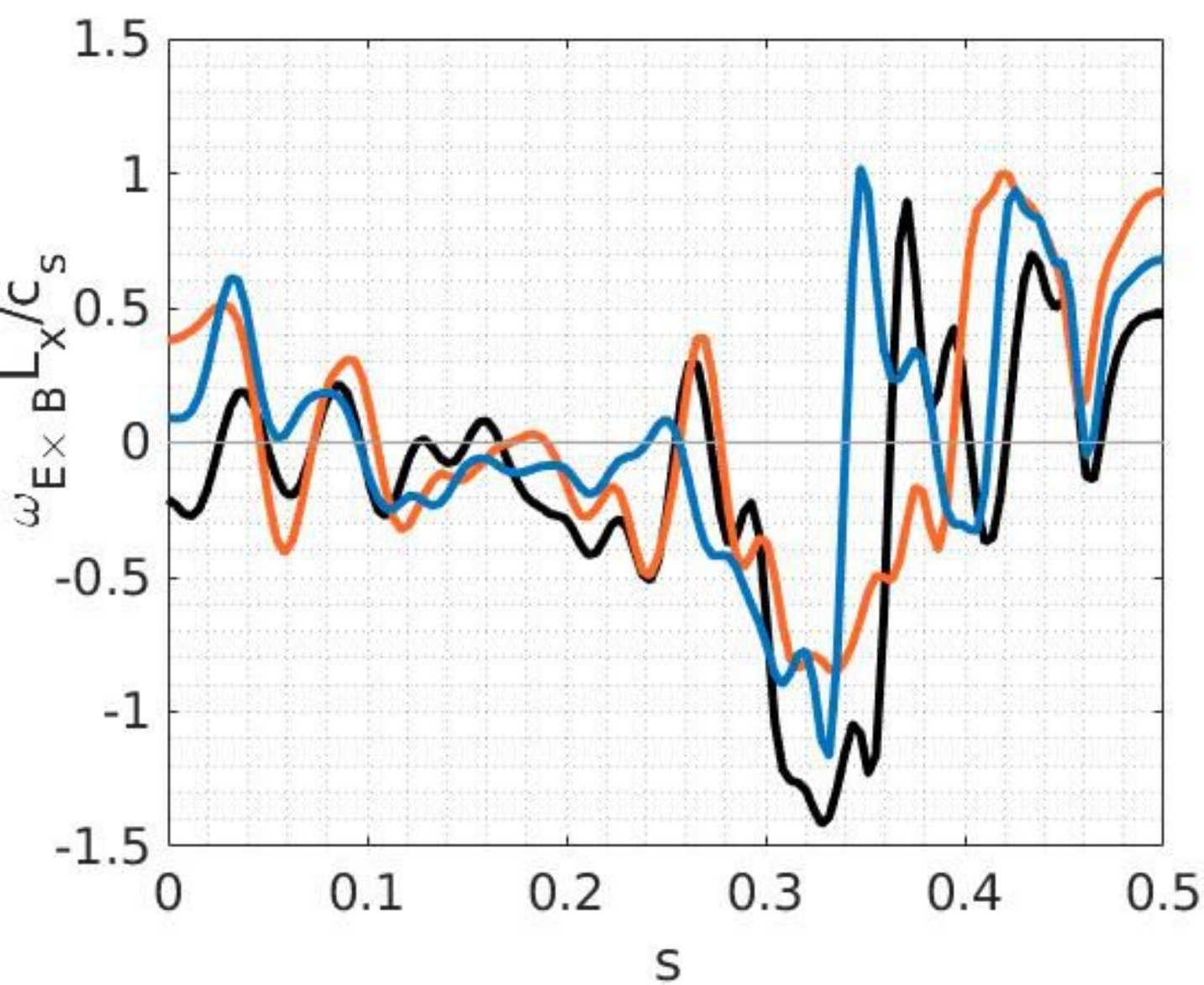}
	\end{subfigure}
	\caption{F.s.a.~profiles at quasi-steady state averaged over a time interval of $c_st/L_x\in[1030,1130]$ for various marker numbers $N_p$ under the adaptive scheme with adaptive rate $\alpha_E=1.92\gamma_{max}$ for the ion temperature (a) relative deviation with respect to adapted background, (b) relative deviation with respect to background at initial time $t=0$, its (c) logarithmic gradient, and the (d) zonal flow shearing rate.}
	\label{fig:prof_np_adp}
\end{figure*}

Characteristic of all simulations is a turbulent burst in the initial phase of the simulation ($0<c_st/L_x<300$) represented by a spike in the radially averaged heat flux $q_H$ and diffusitivity $\chi_H$, expressed by
\begin{eqnarray}
\chi_H(t) &=& \left\langle\left|\frac{q_H(x,t)}{n_{i0}(x)\pp{T_i(x,t)}{x}}\right|\right\rangle_x \nonumber.
\end{eqnarray}
$\chi_H$ is represented in gyro-Bohm units, $\chi_{GB}=\rho_s^2c_s/L_x$ with reference to the radial position of steepest initial profile gradient, i.e. $s=s_0$. \amend All radial averaging in this paper is done outside the heat source profile of Eq.~(\ref{eq:source}), namely $s\in[0.025,0.475] \cup [0.525,0.975]$. \color{black} Fig.~\ref{fig:chi_omega_np} shows how turbulence is quenched for the non-adaptive cases by increasing zonal flow shearing rate $\omega_{E\times B}$. The latter is
estimated by
\begin{eqnarray}
\omega_{E\times B} &=&  \frac{1}{B}\frac{\text{d}^2\phi_{00}}{\text{d}x^2}, \label{eq:omega}
\end{eqnarray}
where, $\phi_{mn}(x)$ is in general the Fourier component of the electrostatic field corresponding to poloidal ($y$) mode $m$ and toroidal ($z$) mode $n$, $\phi_{00}(x)$ thus being the zonal component. As can be seen in Fig.~\ref{fig:omega_np}, the very significant rise in $|\omega_{E\times B}|$ for non-adaptive simulations is dependent on the number of markers, $N_p$, and is thus of numerical origin. We note that even though the rise is reduced by increasing $N_p$, the simulation is far from having converged even for the largest $N_p=512$M considered. We interpret the rise in $|\omega_{E\times B}|$ as resulting from the statistical sampling noise accumulation in the zonal components, which are not physically damped~\cite{Diamond2005}. As expected, the accumulated noise is highest for the case with lowest $N_p$. Corresponding un-physically large $\omega_{E\times B}$ levels lead to large eddy shearing and reduced transport. The time of $\chi_H$ collapse is correlated to $|\omega_{E\times B}|$, reaching a value comparable to $3\gamma_{max}L_x/c_s$, thus the sequence of rises of $\omega_{E\times B}$ in Fig.~\ref{fig:omega_np} correspond to the sequence of falls of $\chi_H$ in Fig.~\ref{fig:chi_np} for the three non-adaptive cases. On the other hand, for the adaptive case, the converged results show that $\omega_{E\times B}$ increases at a much slower rate with time, as was already shown clearly in Fig.~\ref{fig:shear_np_adp} compared to \ref{fig:chi_np}, resulting in a somewhat longer sustained flux. One notes that these converged fluxes nonetheless ultimately drop to zero as seen in Fig.~\ref{fig:chi_np}.

\begin{figure}[H]
	\centering
	\begin{subfigure}[c]{0.49\columnwidth}
        \caption{\label{fig:omega_init}}
	    \centering
		\includegraphics[width=1.0\linewidth]{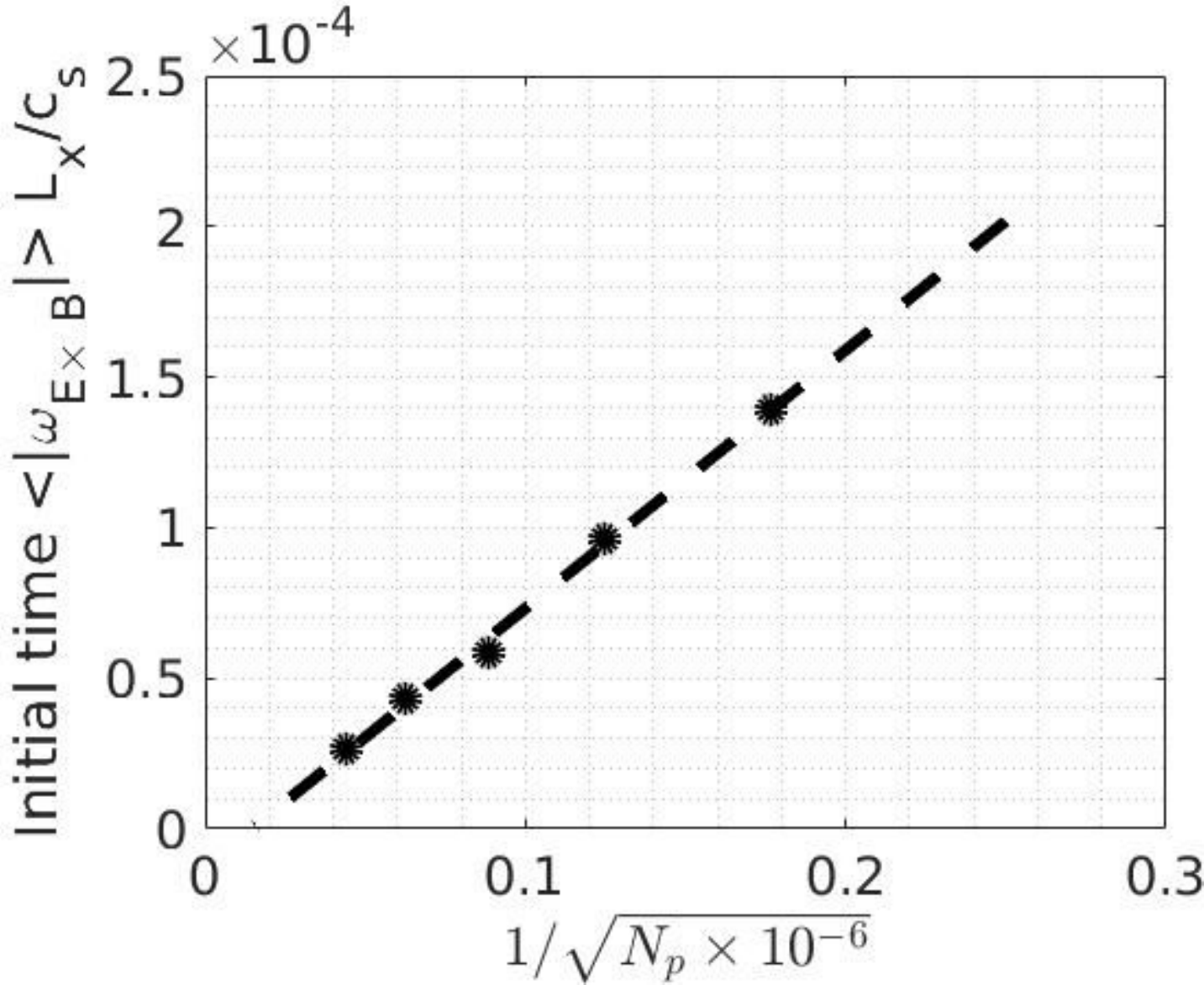}
	\end{subfigure}
	\begin{subfigure}[c]{0.49\columnwidth}
	    \caption{\label{fig:omega_end}}
		\centering
		\includegraphics[width=1.0\linewidth]{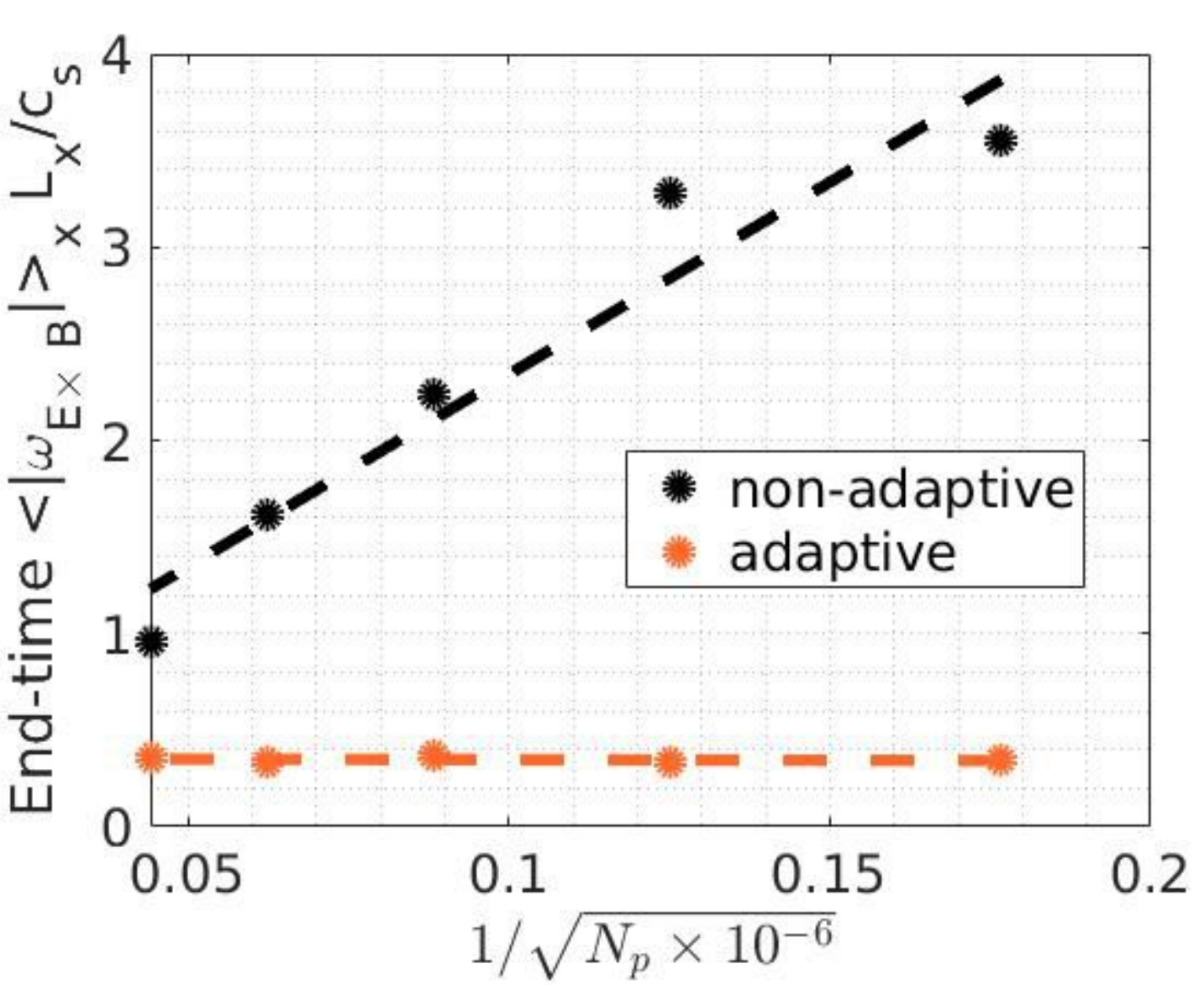}
	\end{subfigure}
	\caption{Radial averaged absolute value of the zonal flow shearing rate $\omega_{E\times B}$ at (a) initial time $c_st/L_x=0$ and (b) end-time $c_st/L_x=753$, as a function of the inverse square root of marker number $N_p$. The adaptive case (orange) adapts at a rate $\alpha_E=1.92\gamma_{max}$. All simulations are initialized with perturbations with toroidal mode $n\ne0$.}
	\label{fig:omega_time}
\end{figure}

To further confirm that low marker numbers lead to an increase in zonal $\omega_{E\times B}$ levels due to noise accumulation, Fig.~\ref{fig:omega_init} shows the radially averaged absolute value of $\omega_{E\times B}$ at the initial time $t=0$ against marker number $N_p$. All simulations are initialized with a density perturbation defined as including only $n\neq0$ Fourier modes. Despite that, due to the finite and random marker number representation of $\delta f$, there is a resulting spurious finite zonal, $(m,n)=(0,0)$, $\omega_{E\times B}$ profile, whose amplitude increases with decreasing number of markers as $\sim 1/\sqrt{N_p}$, as expected due to statistical sampling error. The magnitude of the corresponding zonal flow shearing rate $\omega_{E\times B}$ is then further increased at every time step. Thus, lower marker numbers lead to larger noise accumulation in $\omega_{E\times B}$ with time, which leads to Fig.~\ref{fig:omega_end} for end-time values of $\omega_{E\times B}$. Assuming linear increment, Fig.~\ref{fig:omega_time} indicates that the rate of increase of $\omega_{E\times B}$ is $\dint{\omega_{E\times B}}/\dint{t}\approx2.12\times10^{-3}c_s^2/L_x^2$ for $N_p=256$M. For non-adaptive cases, the general trend of lower zonal $E\times B$ shearing with increasing $N_p$ is apparent. The end-time $\omega_{E\times B}$ value is expected to plateau at higher $N_p$ values, but this limit is not reached for the maximum marker number considered, $N_p=512$M. This converged value would represent the zonal $\omega_{E\times B}$ derived from the physics of the problem, and not the result of the accumulation of noise. On the other hand, the adaptive cases show much lower and similar end-time $\omega_{E\times B}$ values throughout all $N_p$ values considered in the different simulations.

\begin{figure}[H]
	\centering
	\begin{subfigure}[b]{1.0\columnwidth}
        \caption{\label{fig:snr_np}}
	    \centering
		\includegraphics[width=0.7\linewidth]{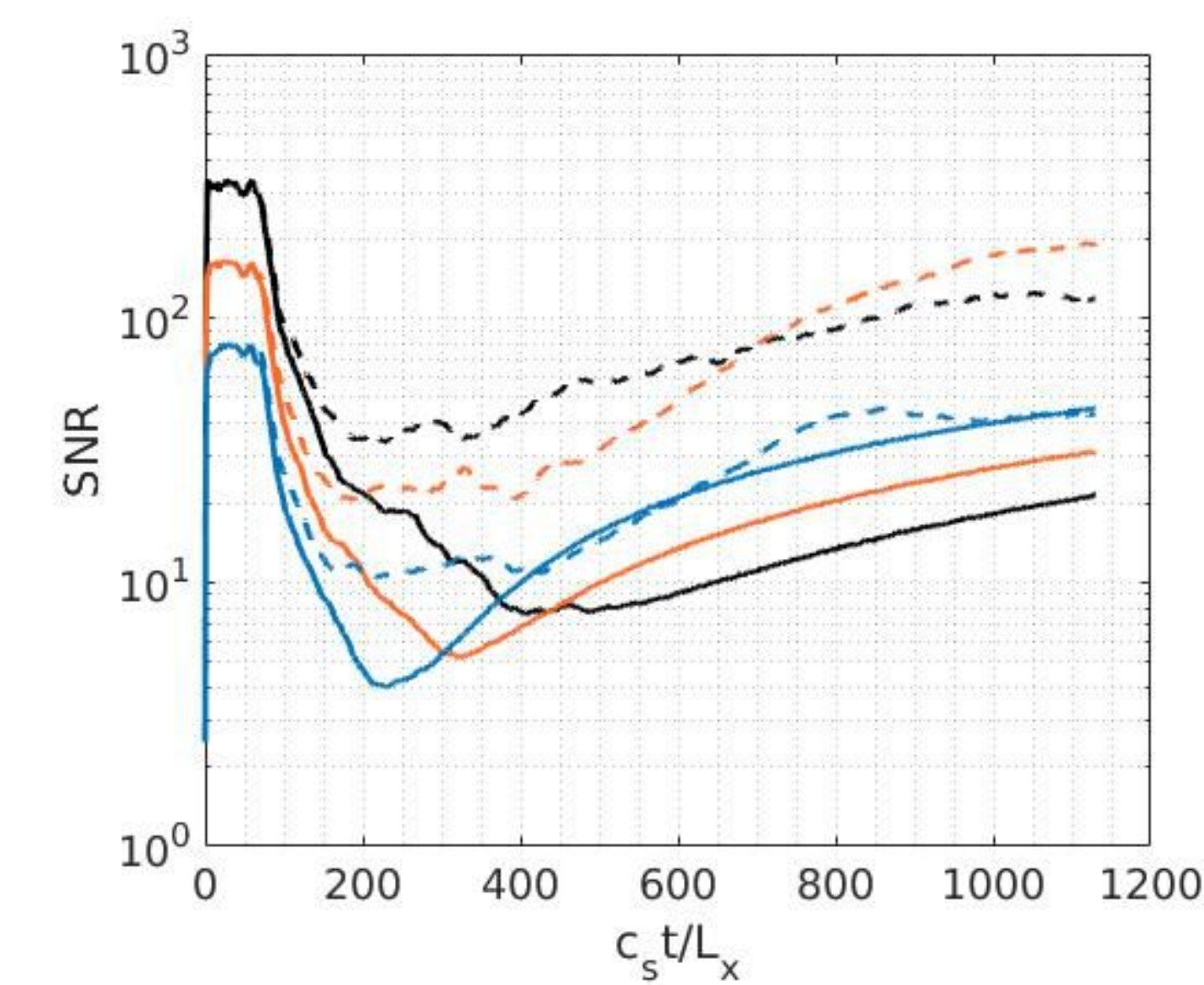}
	\end{subfigure}
	\begin{subfigure}[b]{1.0\columnwidth}
	    \caption{\label{fig:snrnz_np}}
	    \centering
		\includegraphics[width=0.7\linewidth]{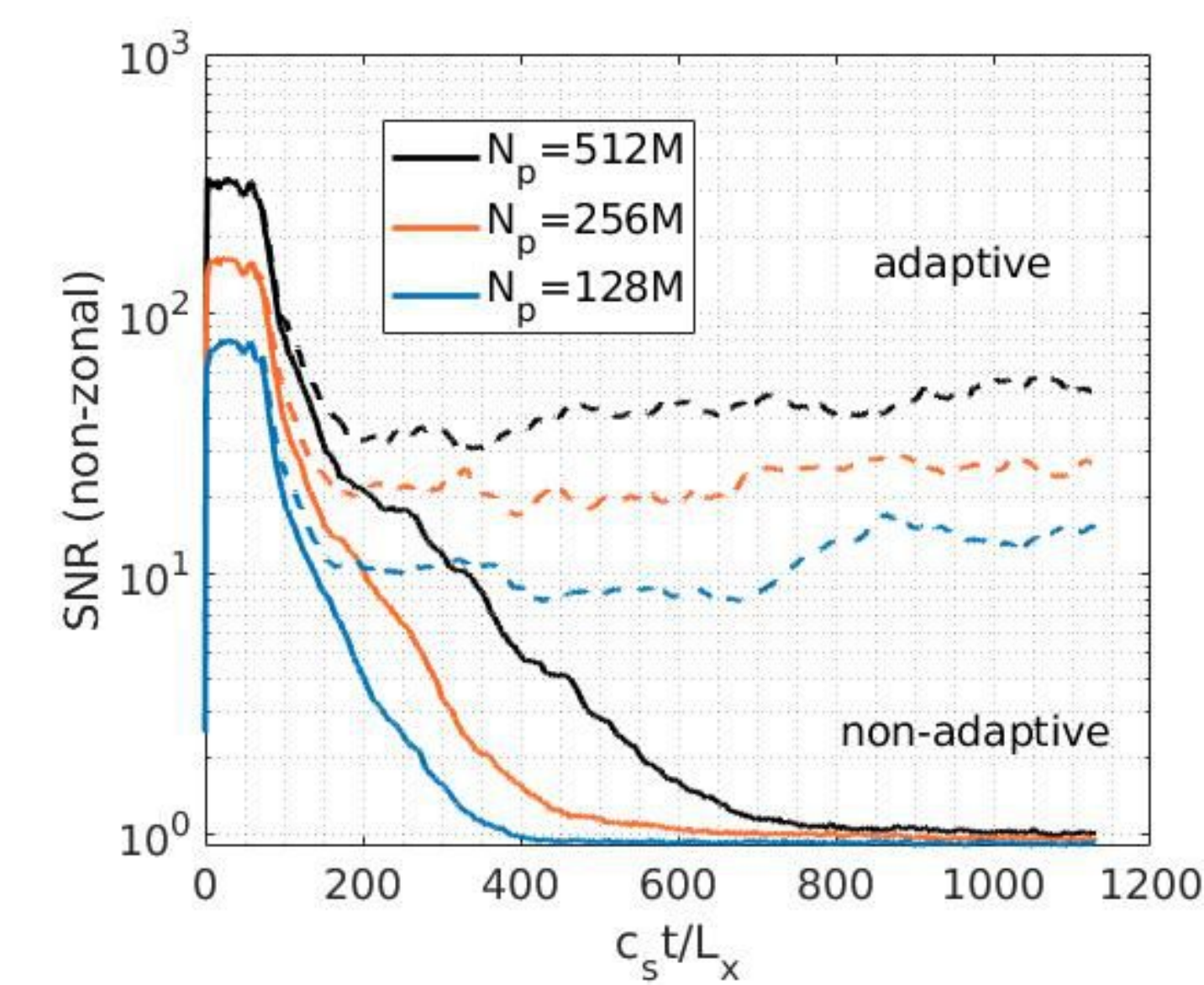}
	\end{subfigure}
	\caption{Signal-to-noise ratio (SNR) time traces for signal (a) including and (b) excluding, the $(m,n)=(0,0)$ mode, for increasing marker number $N_p$, and considering both the non-adaptive (solid line) and adaptive (dashed line) cases.}
	\label{fig:diag_np}
\end{figure}

The potential field $\phi$ of Eq.~(\ref{eq:qn_df}) is solved by first projecting it onto the B-spline basis $\Lambda_{ijk}(\vec{r})=\Lambda_i(x)\Lambda_j(y)\Lambda_k(z)$ followed by taking the Discrete Fourier transform of spline indices in the $y$ and $z$ directions, thus representing the discretized equation in terms of poloidal $m$ and toroidal $n$ mode numbers together with the spline index $i$ related to the remaining radial dimension $x$~\cite{McMillan2010}. By gyrokinetic ordering arguments~\cite{Jolliet2012}, one can apply on the projected right-hand side of Eq.~(\ref{eq:qn_df}) a centered band filter $\mathcal{F}_1$ in $(m,n)$-space involving only keeping modes $|m+nq(x)|<\Delta m=$ constant, as the physically relevant, i.e. $k_\parallel \rho_{th}\ll 1$, modes constituting the signal are nearly field-aligned and thus fall in this band. Here, $k_\parallel=2\pi(m+nq)/(L_zq)$ is the wavevector component along $\vec{B}$. The region $\mathcal{F}_2$ consist of two bands, each on either side of $\mathcal{F}_1$, is defined to represent physically strongly damped modes. Any finite levels of modes in $\mathcal{F}_2$ are thus assumed to provide a measure of the noise level. Estimates of the signal and noise levels are thus respectively provided by
\begin{equation}
\text{signal} = \frac{\sum_i\sum_{(m,n)\in\mathcal{F}_1}|b_i^{(m,n)}|^2}{\sum_i\sum_{(m,n)\in\mathcal{F}_1}1} \hspace{1mm}, \hspace{1mm} \text{noise} = \frac{\sum_i\sum_{(m,n)\in\mathcal{F}_2}|b_i^{(m,n)}|^2}{\sum_i\sum_{(m,n)\in\mathcal{F}_2}1}. \label{eq:snr}
\end{equation}
Here, $\Sigma_i$ sums over all radial indices of spline amplitude $b_i^{(m,n)}$ of the right-hand side. The latter is expressed by

\begin{figure*}
	\centering
	\begin{subfigure}[c]{0.24\textwidth}
	    \caption{\label{fig:reldevTi_np_nadp}}
		\centering
		\includegraphics[width=1.0\linewidth]{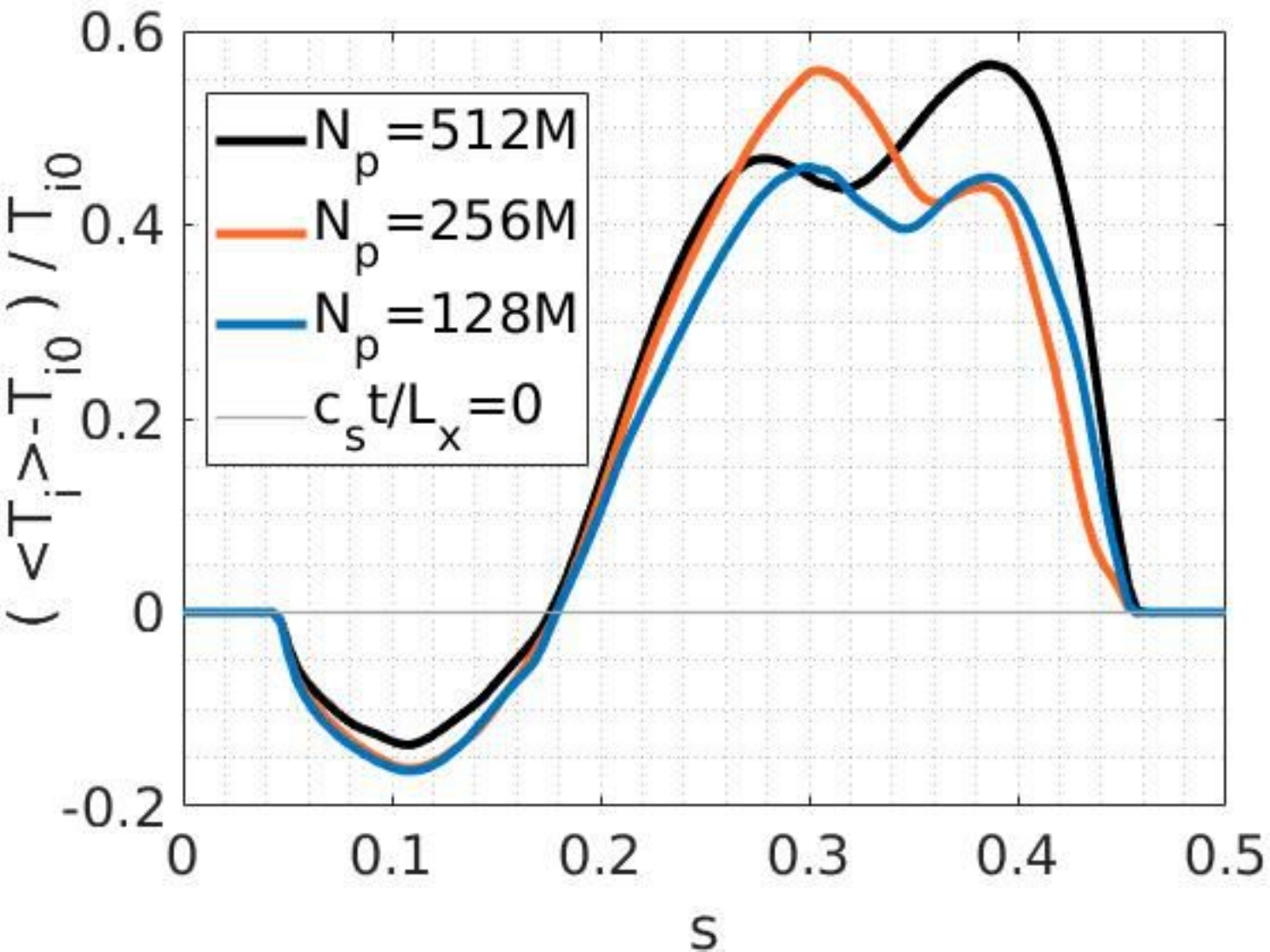}
	\end{subfigure}
	\begin{subfigure}[c]{0.24\textwidth}
	    \caption{\label{fig:kappa_np_nadp}}
		\centering
		\includegraphics[width=1.0\linewidth]{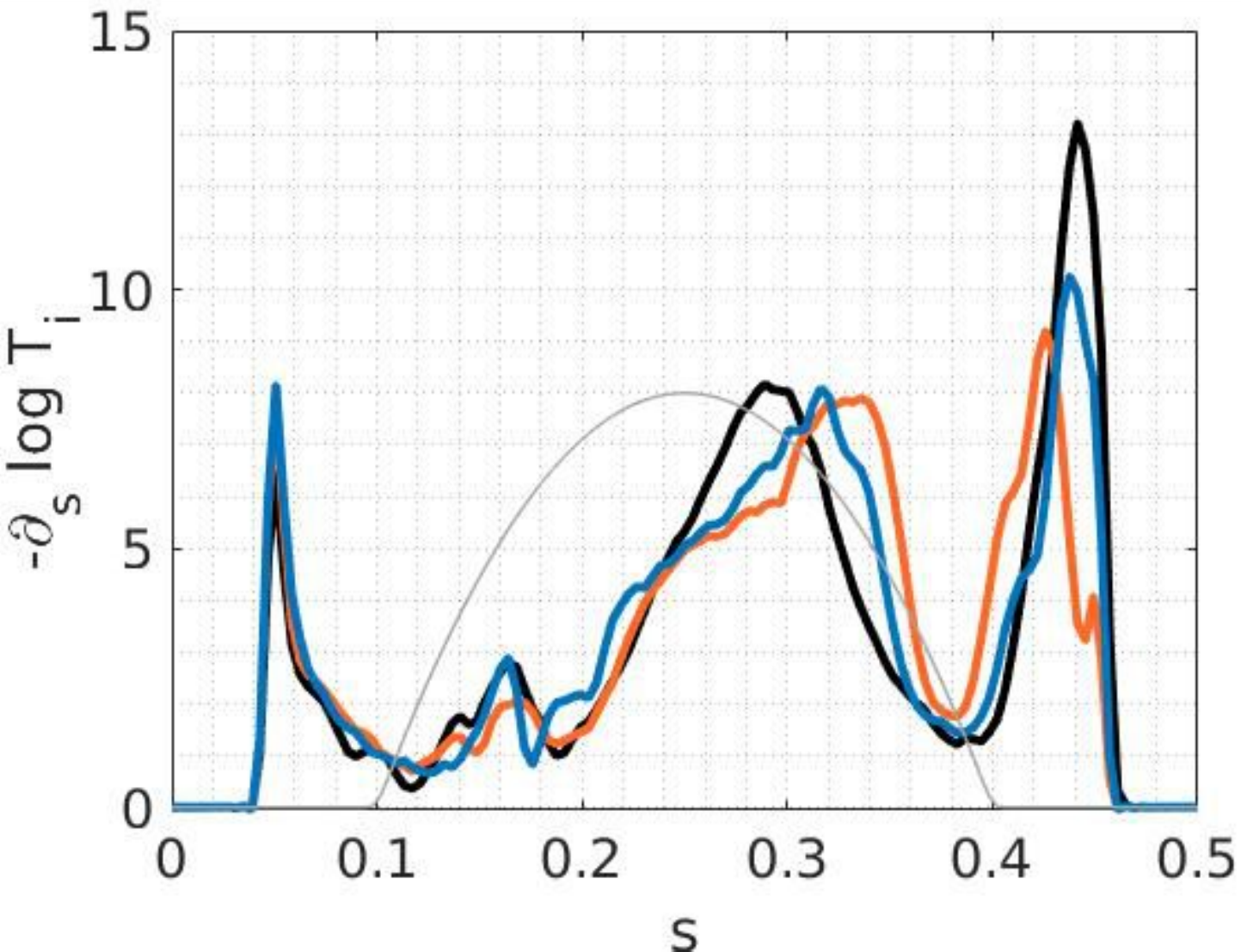}
	\end{subfigure}
	\begin{subfigure}[c]{0.24\textwidth}
	    \caption{\label{fig:shear_np_nadp}}
		\centering
		\includegraphics[width=1.0\linewidth]{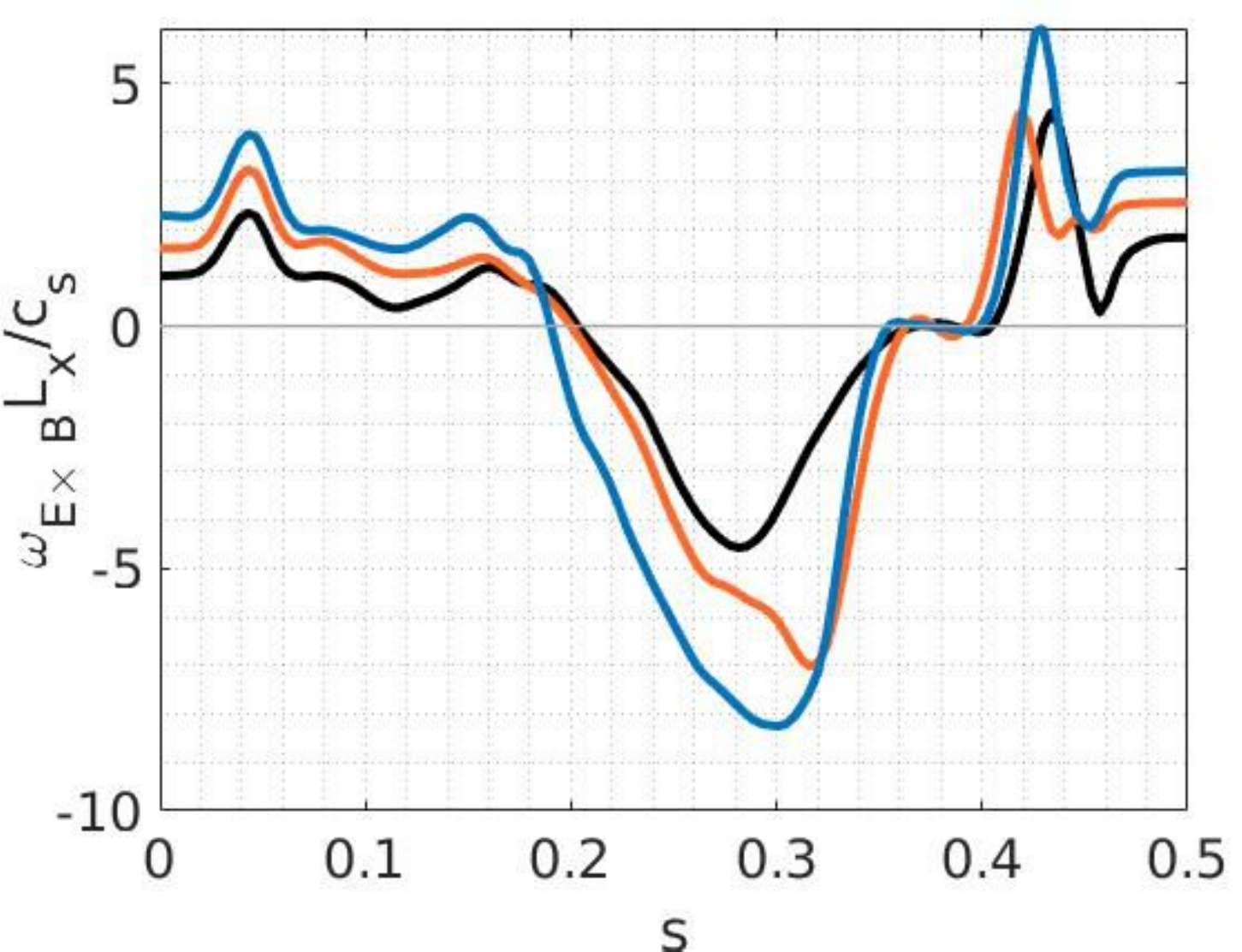}
	\end{subfigure}
	\caption{F.s.a.~profiles at quasi-steady state averaged over a time interval of $c_st/L_x\in[1030,1130]$ for various marker numbers $N_p$ under the non-adaptive scheme for the ion temperature (a) relative deviation with respect to background and its (b) logarithmic gradient, and the (c) zonal flow shearing rate .}
	\label{fig:prof_np_nadp}
\end{figure*}

\begin{eqnarray*}
b_i^{(m,n)} &=& \sum_{j=0}^{N_y-1}\sum_{k=0}^{N_z-1}\exp\left[-2\pi\imag\left(\frac{mj}{N_y}+\frac{nk}{N_z}\right)\right]\times \\
& &\left\{ \int\dint{^3R}\dint{\alpha}\dint{\vp}\dint{\mu}\frac{B_\parallel^\star}{m_i} \Lambda_{ijk}[\vec{R}+\vec{\rho}_L(\mu,\alpha)]\times \right. \\
& & \left. [f_0(\vec{R},\vp,\mu,t)-f_0(\vec{R},\vp,\mu,0)] + \right. \\
& & \left. \sum_p w_p \int\dint{\alpha}\Lambda_{ijk}[\vec{R}_p+\vec{\rho}_L(\mu_p,\alpha)]\right\}.
\end{eqnarray*}

It should be noted that what we call `signal' here also includes the discretization noise present within the filter~\cite{Bottino2007}. Nonetheless, these measures provide a practical estimate of the signal-to-noise ratio (SNR).

From Fig.~\ref{fig:snr_np}, non-adaptive cases start from high SNR values and gradually drop to their respective lowest point after the initial burst $c_st/L_x\sim300$. $N_p$ is reflected in the maximum of SNR values for each case, which latter seem to scale as $1/N_p$. The adaptive cases follow a similar trend, but do not fall as low. From past works~\cite{McMillan2008,Bottino2007}, the rule-of-thumb SNR threshold of $10$ is a value above which the results can be deemed reliable. Therefore, we can see that only the adaptive cases with $N_p=256\text{M},512\text{M}$ meet this criterion throughout the whole simulation. All SNR values eventually rise with time, with the non-adaptive case at $N_p=128$M rising the quickest. This reflects the noise accumulation in the physically undamped zonal component. Indeed, by subtracting the zonal component from the signal, Fig.~\ref{fig:snrnz_np} shows that the non-adaptive case with $N_p=128$M gives the lowest SNR value throughout the simulation. Thus, for this set of parameters, only the results from the adaptive case with $N_p=256$M or $N_p=512$M may be deemed reliable.

Figs.~\ref{fig:prof_np_nadp} and \ref{fig:prof_np_adp} show the f.s.a.~profiles at the end of the simulations for the zonal flow shearing rate $\omega_{E\times B}$, the ion temperature $\langle T_i\rangle$ relative deviation $(\langle T_i\rangle-T_{i0})/T_{i0}$, and its logarithmic gradient for different total number $N_p$ of markers, under the non-adaptive and adaptive cases, respectively. It should be noted that the maximum relative deviation value for $T_i$ of around $60\%$ in Fig.~\ref{fig:reldevTi_np_nadp} for the non-adaptive cases challenges the $\delta f$ assumption of $|\delta f|/|f_0| \ll 1$. For the adaptive case in Fig.~\ref{fig:reldevTi_np_adp} however, the relative deviation of $\langle T_i\rangle$ from the adapted background temperature $T_{i0}(t)$, which remains low at all times, qualify. One notes also that the adaptive cases in fact resulted in f.s.a.~$T_i$ profiles with a larger deviation of $100\%$, from its initial state, as shown in Fig.~\ref{fig:reldev0Ti_np_adp}. This shows that the adaptive scheme appears to allow for simulations with more accurate profile evolution in case of large deviations, not afforded by the standard scheme. One notes also the development of strong $T_i$ gradients at $s=0.05$, just outside the heat source for the non-adaptive case (see Fig.~\ref{fig:prof_np_nadp}). This is suspected to be related to spurious marker accumulation by error in drift calculation, whose magnitude reduces with increasing $N_p$. Under the adaptive scheme, this problem does not occur.

\subsection{Adaptive control variate and noise control} \label{sec:krook}

\begin{figure}[H]
	\centering
	\begin{subfigure}[c]{0.49\columnwidth}
		\caption{\label{fig:chi_nctrl}}
	    \centering
		\includegraphics[width=1.0\linewidth]{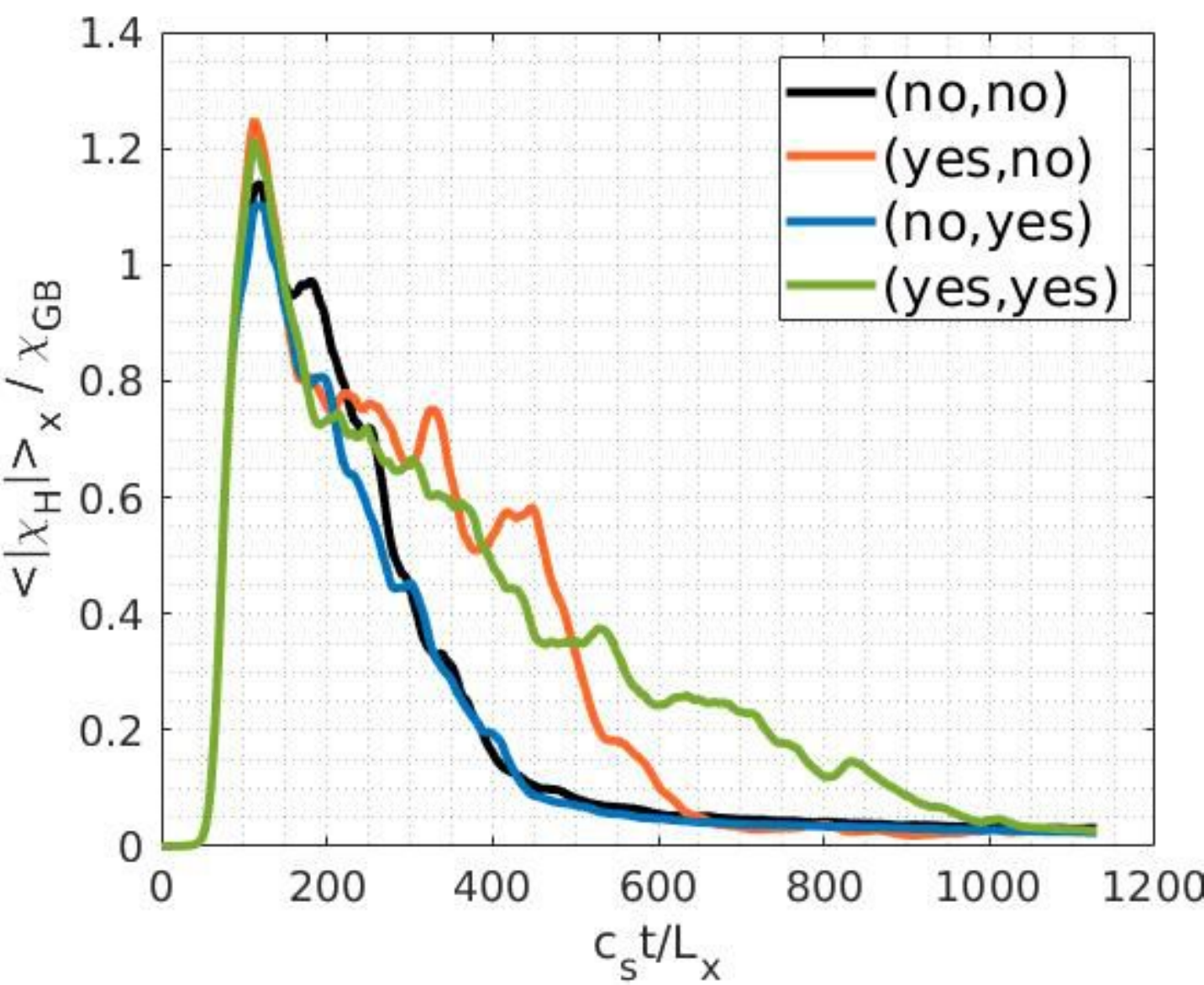}
	\end{subfigure}
	\begin{subfigure}[c]{0.49\columnwidth}
	    \caption{\label{fig:omega_nctrl}}
	    \centering
		\includegraphics[width=1.0\linewidth]{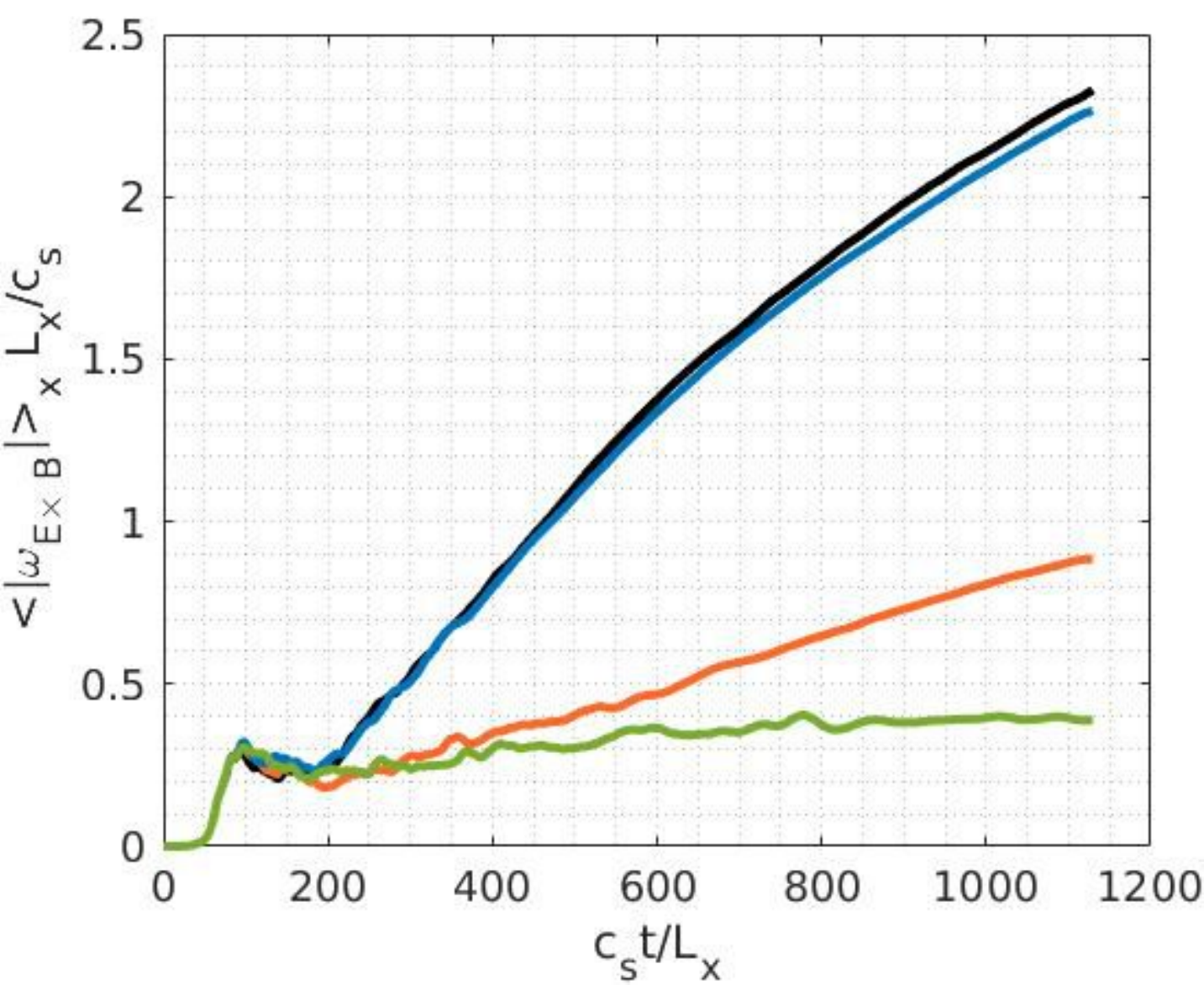}
	\end{subfigure}
	\begin{subfigure}[c]{0.49\columnwidth}
	    \caption{\label{fig:snr_nctrl}}
	    \centering
		\includegraphics[width=1.0\linewidth]{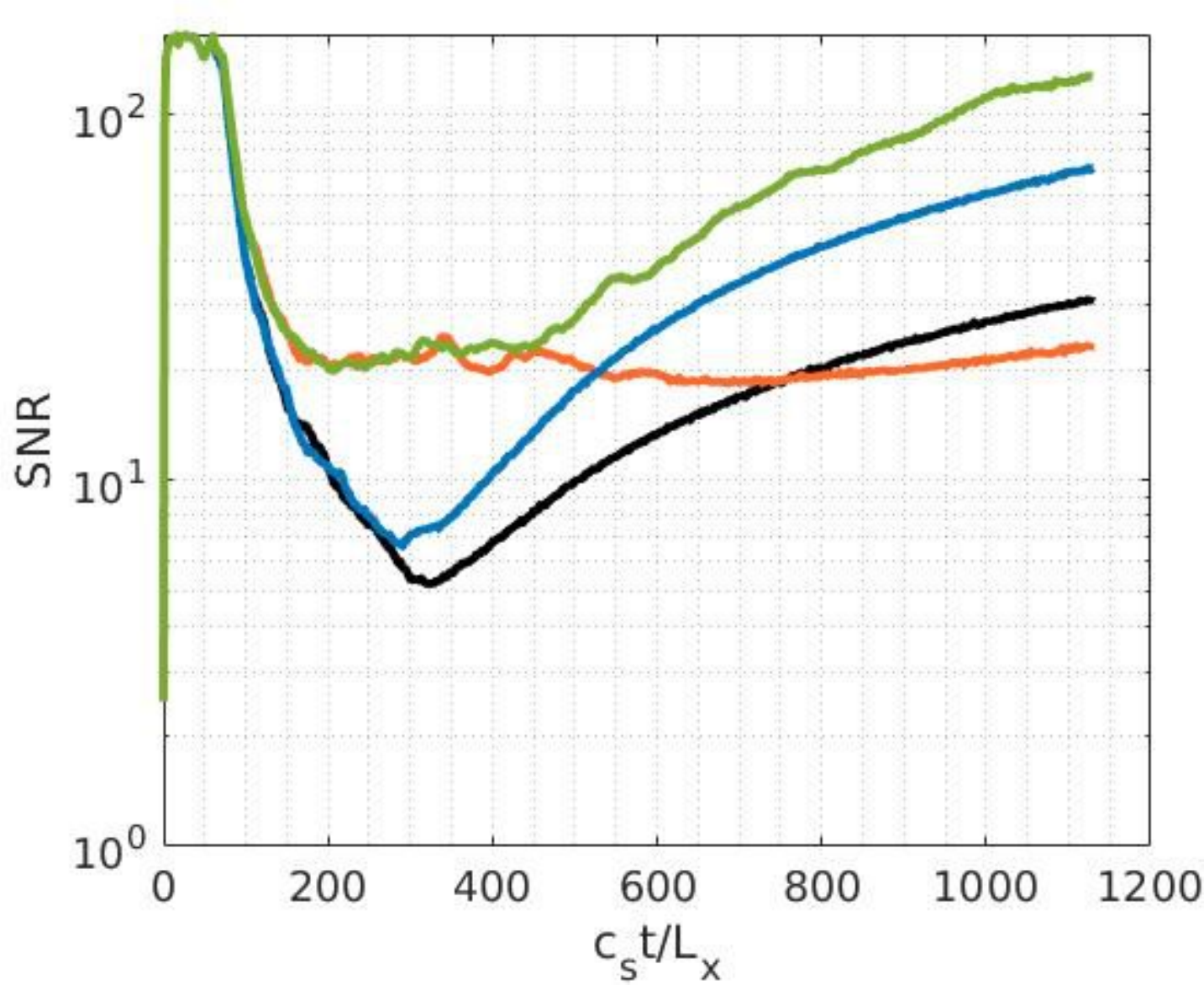}
	\end{subfigure}
	\begin{subfigure}[c]{0.49\columnwidth}
	    \caption{\label{fig:snrnz_nctrl}}
	    \centering
		\includegraphics[width=1.0\linewidth]{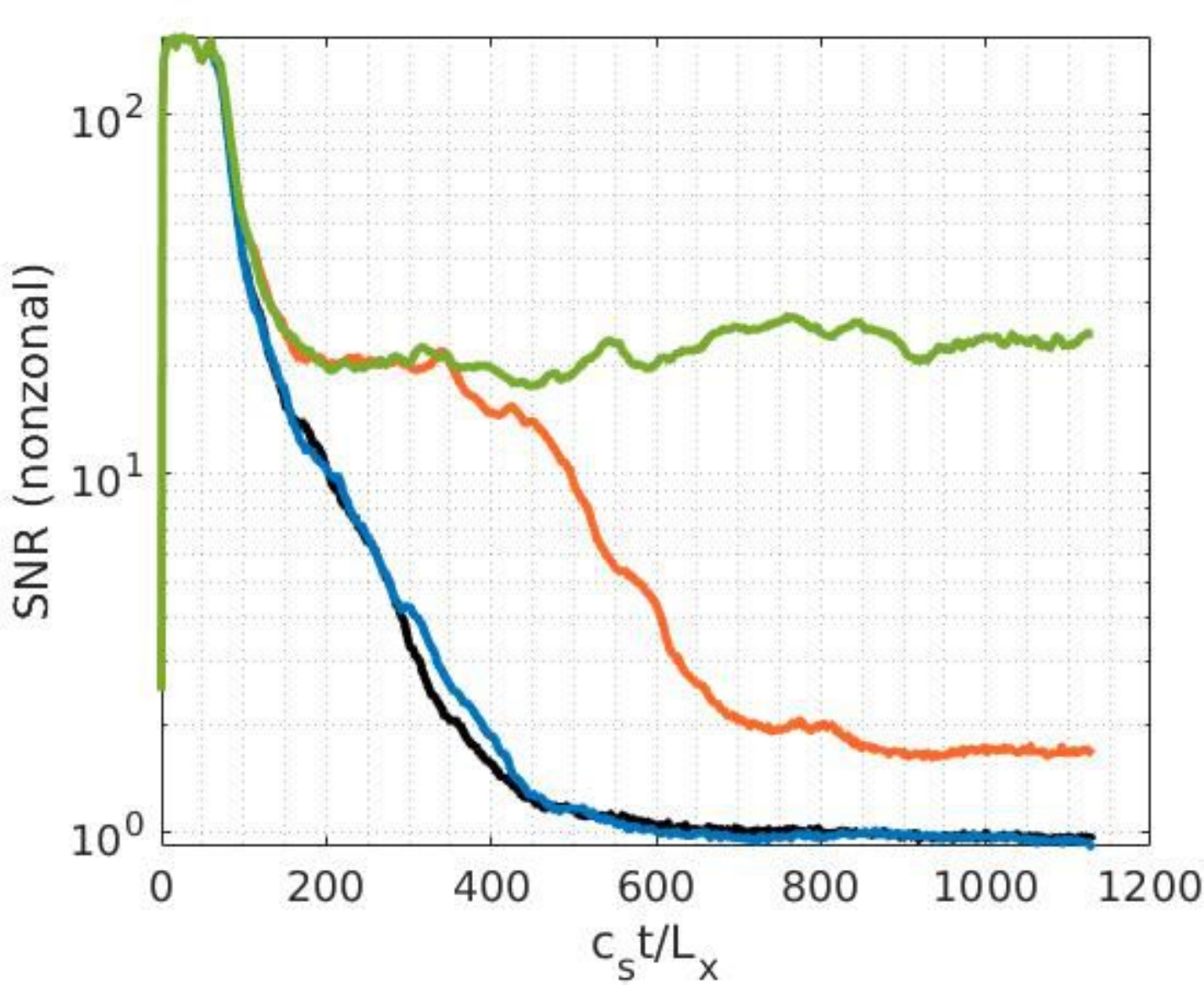}
	\end{subfigure}
	\caption{Diagnostics under four different adaptive scenarios (see description in text) for the radially averaged absolute (a) heat diffusivity $\chi_H$ and (b) zonal flow shearing rate $\omega_{E\times B}$, and the signal-to-noise (SNR) ratio with signal (c) including and (d) excluding, the $(m,n)=(0,0)$ mode. Marker number set to $N_p=256$M, and adaptive rate to $\alpha_E=1.92\gamma_{max}$ where applicable. A moving time-averaging window of half-width $c_st/L_x=10$ has been implemented.}
	\label{fig:nctrl}
\end{figure}

\begin{figure*}
	\centering
	\begin{subfigure}[c]{0.19\textwidth}
	    \caption{(no,no), $\alpha_E=0.00\gamma_{max}$}
		\centering
		\includegraphics[width=1.0\linewidth]{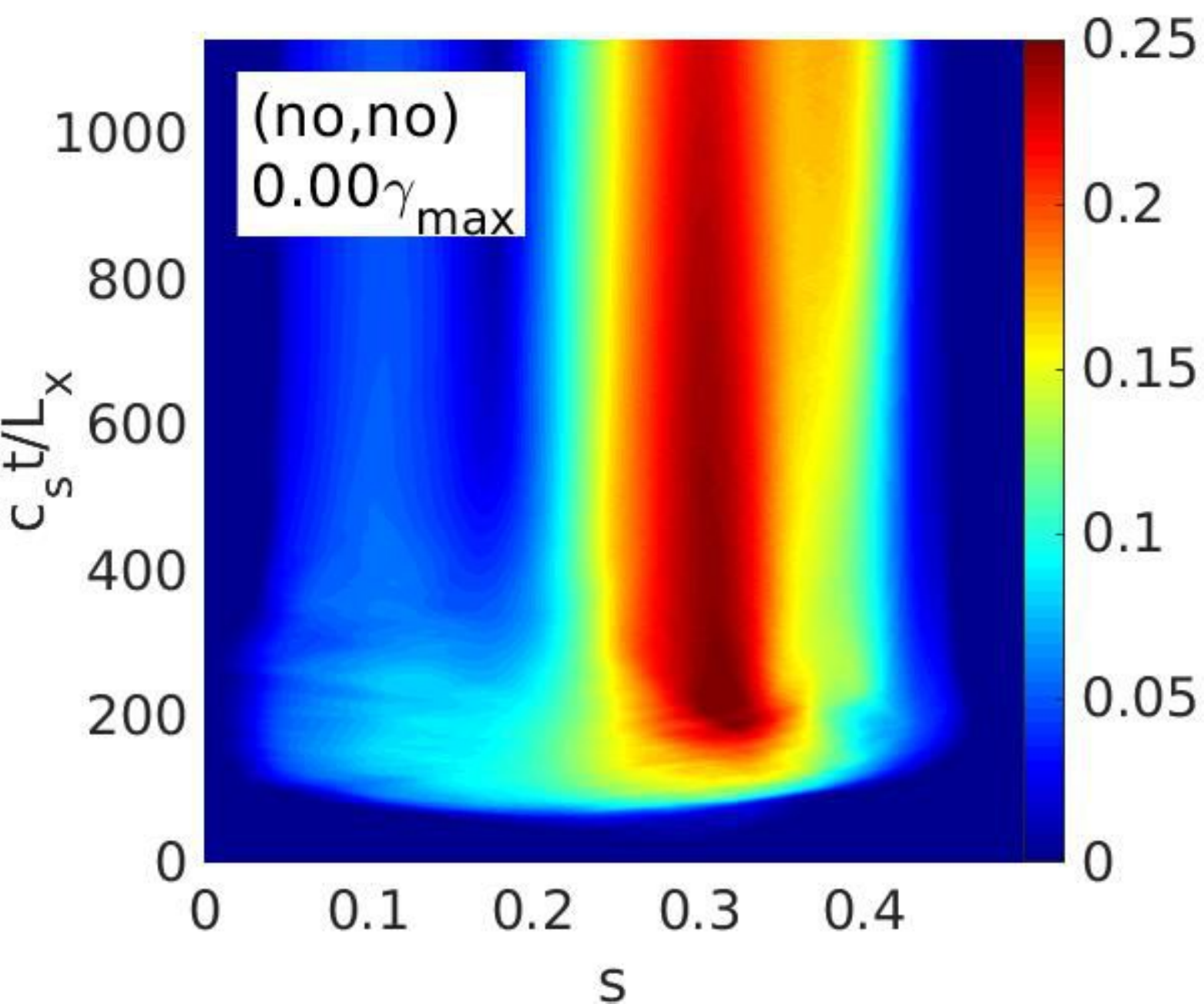}
	\end{subfigure}
	\begin{subfigure}[c]{0.19\textwidth}
	    \caption{(no,yes), $\alpha_E=1.92\gamma_{max}$}
		\centering
		\includegraphics[width=1.0\linewidth]{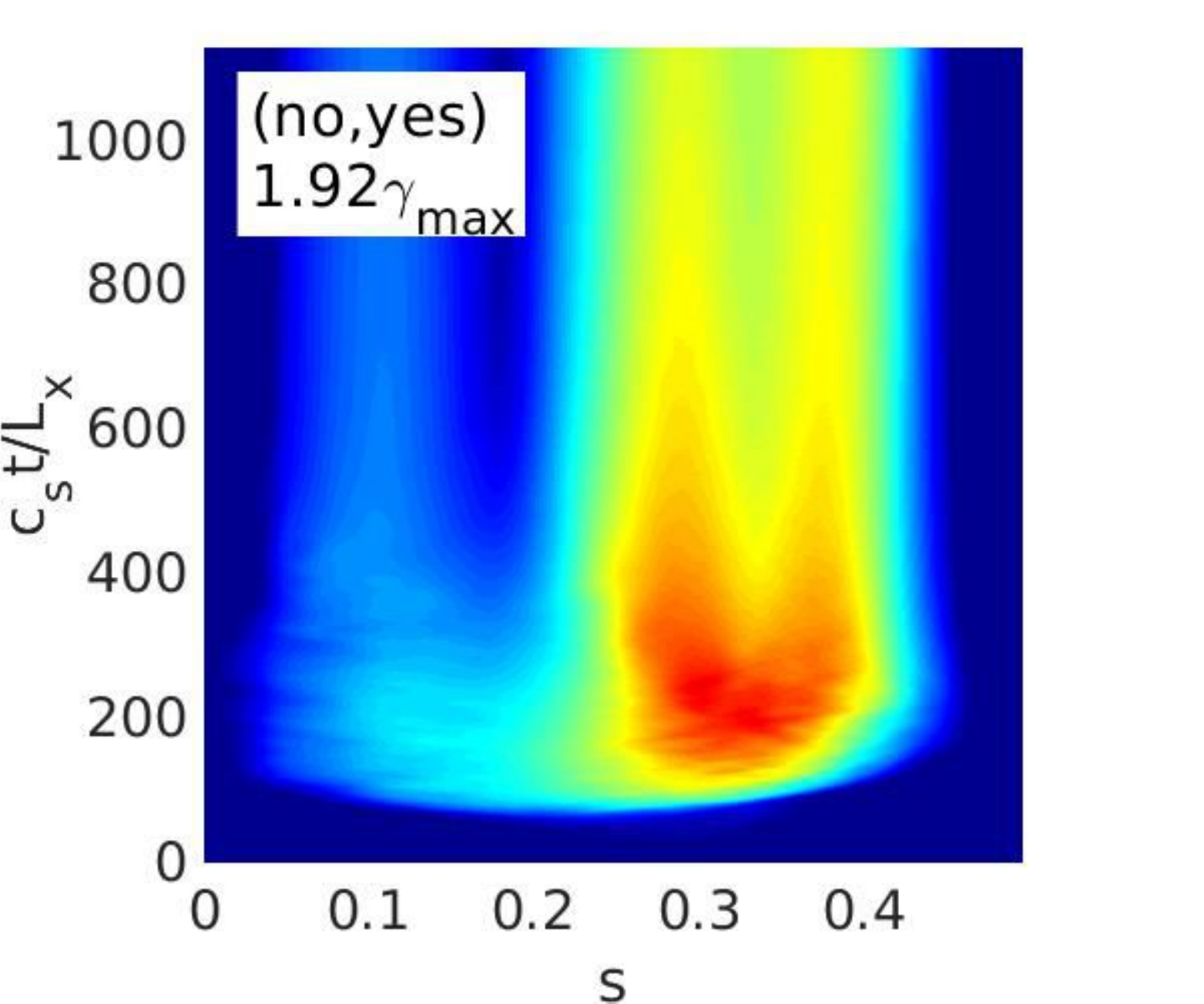}
	\end{subfigure}
	\begin{subfigure}[c]{0.19\textwidth}
	    \caption{(yes,no), $\alpha_E=1.92\gamma_{max}$}
		\centering
		\includegraphics[width=1.0\linewidth]{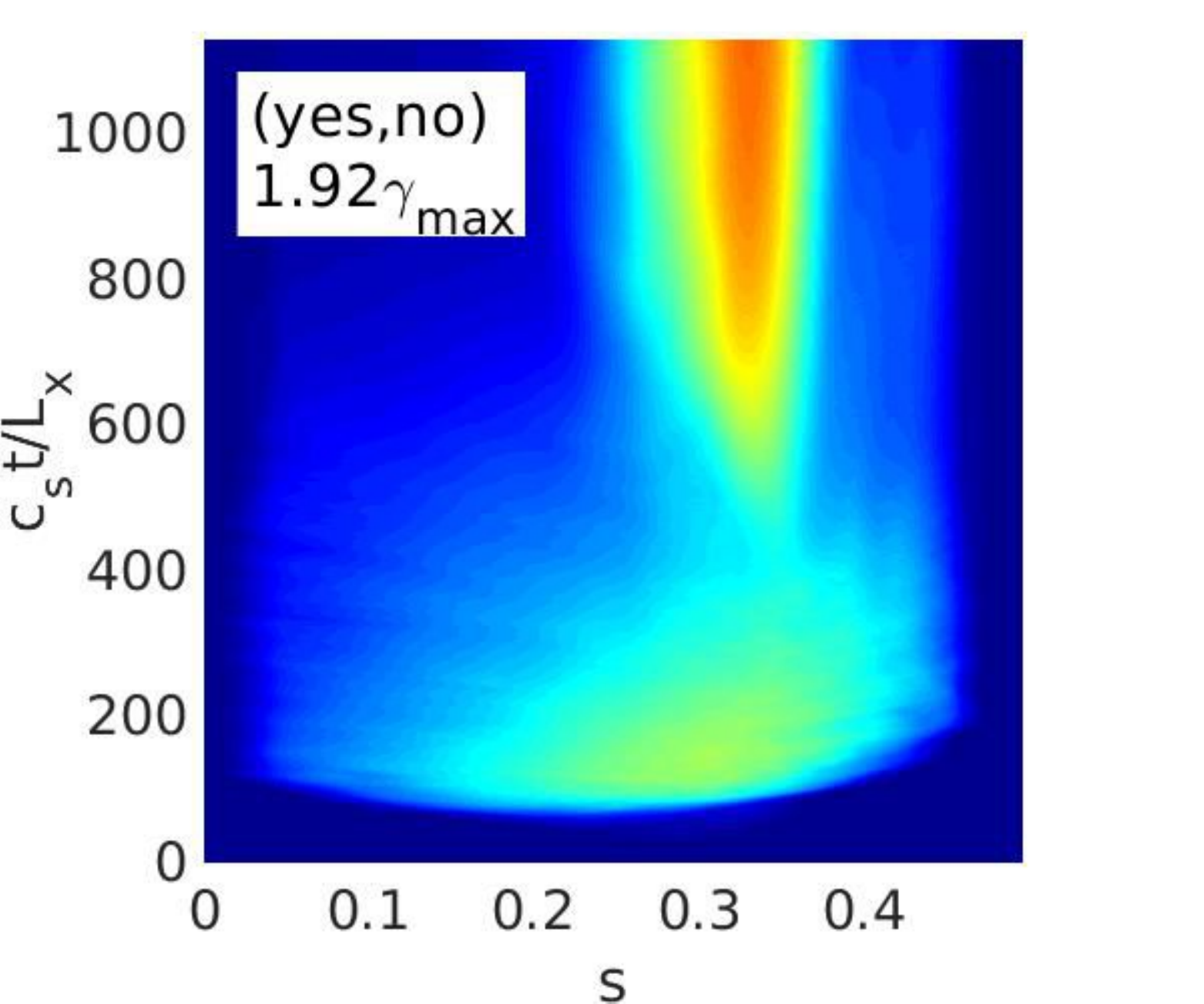}
	\end{subfigure}
	\begin{subfigure}[c]{0.19\textwidth}
	    \caption{(yes,no), $\alpha_E=0.12\gamma_{max}$}
		\centering
		\includegraphics[width=1.0\linewidth]{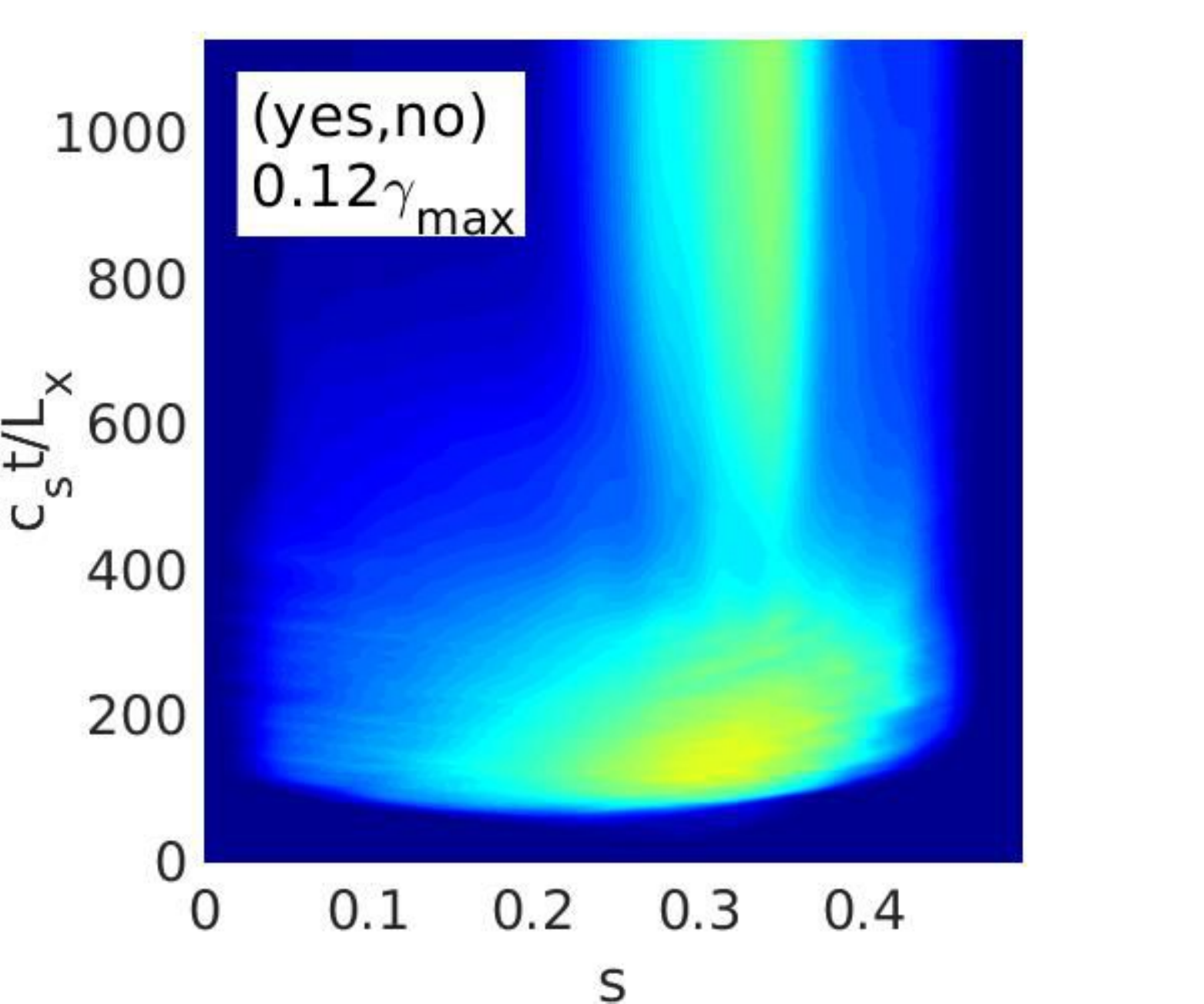}
	\end{subfigure}
	\begin{subfigure}[c]{0.19\textwidth}
	    \caption{(yes,yes), $\alpha_E=1.92\gamma_{max}$}
		\centering
		\includegraphics[width=1.0\linewidth]{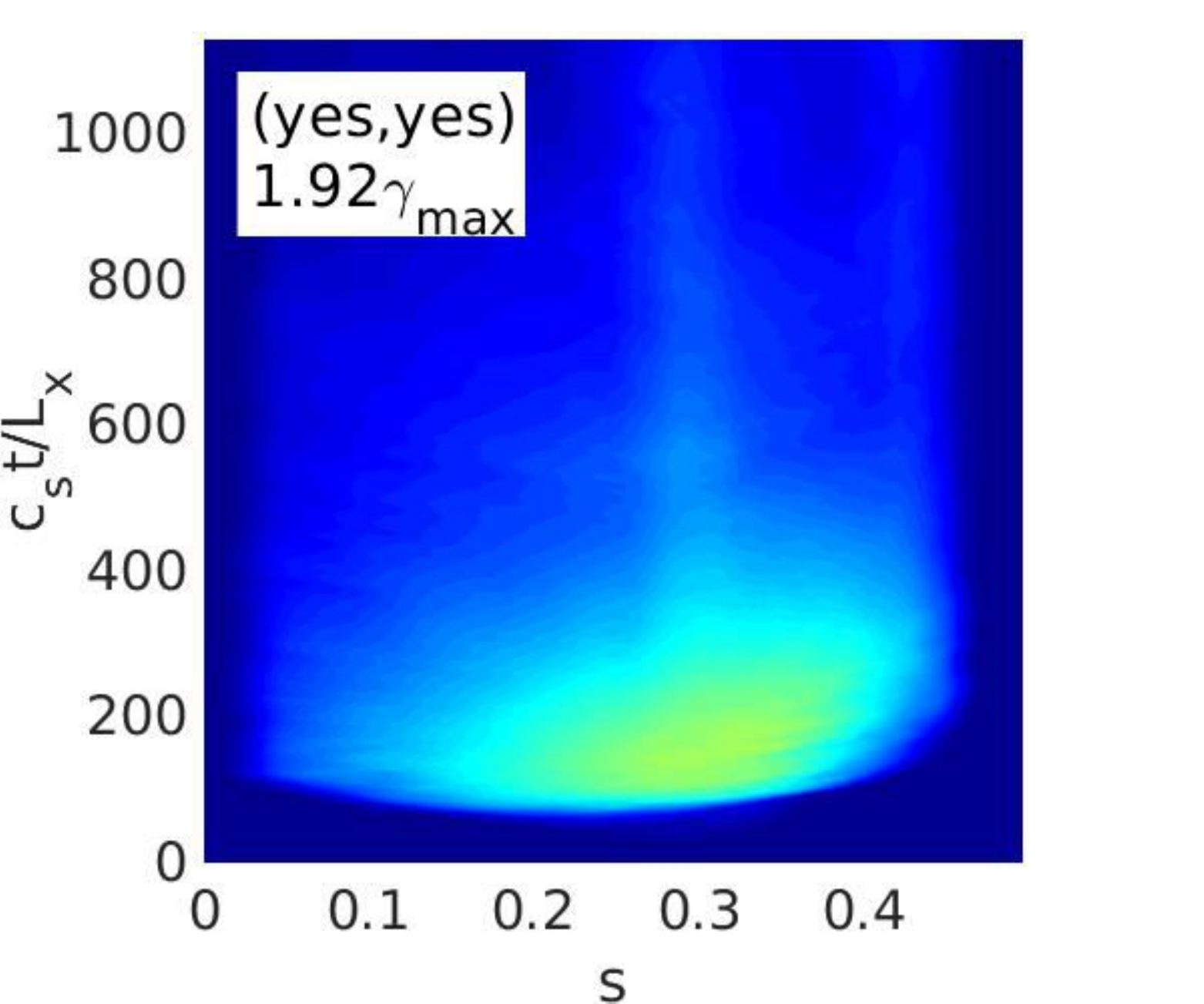}
	\end{subfigure}
	\caption{Local f.s.a.~weight standard deviation $\sigma_w$ through time under four different adaptive scenarios (see description in text), illustrated with the cases of (a) (no,no), (b) (no,yes) with $\alpha_E=1.92\gamma_{max}$, (c) (yes,no) with $\alpha_E=1.92\gamma_{max}$, (d) (yes,no) with $\alpha_E=0.12\gamma_{max}$, and (e) (yes,yes) with $\alpha_E=1.92\gamma_{max}$. $\sigma_w$ is defined by $\sqrt{\Sigma_{p\in i} w_p^2/N_i - (\Sigma_{p\in i} w_p/N_i)^2}$, with $N_i=N_p/N_x$ the average number of markers of the $i^{th}$ radial bin, taken to be uniform for all bins. Marker number set to $N_p=256$M. All figures share the same colour scale.}
	\label{fig:sigmaw_nctrl}
\end{figure*}

Tentatively, two mechanisms contribute to the improvement of SNR with the adaptive scheme: (a) the adapted $f_0(t)$ as a good control variate; (b) the noise control operator $S_n$ of the non-adapted scheme, which tends to bring back $f$ towards the initial distribution $f_{00}$, whereas in the adapted scheme it tends to relax $f$ towards the time-evolved $f_0$, which is closer to the time-averaged $f$, especially at late times. To study the relative importance of the adaptive scheme and the noise control, we varied in this section the control variate $f_0$, both in the framework of the adaptive delta-f scheme, Eq.~(\ref{eq:delta-f}), and in the reference function $f_n$ of the noise control operator i.e.
\begin{eqnarray}
S_n &=& -\gamma_n(f-f_n). \nonumber
\end{eqnarray}

All simulations begin with $f = f_{M0} + \delta f(t=0)$, where $\delta f(t=0)$ represents a small perturbation and the control variate $f_0$ is taken to be $f_M$ (see Eq.~(\ref{eq:f0})). By choosing different $f_0$ and $f_n$, four different adaptive scenarios can be constructed, depending on whether the adaptive scheme is used ($f_0=f_{M0}$ or $f_0=f_M(t)$) and whether adaptive noise control scheme is used ($f_n=f_{M0}$ or $f_n=f_M(t)$). Below, except for scenario $3$, $f_M(t)$ adapts via $T_{i0}$ from Eq.~(\ref{eq:f0}) according to Eq.~(\ref{eq:adp_relax}).

\begin{enumerate}
	\item non-adaptive $\delta f$ scheme, non-adaptive $f_n$, labeled (no,no)
	\begin{itemize}
		\item $(f_0,f_n)=(f_{M0},f_{M0})$
	\end{itemize}
	\item adaptive $\delta f$ scheme, non-adaptive $f_n$, labeled (yes,no)
	\begin{itemize}
		\item $(f_0,f_n)=(f_M(t),f_{M0})$
	\end{itemize}
	\item non-adaptive $\delta f$ scheme, adaptive $f_n$, labeled (no,yes)
	\begin{itemize}
		\item $(f_0,f_n)=(f_{M0},f_M(t))$
		\item $f_n=f_0(t)$ is adaptive according to
		\begin{eqnarray}
		\pp{}{t}\left(\frac{3}{2}n_{i0}T_{i0}\right) &=& \alpha_E\left\langle\int\dint{\vp}\dint{\mu}\frac{2\pi B_\parallel^\star}{m_i}[\delta f - (f_M(t)-f_{M0})]\right\rangle \nonumber \\
		& &
		\end{eqnarray}
		\item the adaptive scheme is run in the background to update $f_n=f_M(t)$, but the control variate $f_0=f_{M0}$ remains time-independent.
	\end{itemize}
	\item adaptive $\delta f$ scheme, adaptive $f_n$ labeled (yes,yes)
	\begin{itemize}
		\item $(f_0,f_n)=(f_M(t),f_M(t))$
	\end{itemize}
\end{enumerate}

Viewing the adaptive scheme as essentially a means to reduce noise, considering the different scenarios described allows one to determine which strategy is the most effective in this respect. Fig.~\ref{fig:nctrl} shows the effect of different adaptive scenarios on the simulation results. The adaptation of $f_0$ and/or $f_n$ for the different scenarios are done at the same rate $\alpha_E$. It can be seen that the full adaptive scheme with adaptive $f_0$ and $f_n$ [scenario (yes,yes)], implying weight transfer from $\delta f$ to $f_0$ of Eq.~(\ref{eq:dw_adp}) and noise control only on the fluctuating part $\delta f$ respectively, is necessary for effective noise control, as indicated by a reasonably high SNR value and a zonal flow shearing rate $\omega_{E\times B}$ value that does not increase indefinitely. For scenario (yes,no) with only the adaptive $f_0$, noise control is effective only at an early stage when $f$ is close to $f_{M0}$. As the former deviates away from the latter, $S_n$ acts as a weak source, thus enlarging the $\delta f$ component. The larger the portion of $f$ that is represented by markers, the more noise accumulates. The weak improvement of scenario (no,yes), which is the standard delta-f scheme with a time-dependent reference function $f_n=f_M(t)$, indicates that the improvement from weight transfer from $\delta f$ to $f_0$ far out-weights an adaptive $S_n$.

Noise can also be approached from the standard deviation of the f.s.a.~weights $\sigma_w$, as shown in Fig.~\ref{fig:sigmaw_nctrl}. Based on this measure, the full adaptive scheme (yes,yes) once again gives the best results, with low values of $\sigma_w$ right after the burst at around $c_st/L_x=150$. One can see that $\sigma_w$ as already plateaued for the non-adaptive case (no,no), whereas for the case (yes,no), the $S_n$ acting as weak source continues to relax the distribution towards that at initial time. This is proved to be the case when see that a lower $\alpha_E$ value gives a smaller yet increasing $\sigma_w$ value. Finally, the scenario (no,yes) with an adaptive noise control is able to continuously decrease $\sigma_w$ values, but these values remain high after the burst, which may affect the results at late times.

\amend A note on the inclusion of the $f_0(t)-f_0(t=0)$ term in Eq.~(\ref{eq:qn_df}) is in order. While some cases studied in this paper do not involve a time-dependent background density, with reference to the full adaptive scheme [scenario (yes,yes)] described above, the exclusion of this extra term leads to a lower heat diffusivity, and a $17\%$ increase in the zonal flow shearing rate at quasi-steady state, which in turn These resulted in a lower final ion temperature deviation. Nonetheless, similar improvement in SNR and local f.s.a. weight standard deviation (see Fig.~\ref{fig:sigmaw_nctrl}) have been observed.
\color{black}

\subsection{Noise control strength and adaptive rate variations} \label{sec:variations}

\begin{figure}[H]
	\centering
	\begin{subfigure}[c]{0.49\columnwidth}
	    \caption{\label{fig:chi_gamma1}}
	    \centering
		\includegraphics[width=1.0\linewidth]{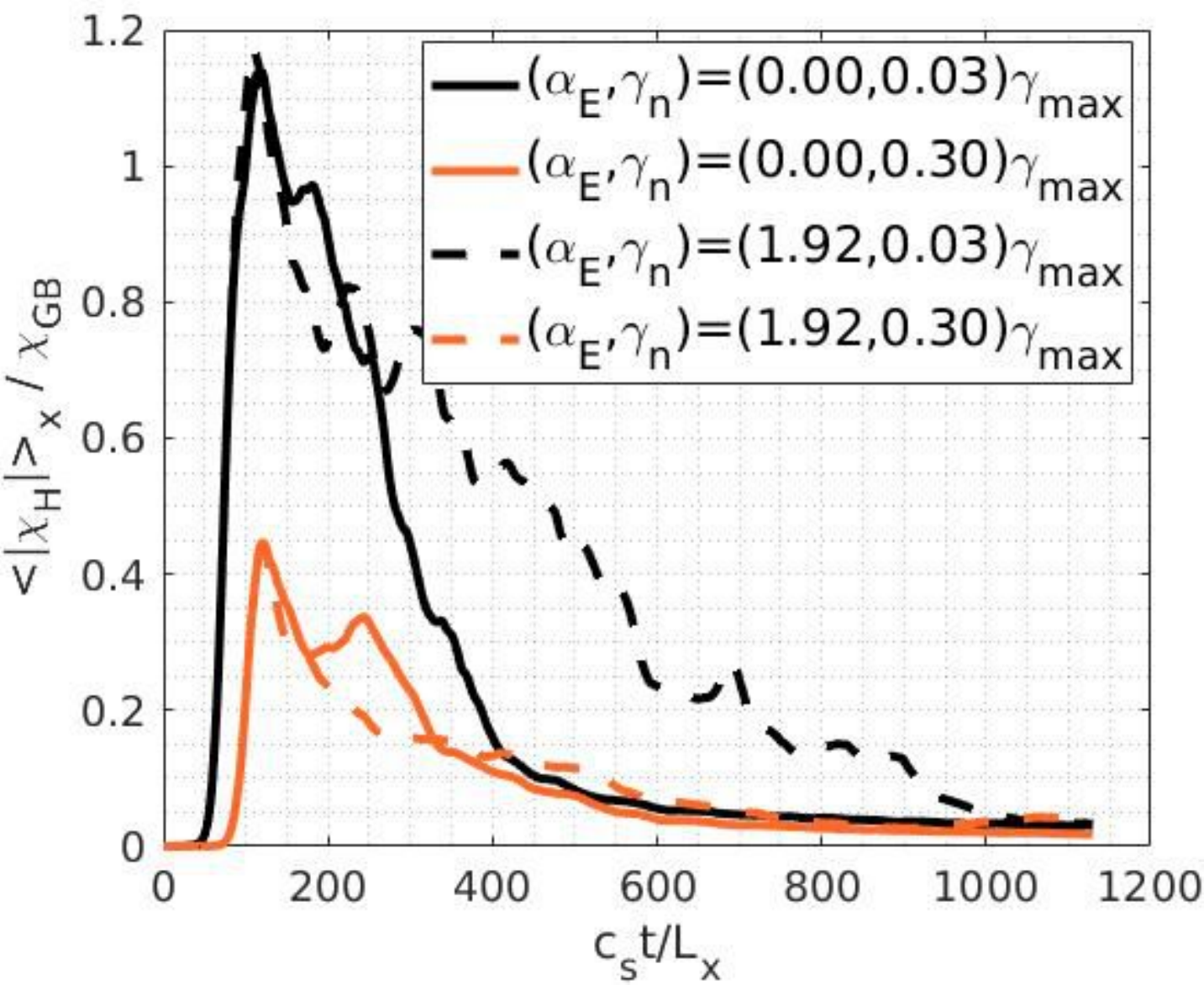}
	\end{subfigure}
	\begin{subfigure}[c]{0.49\columnwidth}
	    \caption{\label{fig:omega_gamma1}}
	    \centering
		\includegraphics[width=1.0\linewidth]{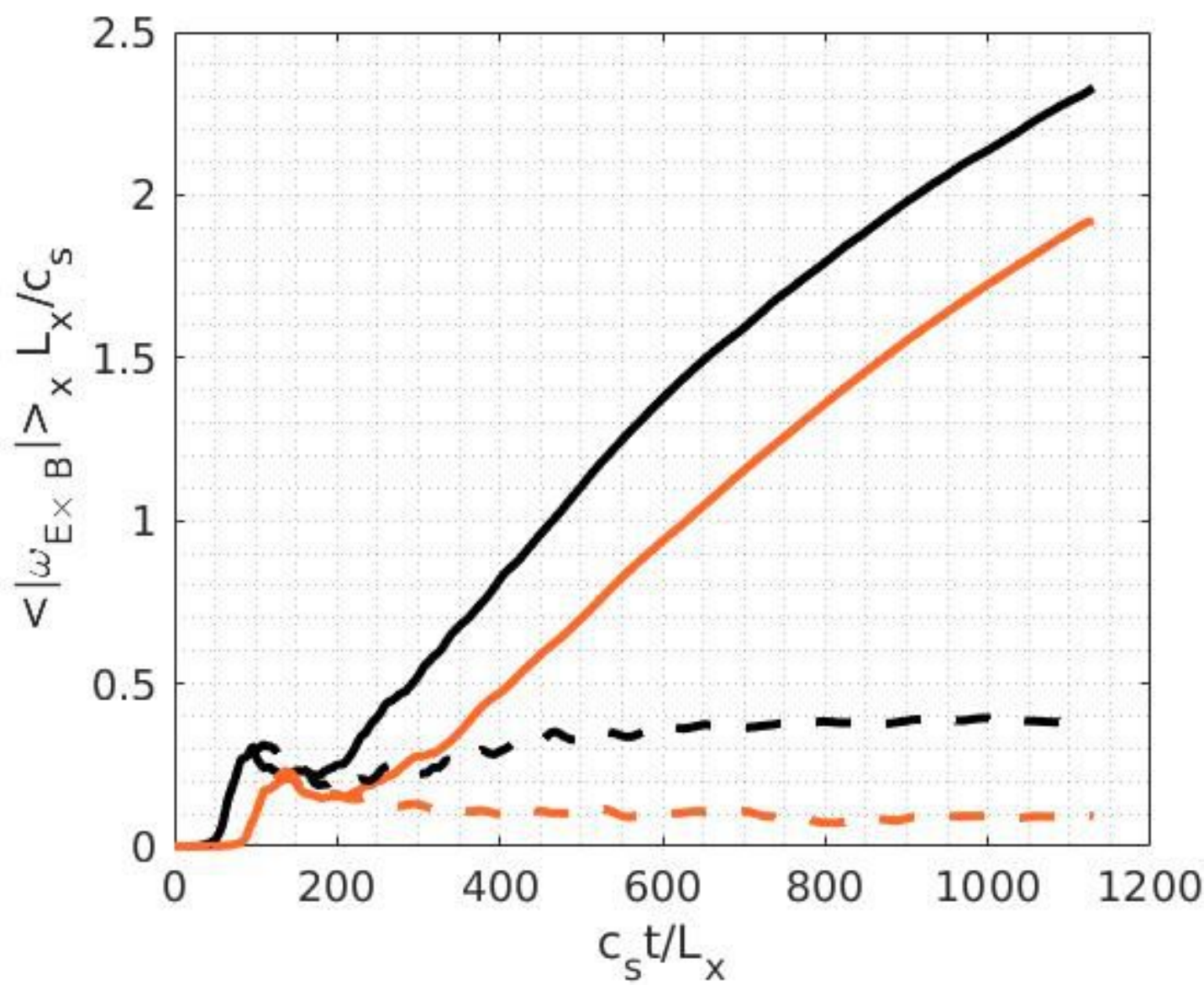}
	\end{subfigure}
	\begin{subfigure}[c]{0.49\columnwidth}
	    \caption{\label{fig:snrnz_gamma1}}
	    \centering
		\includegraphics[width=1.0\linewidth]{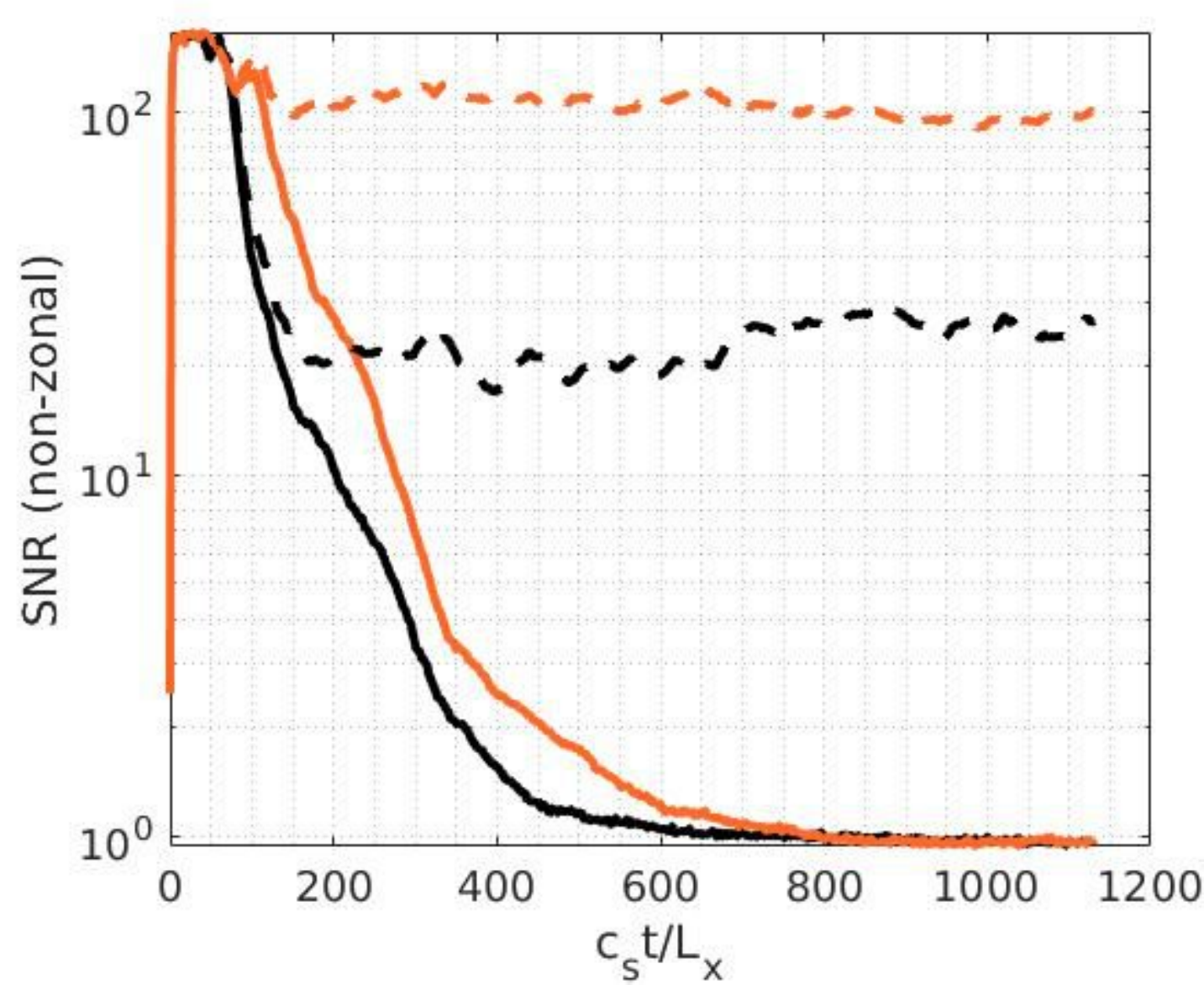}
	\end{subfigure}
	\begin{subfigure}[c]{0.49\columnwidth}
        \caption{\label{fig:sigmaw_gamma1}}
	    \centering
		\includegraphics[width=1.0\linewidth]{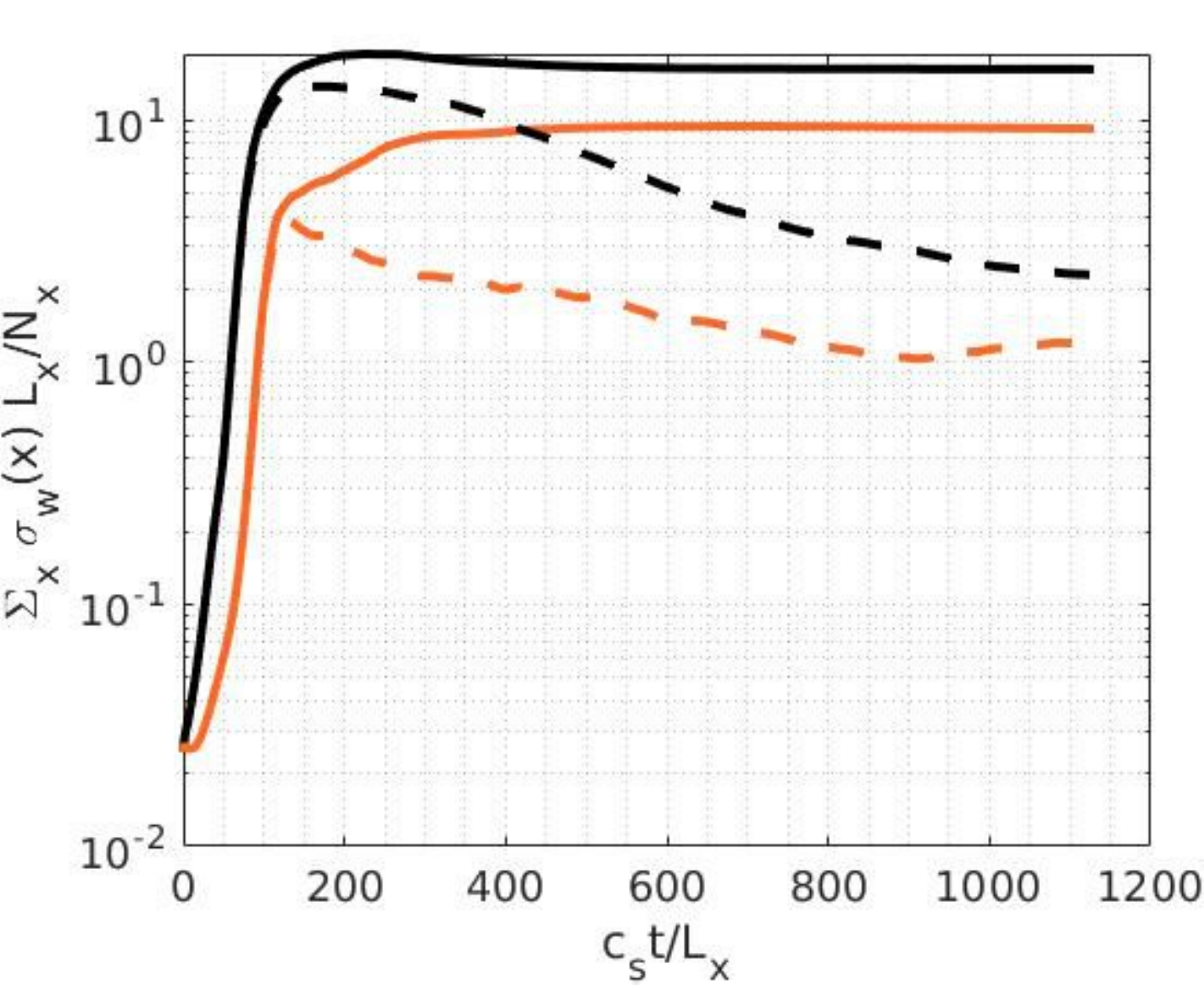}
	\end{subfigure}
	\caption{Diagnostics for various noise control strength $\gamma_n$ considering both the non-adaptive and adaptive cases, for the radially averaged absolute (a) heat diffusivity $\chi_H$ and (b) zonal flow shearing rate $\omega_{E\times B}$, and the (c) signal-to-noise (SNR) ratio with signal excluding the $(m,n)=(0,0)$ mode, and the (d) sum of f.s.a. weight standard deviation. Marker number set to $N_p=256$M. The sum of f.s.a.~weight standard deviation is calculated by summing the standard deviations of f.s.a.~weights from each radial bin, and multiplying by the sum by $L_x/N_x$.}
	\label{fig:gamma1}
\end{figure}

The purpose of the conservative noise control is to reduce the weight standard deviation at the expense of \amend introducing numerical diffusion, thus affecting the validity of the simulation by adding artificial damping on the main instability drive and zonal flows. \color{black} From Fig.~\ref{fig:chi_gamma1} and \ref{fig:omega_gamma1}, it can be seen that at a larger $\gamma_n$ value, both $\chi_H$ and $\omega_{E\times B}$ values are lower. Therefore, its amplitude $\gamma_n$ should be adjusted just high enough to maintain a good SNR value throughout the simulation, taken in this work to be $10$. Fig.~\ref{fig:snrnz_gamma1} shows that $\gamma_n=0.03\gamma_{max}$ for the adaptive case is just enough, and it is this value of $\gamma_n$ that is used in all other sections of the paper. For the non-adaptive case, it is seen that a high $\gamma_n$ value only postpones the eventual decrease in SNR, implying that noise control alone is insufficient to prevent simulations being drowned in noise. Finally, the reduced weight standard deviation is shown in Fig.~\ref{fig:sigmaw_gamma1}. It is shown that $S_n$ alone is insufficient to control noise to acceptable levels, while the adaptive scheme is able to do so even in the presence of minimal noise control relaxation rate $\gamma_n$.

\begin{figure*}
	\centering
	\begin{subfigure}[c]{0.19\textwidth}
	    \caption{$\alpha_E=0.00\gamma_{max}$}
		\centering
		\includegraphics[width=1.0\linewidth]{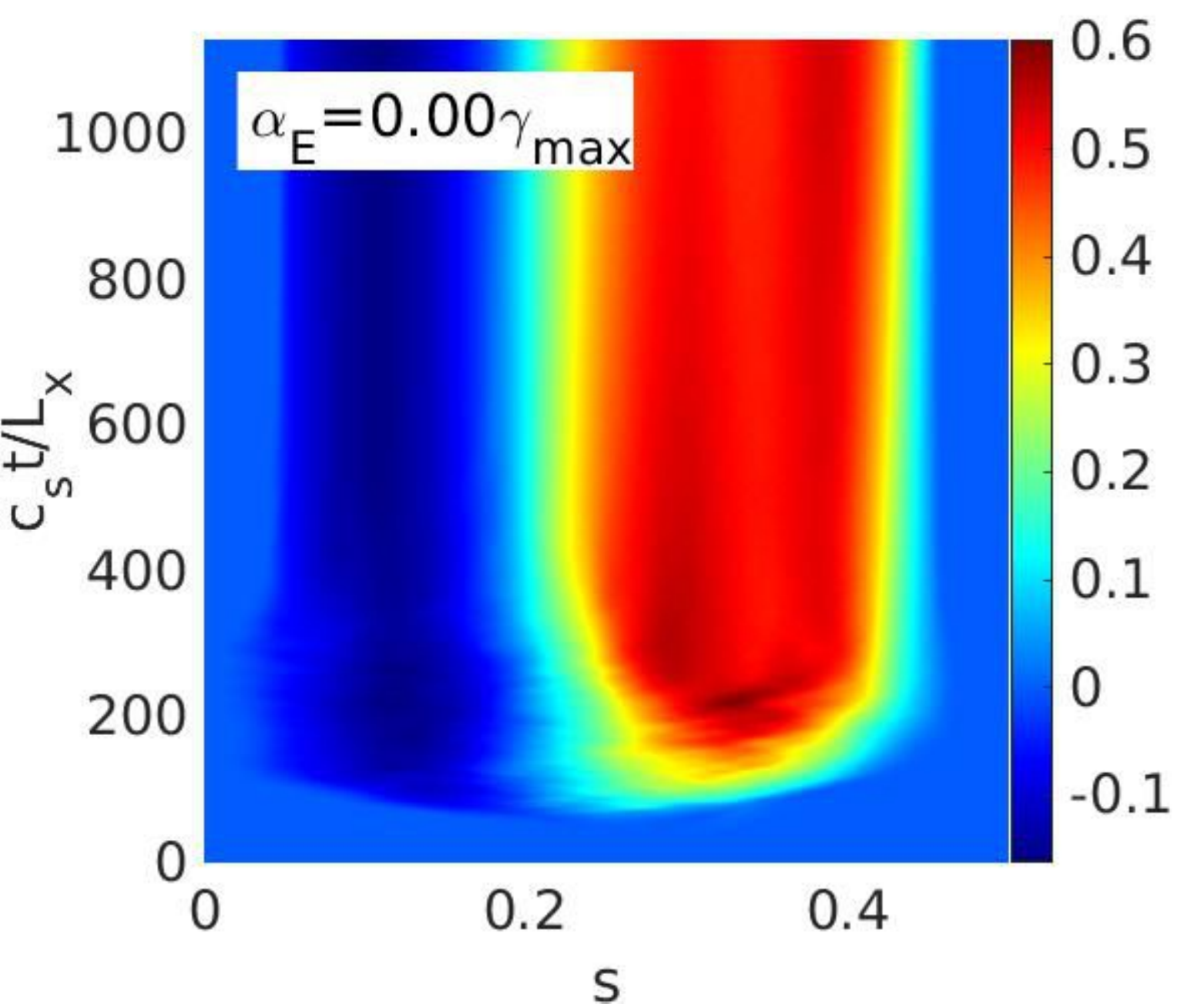}
	\end{subfigure}
	\begin{subfigure}[c]{0.19\textwidth}
	    \caption{$\alpha_E=0.03\gamma_{max}$}
		\centering
		\includegraphics[width=1.0\linewidth]{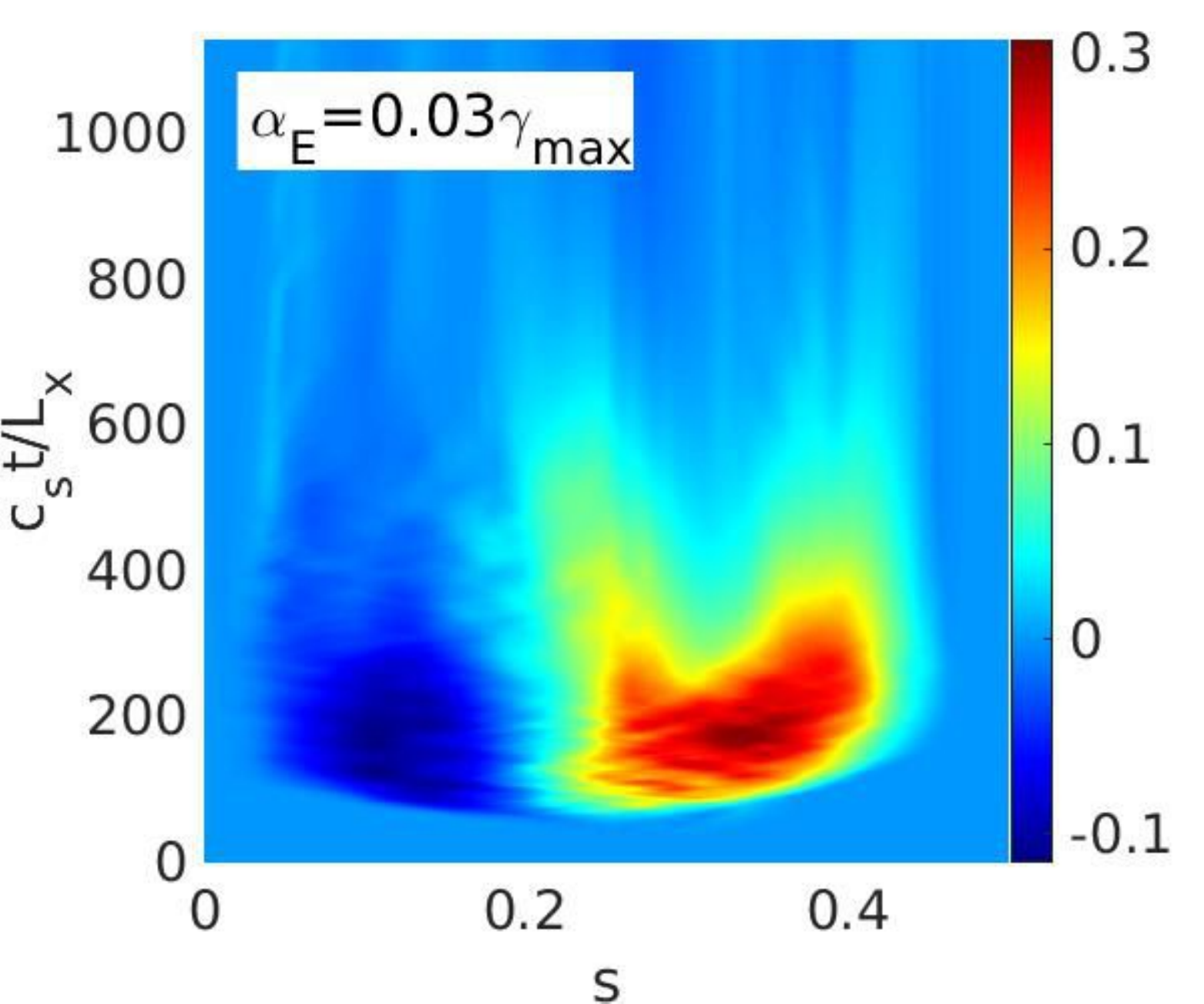}
	\end{subfigure}
	\begin{subfigure}[c]{0.19\textwidth}
	    \caption{$\alpha_E=0.12\gamma_{max}$}
		\centering
		\includegraphics[width=1.0\linewidth]{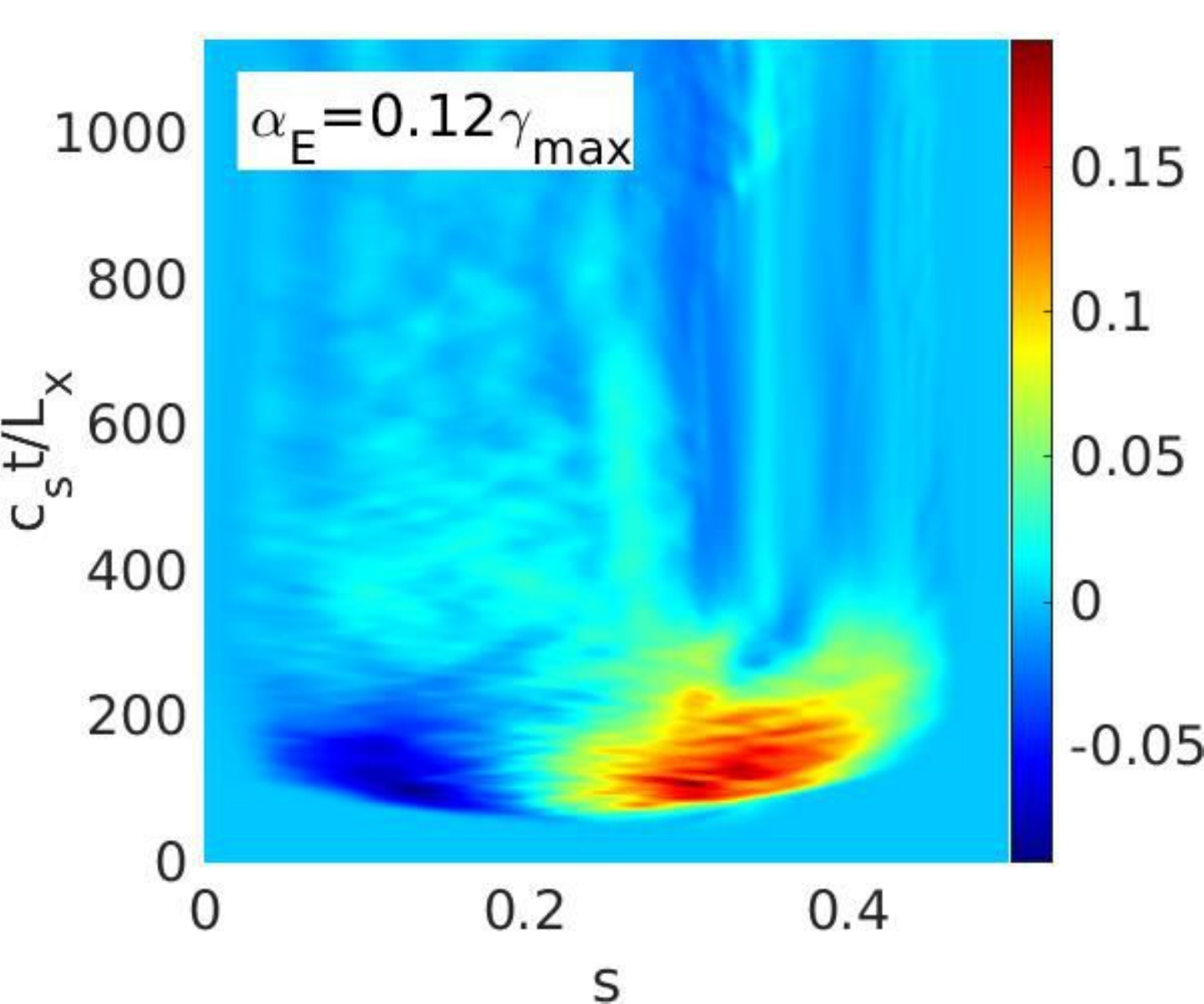}
	\end{subfigure}
	\begin{subfigure}[c]{0.19\textwidth}
	    \caption{$\alpha_E=0.48\gamma_{max}$}
		\centering
		\includegraphics[width=1.0\linewidth]{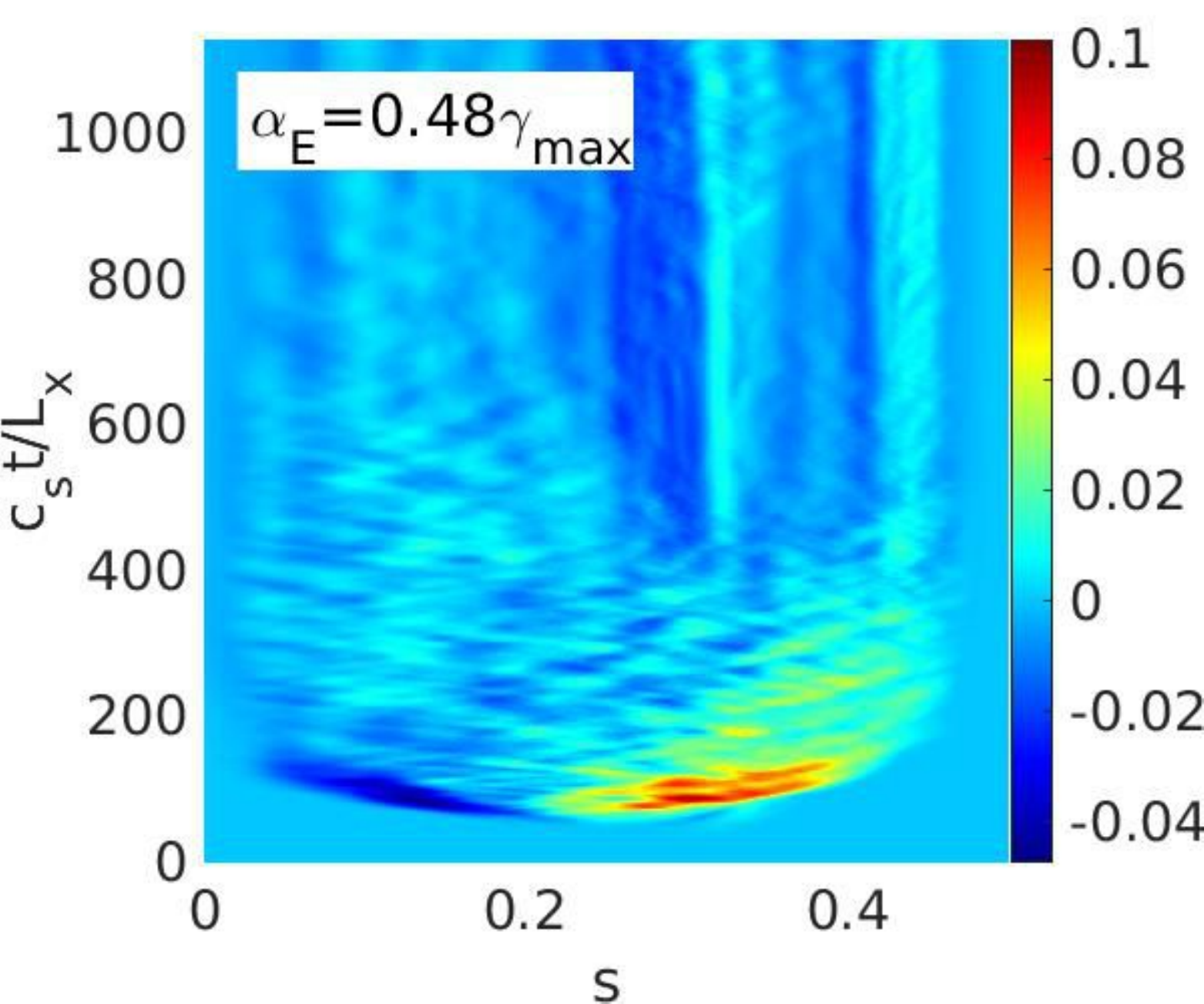}
	\end{subfigure}
	\begin{subfigure}[c]{0.19\textwidth}
	    \caption{$\alpha_E=1.92\gamma_{max}$}
		\centering
		\includegraphics[width=1.0\linewidth]{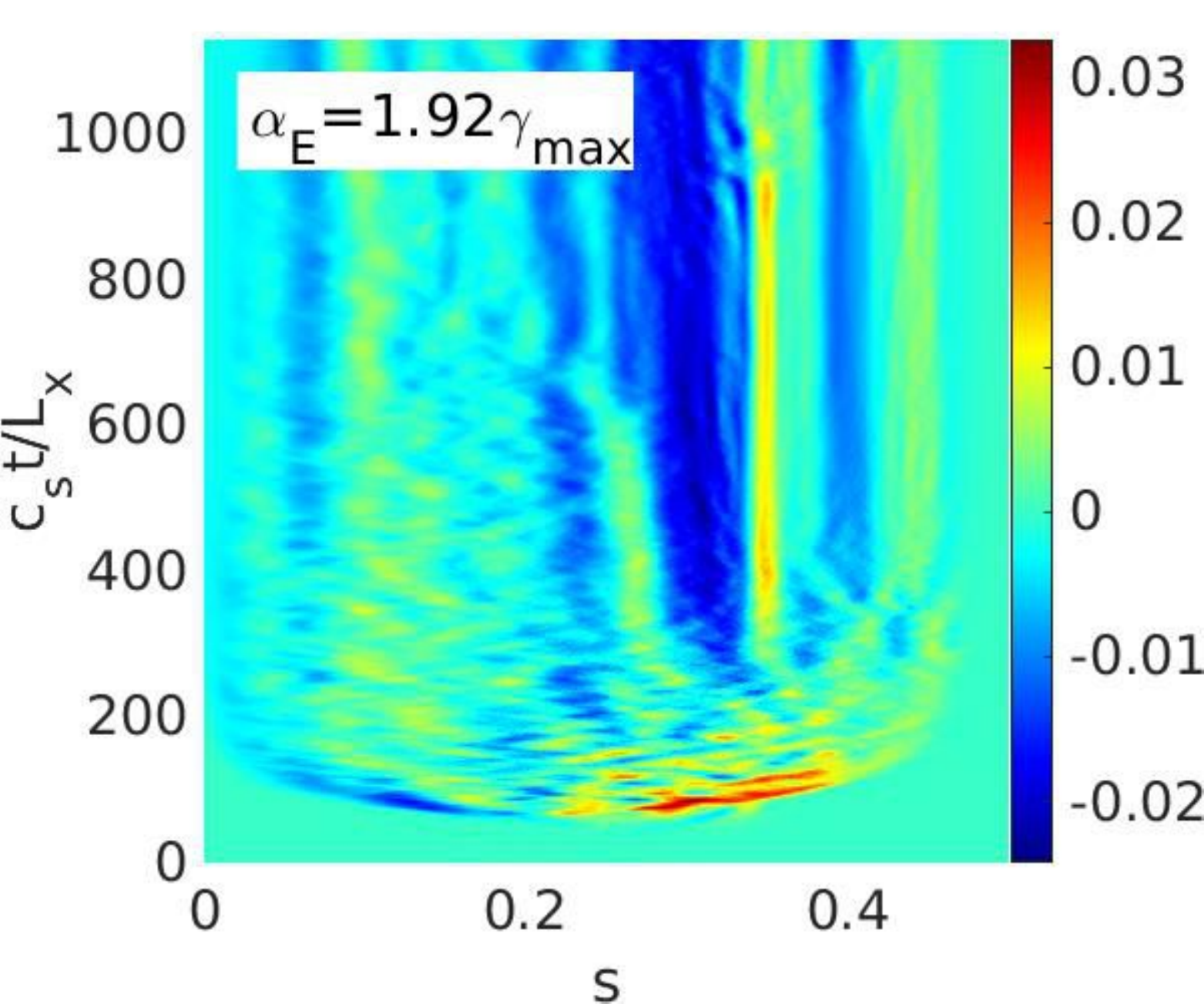}
	\end{subfigure}
	\caption{Time evolution of f.s.a.~ion temperature relative deviation from its time-evolved adapted $T_{i0}(t)$, $(\langle T_i\rangle(x,t)-T_{i0}(x,t))/T_{i0}(t)$ under various adaptive rates $\alpha_E$. Marker number set to $N_p=256$M.}
	\label{fig:relTi_alpha}
\end{figure*}

\begin{figure}[H]
	\centering
	\begin{subfigure}[c]{1.0\columnwidth}
	    \caption{\label{fig:omega_alpha}}
	    \centering
		\includegraphics[width=0.6\linewidth]{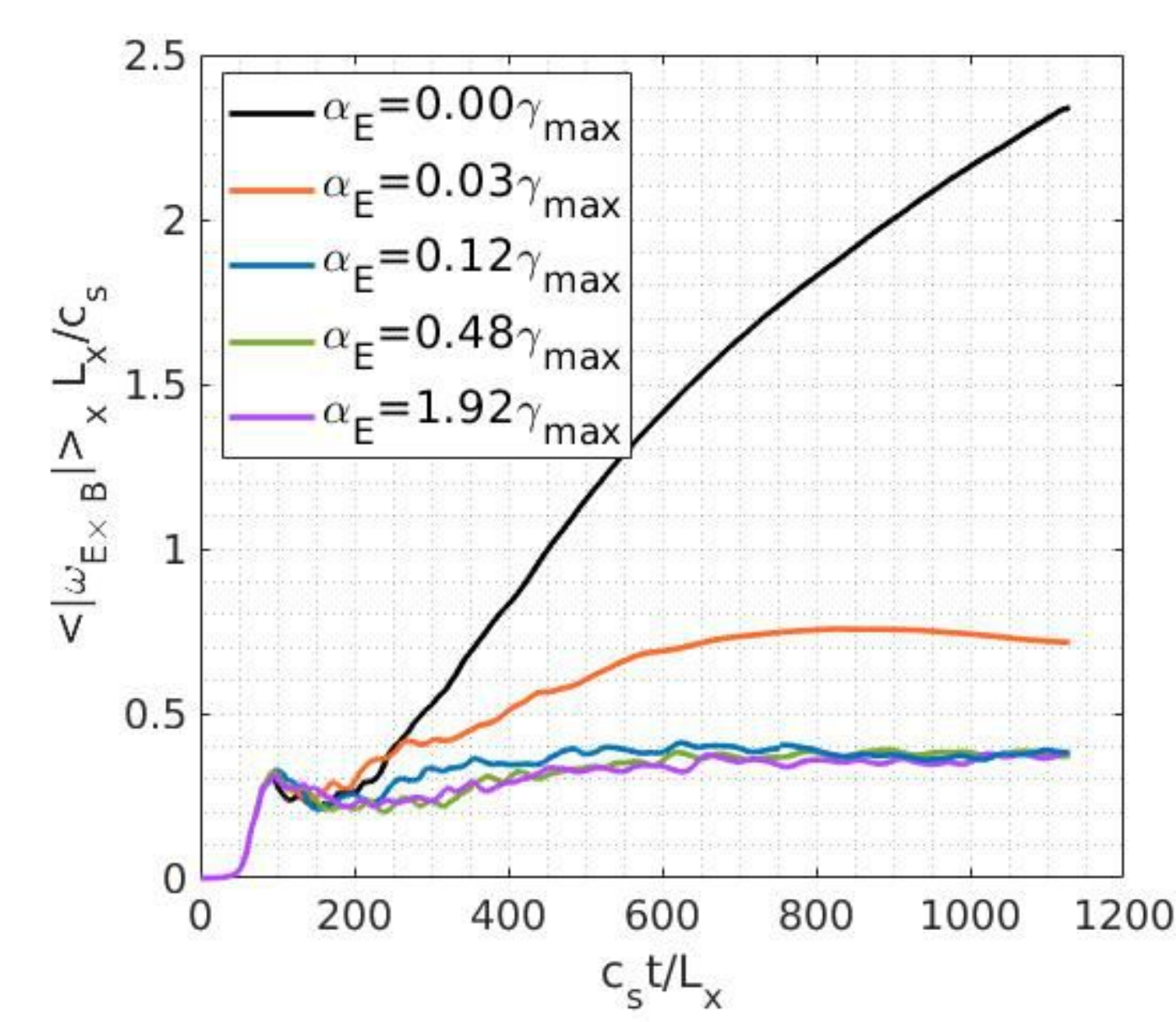}
	\end{subfigure}
	\begin{subfigure}[c]{1.0\columnwidth}
	    \caption{\label{fig:snrnz_alpha}}
	    \centering
		\includegraphics[width=0.6\linewidth]{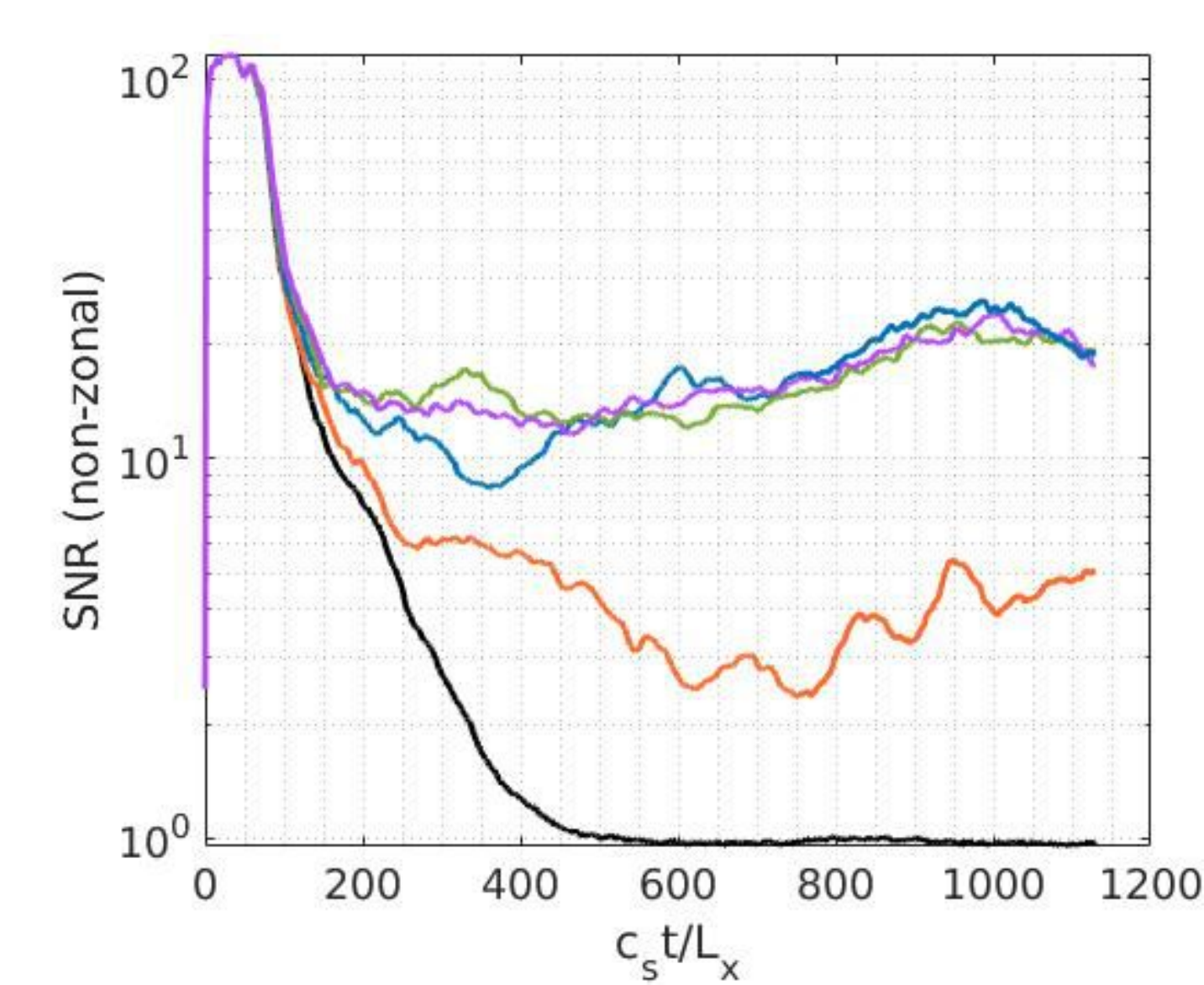}
	\end{subfigure}
	\caption{Time traces of the (a) radially averaged absolute zonal shearing rate $\omega_{E\times B}$ and the (b) signal-to-noise ratio (SNR) with signal excluding the $(m,n)=(0,0)$ mode, with varying adaptation rates $\alpha_E$. Marker number set to $N_p=256$M.}
	\label{fig:alpha}
\end{figure}

Turning now to the choice of the adaptive rate $\alpha_E$, from Eqs.~(\ref{eq:adp_relax}) and (\ref{eq:dw_adp}), the greater the $\alpha_E$ value, the greater the rate of transfer of the second velocity moment of the f.s.a.~$\delta f$ to the control variate, in this case, $f_0=f_M(t)$. Fig.~\ref{fig:relTi_alpha} shows that the maximum relative deviation of $T_i$ from its time evolved adapted profile $T_{i0}(x,t)$ is lower with higher values of $\alpha_E$. Also the decrease of the relative deviation after the initial burst is faster with increasing $\alpha_E$. More specifically, fig.~\ref{fig:omega_alpha} and \ref{fig:snrnz_alpha} show that for $\alpha_E=0.12\gamma_{max}$ or higher, the simulation under given parameters is sufficient in terms of low zonal flow shearing rate $\omega_{E\times B}$ and high enough SNR values respectively. \amend Therefore, under the parameters studied in this paper, any $\alpha_E$ value satisfying $\alpha_E>0.12\gamma_{max}$ and Eq.~(\ref{eq:Nalpha}) gives results with the lowest noise accumulation. \color{black}

\subsection{Adjusting the f.s.a.~potential term} \label{sec:fsa}

In sheared-slab geometry with adiabatic electrons on the magnetic surfaces, ITG turbulence is strongly suppressed by zonal flows, resulting in a quasi-steady state with relatively low heat fluxes. In a real tokamak, much stronger heat fluxes and large relative fluctuation amplitudes are present in the plasma edge. In order to emulate such a situation but staying in slab geometry, the f.s.a.~potential term $\langle\phi\rangle$ of the adiabatic electron response of Eq.~(\ref{eq:qn_df}) is tuned by defining a multiplicative parameter $\lambda$:
\amend
\begin{eqnarray}
& &
\frac{en_0}{T_e}\left(\phi-\lambda\langle\phi\rangle\right)-\nabla_\perp\cdot\left(\frac{m_in_0}{eB^2}\nabla_\perp\phi\right) \nonumber \\
&=&\int\dint{^3R}\dint{\alpha}\dint{\vp}\dint{\mu}\,\frac{B_\parallel^\star}{m_i}\delta[\vec{r}-(\vec{R}+\vec{\rho}_L(\mu,\alpha))]\times \nonumber \\
& &[f_0(\vec{R},\vp,\mu,t)-f_0(\vec{R},\vp,\mu,0) + \delta f] \label{eq:qn_lambda}.
\end{eqnarray}
\color{black}

With $\lambda=0.00$ \cite{Hatch2014,Watanabe2002}, the electrons respond adiabatically in all directions, i.e. not only in the magnetic surface, but also radially. This results in much lower $E\times B$ flows and thus higher turbulent heat fluxes.

\begin{figure}[H]
	\centering
	\begin{subfigure}[c]{0.49\columnwidth}
	    \caption{\label{fig:chi_nadp_lambda}}
		\centering
		\includegraphics[width=1.0\linewidth]{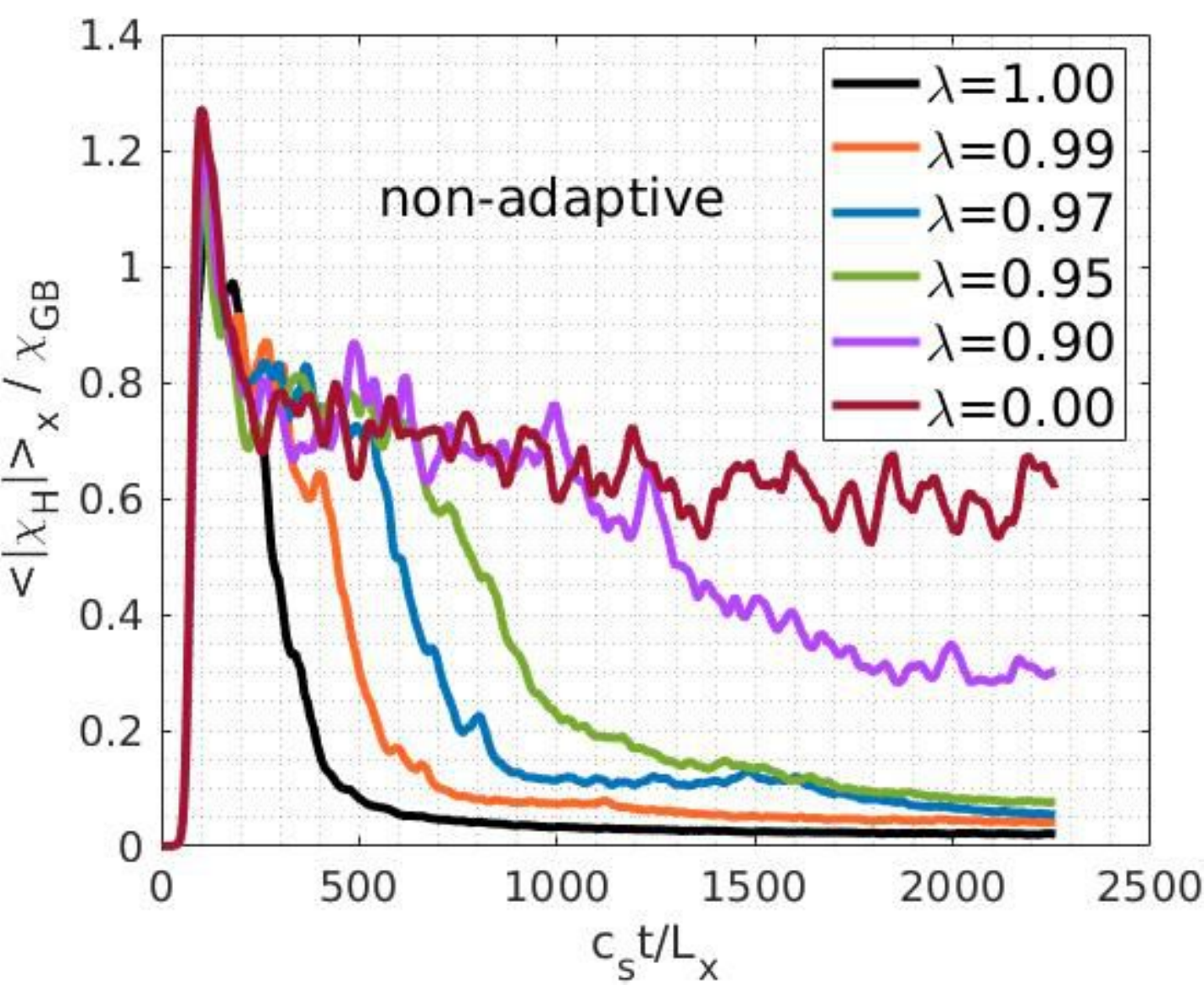}
	\end{subfigure}
	\begin{subfigure}[c]{0.49\columnwidth}
	    \caption{\label{fig:chi_adp_lambda}}
		\centering
		\includegraphics[width=1.0\linewidth]{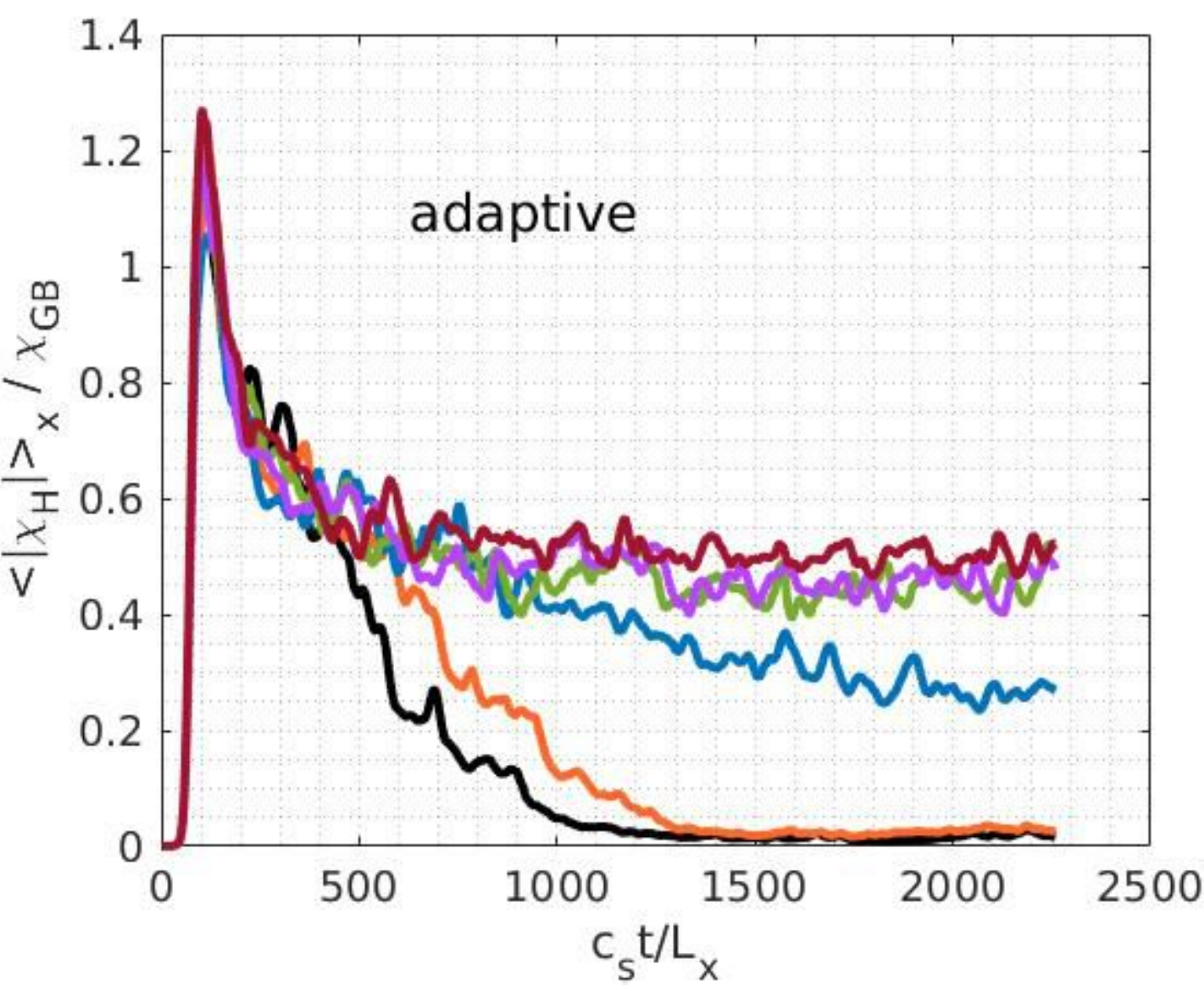}
	\end{subfigure}
	\begin{subfigure}[c]{0.49\columnwidth}
	    \caption{\label{fig:omega_nadp_lambda}}
		\centering
		\includegraphics[width=1.0\linewidth]{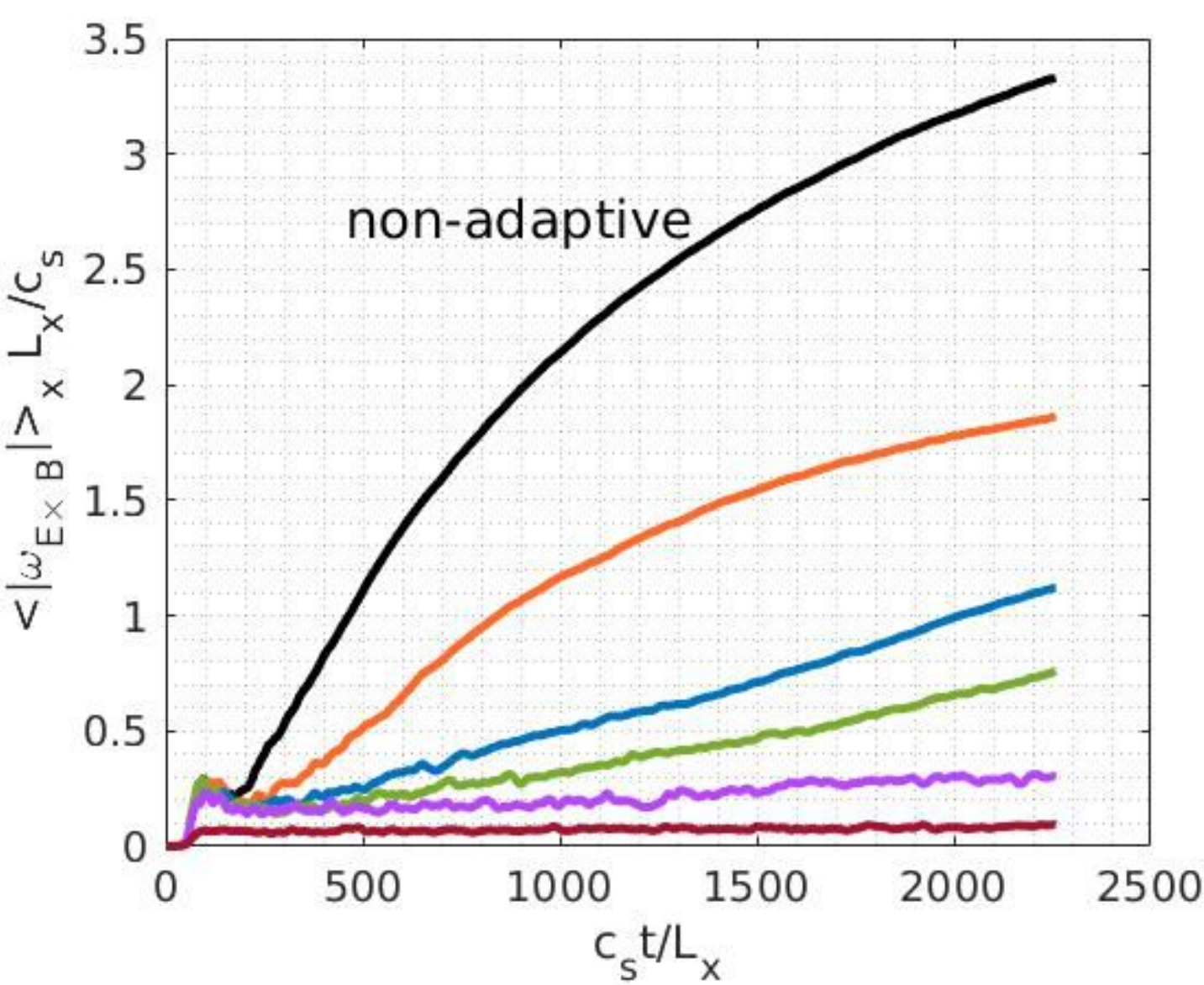}
	\end{subfigure}
	\begin{subfigure}[c]{0.49\columnwidth}
	    \caption{\label{fig:omega_adp_lambda}}
		\centering
		\includegraphics[width=1.0\linewidth]{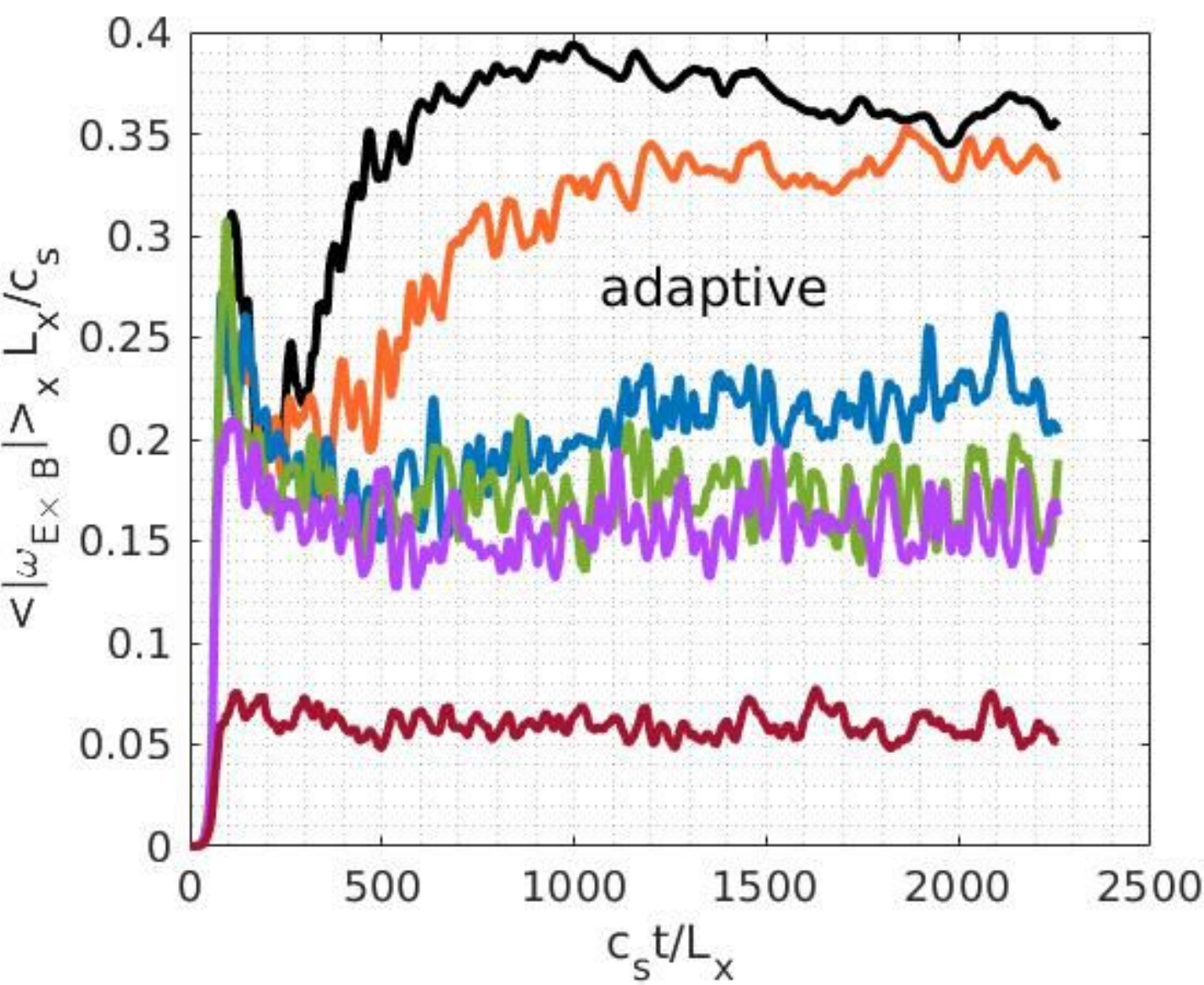}
	\end{subfigure}
	\caption{Time traces of radial averaged absolute value of heat diffusivity $\chi_H$, (a) and (b), and zonal flow shearing rate $\omega_{E\times B}$, (c) and (d), for the non-adaptive and adaptive cases respectively, under various tuning parameter $\lambda$ (see Eq.~(\ref{eq:qn_lambda})). The adaptive rate is set to $\alpha_E=1.92\gamma_{max}$. A moving time-averaging window of half-width $c_st/L_x=10$ has been implemented. Total number of markers is set to $N_p=256$M.}
	\label{fig:chi_omega_lambda}
\end{figure}

\begin{figure}[H]
	\centering
	\begin{subfigure}[c]{1.0\columnwidth}
	    \caption{\label{fig:omega_l095}}
	    \centering
		\includegraphics[width=0.6\linewidth]{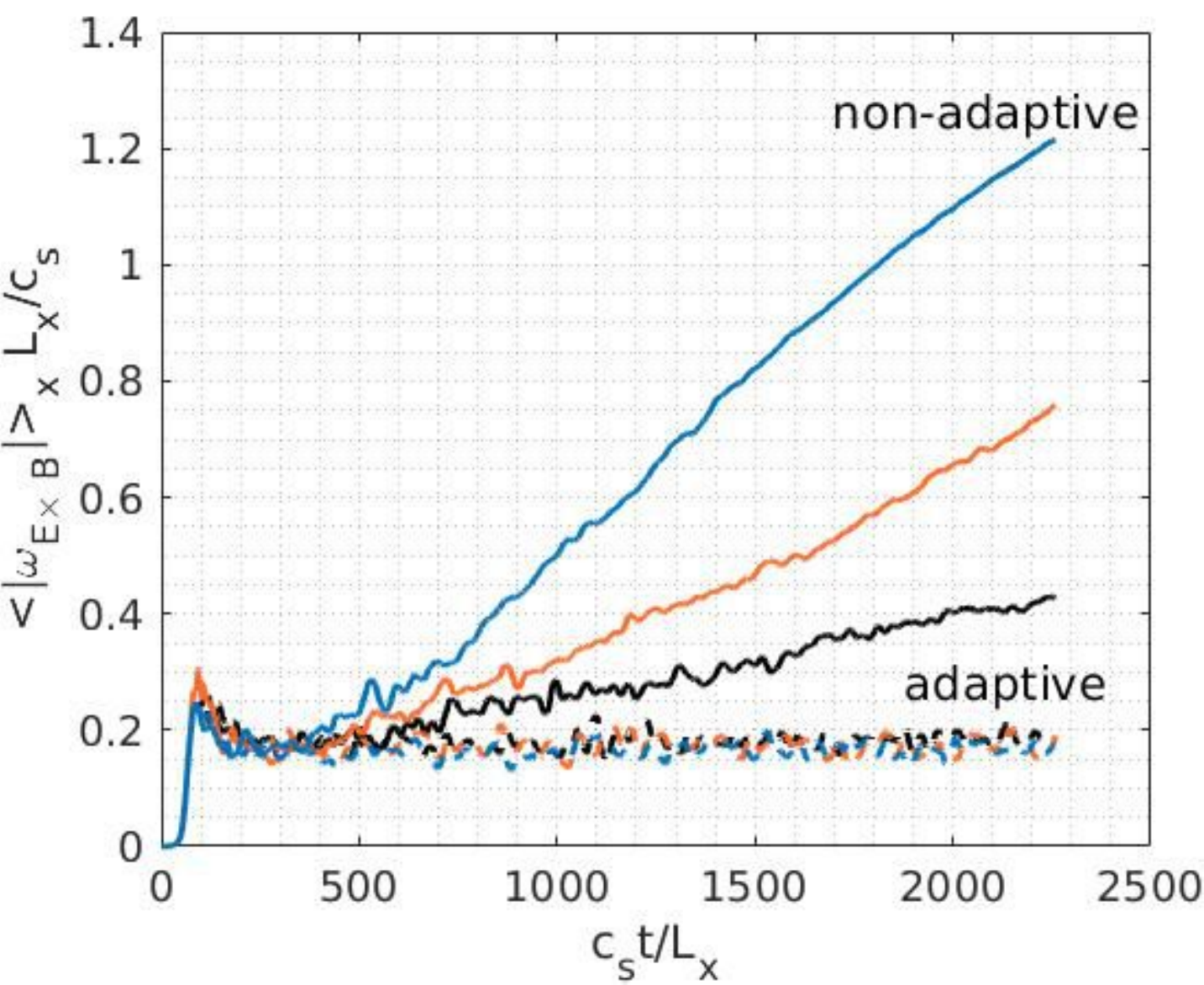}
	\end{subfigure}
	\begin{subfigure}[c]{1.0\columnwidth}
	    \caption{\label{fig:snrnz_l095}}
	    \centering
		\includegraphics[width=0.6\linewidth]{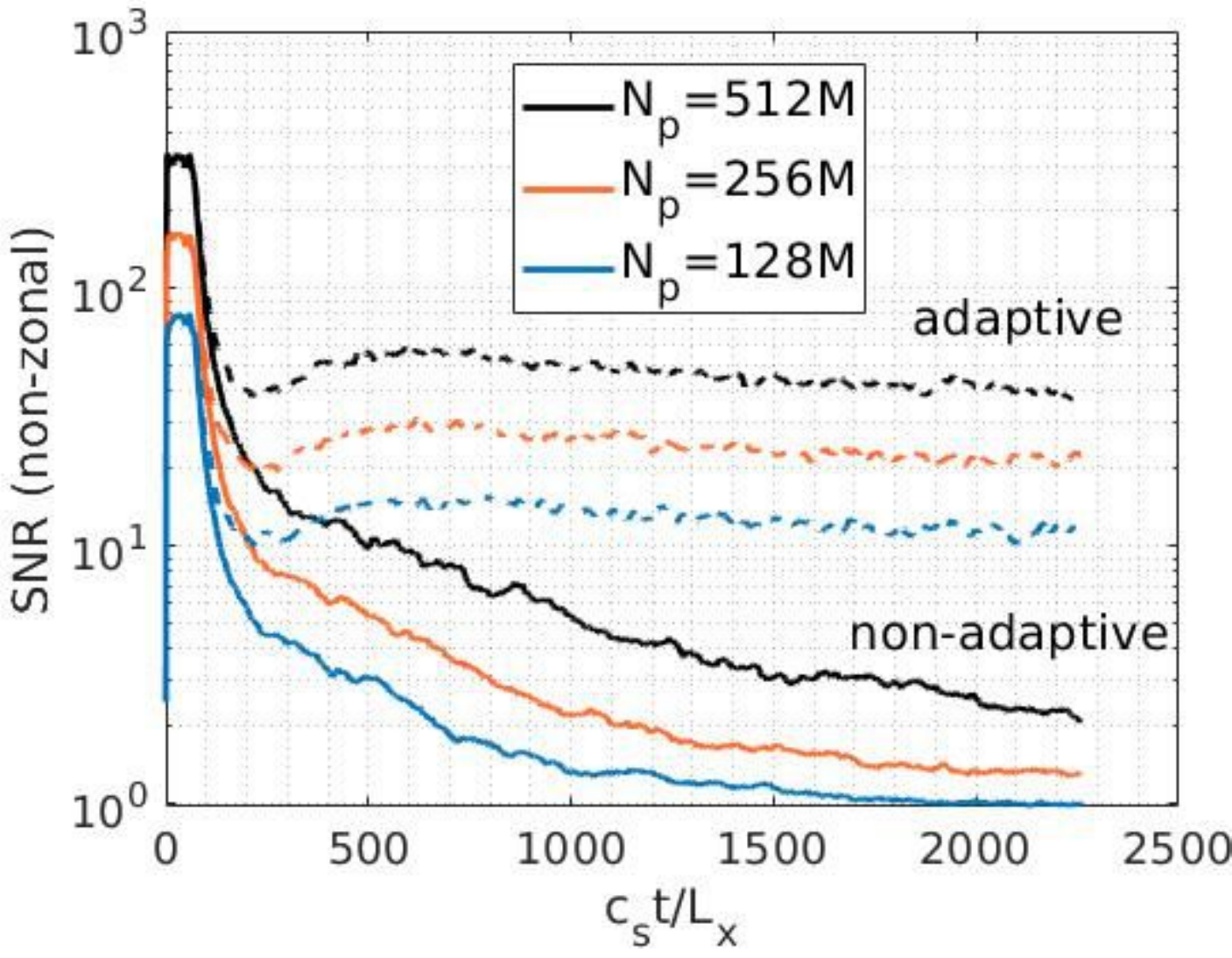}
	\end{subfigure}
	\caption{Time traces of the (a) radially averaged absolute zonal flow shearing rate $\omega_{E\times B}$, and the (b) signal-to-noise ratio (SNR) with signal excluding the $(m,n)=(0,0)$ mode. The tuning parameter is set to $\lambda=0.95$, and the adaptive rate is set to $\alpha_E=1.92\gamma_{max}$ where applicable. A moving time-averaging window of half-width $c_st/L_x=10$ has been implemented.}
	\label{fig:omega_snrnz_l095}
\end{figure}

Fig.~\ref{fig:chi_omega_lambda} shows the effect of tuning $\lambda$ on the heat diffusitivity $\chi_H$ and zonal shearing rate $\omega_{E\times B}$. One can see that fluxes are sustained longer and higher due to a lower zonal flow shearing rate $\omega_{E\times B}$ from a greater attenuation of $\langle\phi\rangle$. This trend also exists for the adaptive cases, though $\omega_{E\times B}$ levels there are generally low as compared to the  non-adaptive cases, see Fig.~\ref{fig:omega_adp_lambda}. From Fig.~\ref{fig:chi_adp_lambda} for the adaptive cases, the value of $\lambda=0.95$ seems to be just sufficient to sustain the flux. Therefore,  looking at the case of $\lambda=0.95$ specifically, Fig.~\ref{fig:omega_snrnz_l095} shows that the result trend is similar to that of Fig.~\ref{fig:chi_omega_np}, albeit with lower levels of $\omega_{E\times B}$. Under this tuning, the non-adaptive cases (continuous lines) seem to evolve towards low $\chi_H$ values at long times, whereas the adaptive cases (dashed lines) are maintained at a higher $\chi_H$ value as compared to Fig.~\ref{fig:chi_np}. The adaptive cases seem to have converged already with $N_p=128$M markers whereas the non-adaptive case is still subject to collapse even with $N_p=512$M markers.

\begin{figure}[H]
	\centering
	\includegraphics[width=0.8\linewidth]{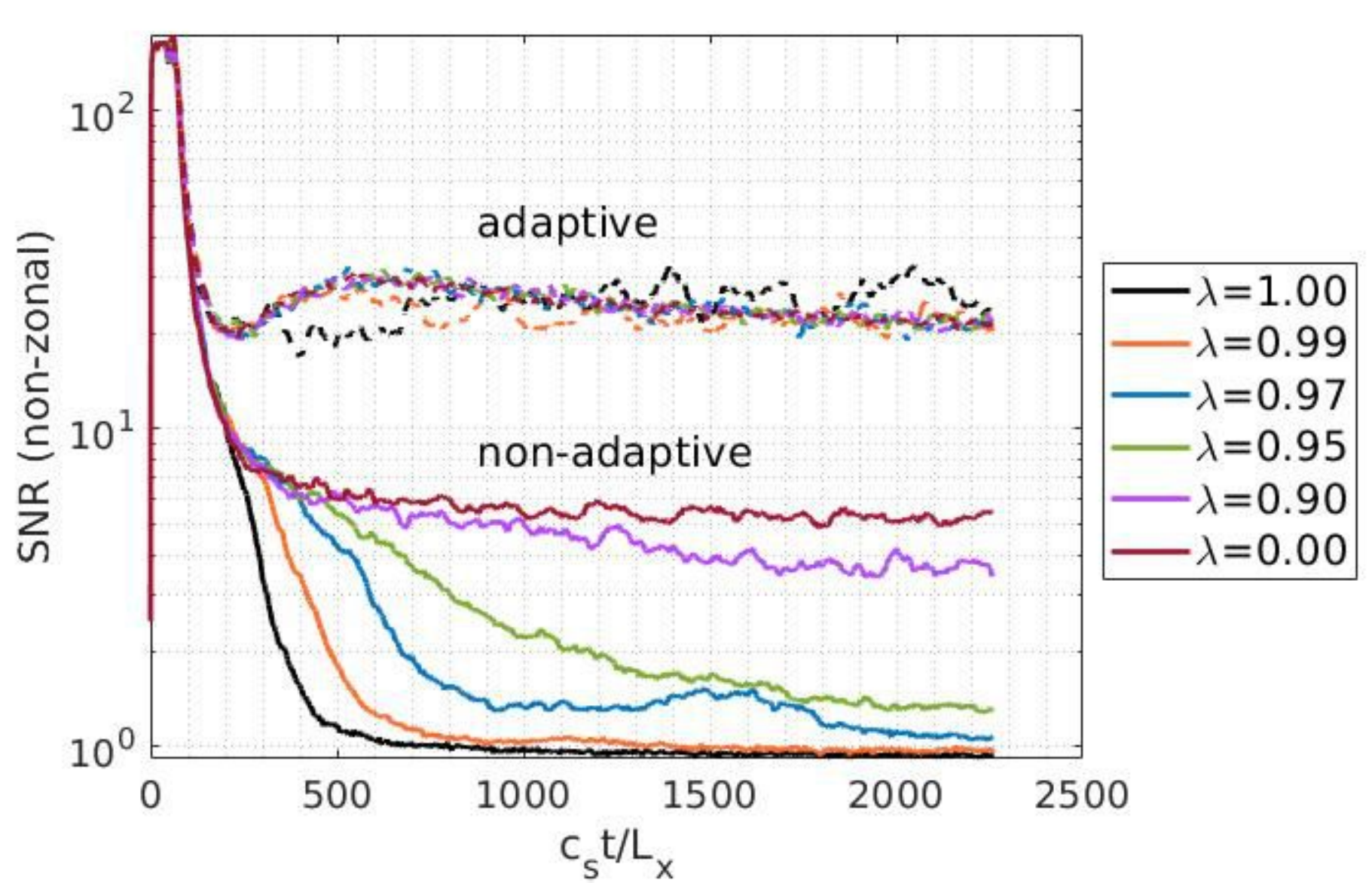}
	\caption{Signal-to-noise ratio (SNR) time traces with signal excluding the $(m,n)=(0,0)$ mode, under various tuning parameter $\lambda$ (see Eq.~(\ref{eq:qn_lambda})) considering the non-adaptive and the adaptive cases. Marker number set to $N_p=256$M, and adaptive rate to $\alpha_E=1.92\gamma_{max}$ where applicable.}
	\label{fig:snrnz_lambda}
\end{figure}

Fig.~\ref{fig:snrnz_lambda} shows that $\lambda$ only affects the SNR values of the non-adaptive cases. Taking the standard non-adaptive case of $\lambda=1.00$ as reference, the effect of higher attenuation of $\langle\phi\rangle$ only delays the eventual collapse of SNR values for each case, except for the case of complete $\langle\phi\rangle$ suppression, in which high SNR value and sustained flux (see Fig.~\ref{fig:chi_nadp_lambda}) are achieved. \amend For a fixed value of $\lambda=0.95$, besides the delayed fall of SNR values for the non-adaptive cases, Fig.~\ref{fig:snrnz_l095} reflects the SNR value proportional to $N_p$ relation, as was already shown in Fig.~\ref{fig:snrnz_np}. This convergence, instead of $\sqrt{N_p}$, is a result of taking the noise as a quadratic measure, see Eq.~(\ref{eq:snr}). \color{black}

\begin{figure}[H]
	\centering
		\begin{subfigure}[c]{0.49\columnwidth}
		    \caption{}
			\centering
			\includegraphics[width=1.0\linewidth]{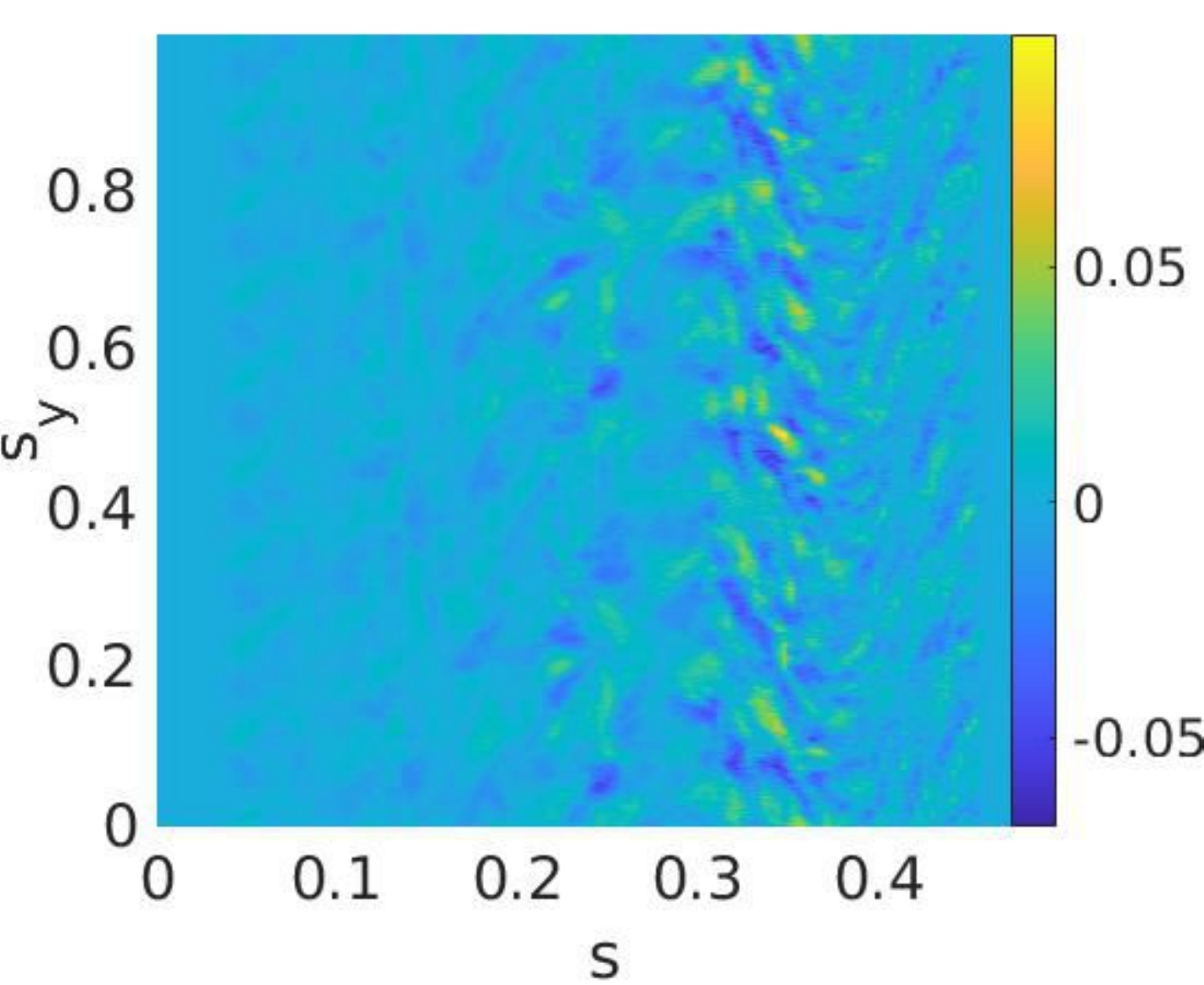}
		\end{subfigure}
		\begin{subfigure}[c]{0.49\columnwidth}
		    \caption{}
			\centering
			\includegraphics[width=1.0\linewidth]{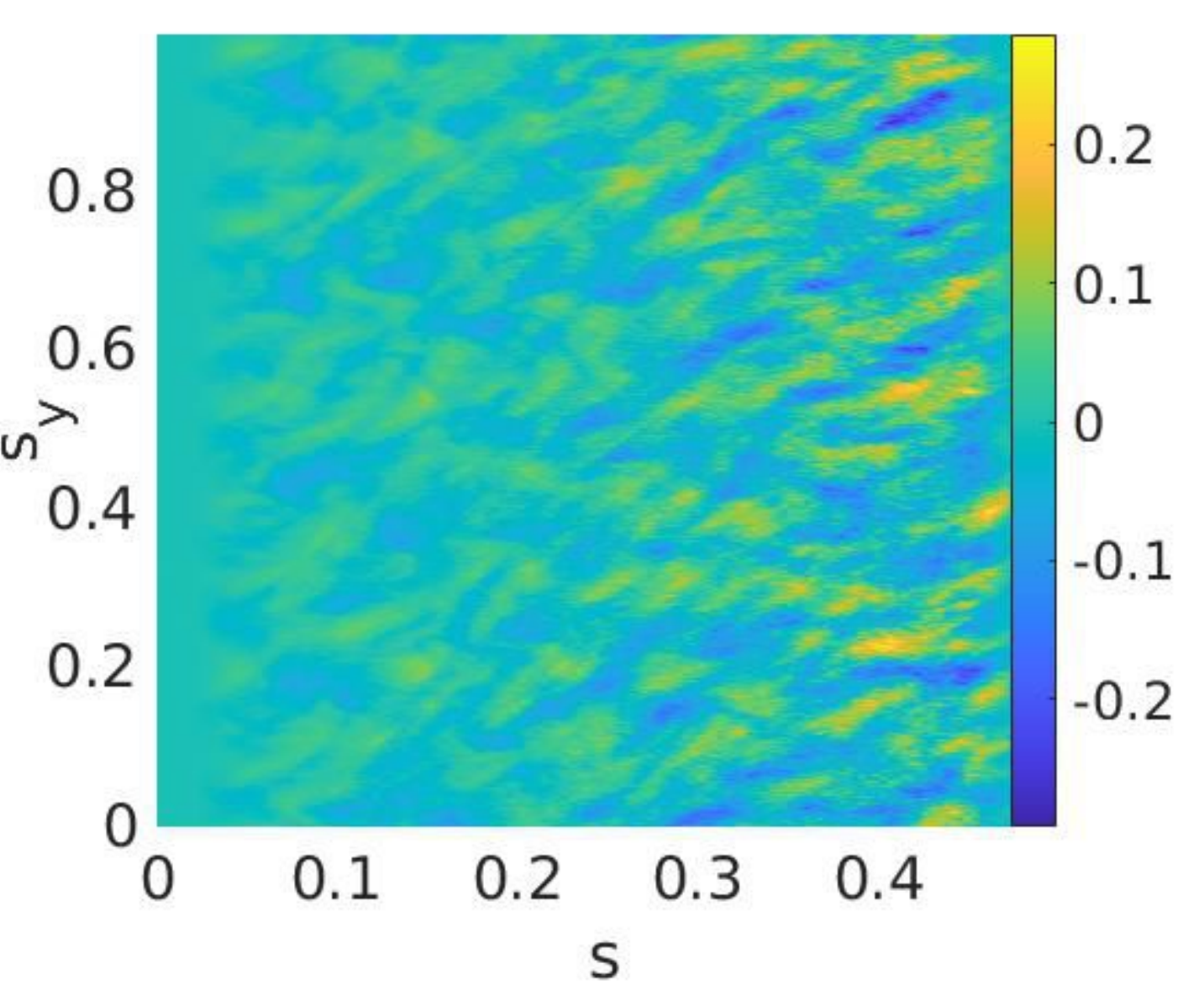}
		\end{subfigure}
		\caption{Time snapshot at $c_st/L_x=2259$ with tuning parameter $\lambda$ at (a) 1.00 and (b) 0.95, of the ion temperature $T_i$ relative non-zonal deviation on each magnetic surface, expressed by $(\delta T_i-\langle\delta T_i\rangle) / \langle T_i\rangle$, where $T_i=T_{i0}+\delta T_i$, integrated over the toroidal $z$ direction. $s_y$ is the normalized poloidal $y$ axis. Both quasi-steady state cases have adaptive rate and marker number set to $\alpha_E=1.92\gamma_{max}$ and $N_p=256$M, respectively.}
		\label{fig:relfluc_space}
\end{figure}

We now consider simulations with $\lambda=0.95$, which demonstrated high sustained flux for the adaptive cases for the following analysis. Fig.~\ref{fig:relfluc_space} shows that the fixed-time non-zonal $T_i$ relative deviation across each magnetic surface increases towards the low-end of the $\langle T_i\rangle$ profile at quasi-steady state. The case with $\lambda=0.95$ has relative deviation at least twice that of $\lambda=1.00$, indicating the expected higher levels of turbulence. Since the adaptive scheme implemented in this work adapts its control variate $T_{i0}$ by its f.s.a.~values as shown in Eq.~(\ref{eq:adp_relax}), it is not expected to further improve noise reduction for edge plasma simulations involving relative non-zonal deviation much higher than $25\%$. However, a similar adaptive scheme could still be used for noise control, provided that $f_0$ is now a function of all spatial dimensions. This would allow for a transfer of non-zonal components of $\delta f$ to $f_0$, though with increased noise levels due to lower $N_p$ per spatial bin when implementing Eq.~(\ref{eq:adp_relax}).

\begin{figure}[H]
	\centering
	\includegraphics[width=0.7\linewidth]{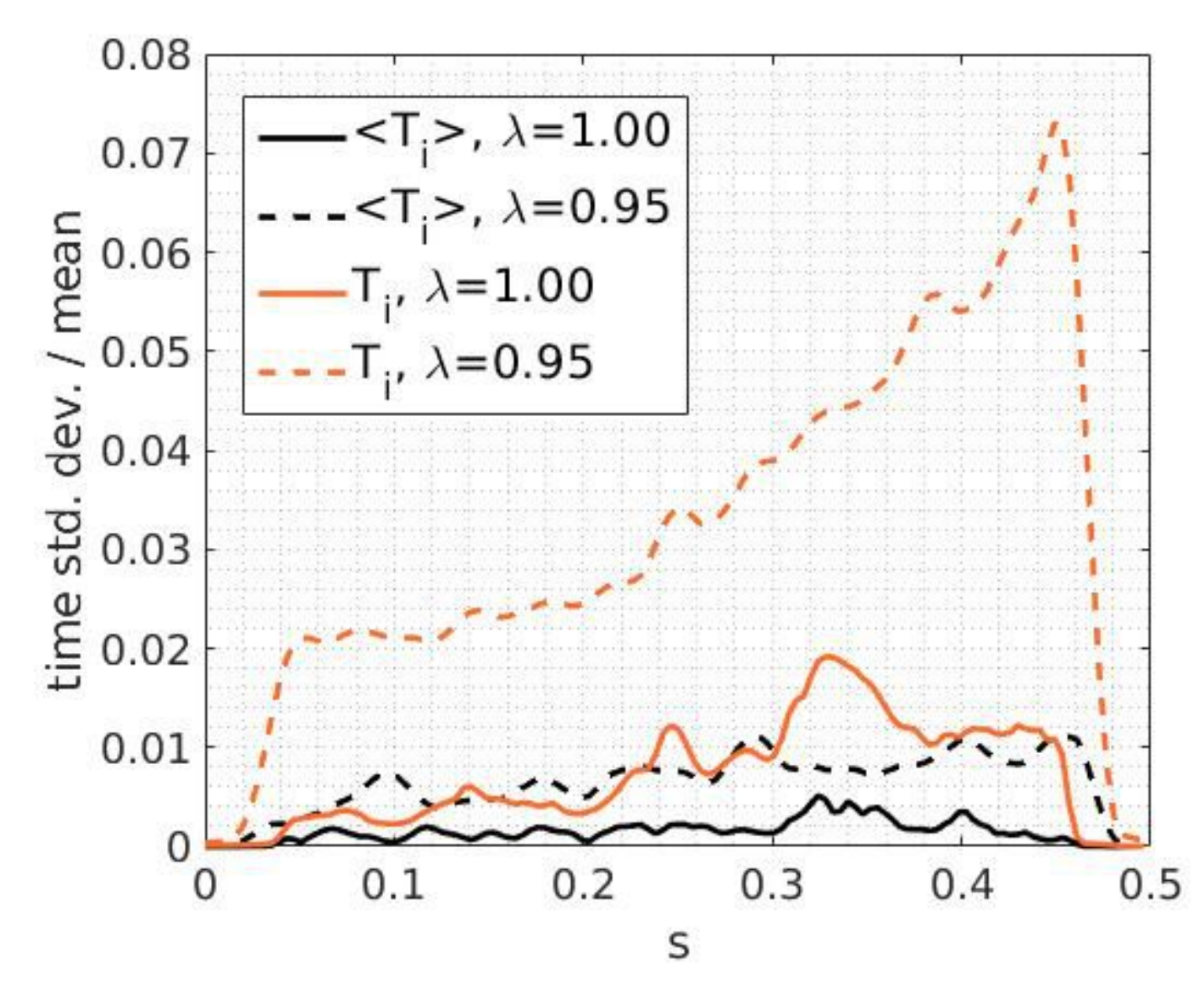}
	\caption{Relative fluctuation averaged over $c_st/L_x\in[1807,2259]$ of ion temperature $T_i$ and its f.s.a.~profile $\langle T_i\rangle$, as measured by its standard deviation over mean, see Eq.~(\ref{eq:black_red}), for the cases of $\lambda=1.00, 0.95$ with adaptive rate $\alpha_E=1.92\gamma_{max}$ and marker number $N_p=256$M.}
	\label{fig:relfluc_time}
\end{figure}

Under the same simulation parameters, Fig.~\ref{fig:relfluc_time} further shows that the relative fluctuation of $T_i$ evaluated in an end-time window is derived mostly from its non-zonal variations. There, the curves are calculated as follows. Let the $j$-th flux tube on flux surface $x$ occupy the space:
\begin{eqnarray}
(y,z)_j &\in& \begin{bmatrix}
\frac{B_y(x)z}{B_z}\le y-j\Delta y<\frac{B_y(x)z}{B_z}+\Delta y & \\
0\le z<L_z
\end{bmatrix} \nonumber.
\end{eqnarray}

Visually, these $N_y$ flux tubes are the straight blue lines of Fig.~\ref{fig:Bgeometry} of $y-$width $\Delta y$, spanning each $x=$constant plane. The $j$-th flux tube gives the value $T_i(x,t;j)$. It is assumed that the plasma reaches thermal equilibrium instantly along the flux tube. Finally, let $\langle\cdot\rangle_t$ be the averaging in time for $t\in[t_1,t_2]$, and $\langle\cdot\rangle_{f.t.}$ be the flux-tube-average on the flux surface $x$, i.e.
\begin{eqnarray}
\langle T_i(x,t;j)\rangle_{f.t.} &=& \frac{1}{N_y}\sum_{j=1}^{N_y}T_i(x,t;j). \nonumber
\end{eqnarray}
Then, referring to Fig.~\ref{fig:relfluc_time},
\begin{eqnarray}
\text{black} &:& \frac{\sqrt{\langle\langle T_i\rangle^2(x,t)\rangle_t - \langle\langle T_i\rangle(x,t)\rangle_t^2}}{\langle\langle T_i\rangle(x,t)\rangle_t} \nonumber \\
\text{\amend orange \color{black}} &:& \frac{\sqrt{\langle\langle T_i^2(x,t;j)\rangle_{f.t.}\rangle_t - \langle\langle T_i(x,t;j)\rangle_{f.t.}\rangle_t^2}}{\langle\langle T_i(x,t;j)\rangle_{f.t.}\rangle_t} \label{eq:black_red}.
\end{eqnarray}

For each fixed $\lambda$, the relative fluctuation when non-zonal variations are included gives a value at least twice as high as that of the case when only the f.s.a.~values are considered. Consistent with Fig.~\ref{fig:relfluc_space}, lower $\lambda$ value gives higher fluctuation levels. These results summarily show that non-zonal fluctuations are dominant at quasi-steady state under current simulation parameters.

In conclusion, simulations using the adaptive scheme are shown to be better than the non-adaptive ones under all scenarios considered. To further test the advantage gained from the adaptive scheme under high fluctuation level scenarios would require simulating ITG instabilities in toroidal geometry, which could be done, for example, in the code ORB5~\cite{Lanti2020}. There, toroidal effects naturally results in higher fluxes and fluctuation levels, while zonal flows are an important factor determining the turbulent flux levels.

\section{Conclusion} \label{sec:conclusion}

In this work, the advantage gained using a simple adaptive control variate for the $\delta f$ scheme has been demonstrated in cases with high $T_i$ gradients and simple physics in sheared-slab geometry. The necessary implementation of the boundary conditions and a stationary heat source has been done to further ensure simulations reach quasi-steady state in reasonable integration time. The mechanism of the adaptive scheme has been described in detail. Namely, the adaptation of $T_i$ of the f.s.a.~Maxwellian control variate $f_0=f_M$ via a relaxation equation, through which a fraction the ion kinetic energy density derived from the marker represented $\delta f$ is periodically transfered to $f_0$.

For the cases considered, the adaptive scheme has shown to reduce the ion temperature relative deviation with increasing adaptive rate. Maximum relative deviation of $1\%$ has been achieved from $100\%$ of the non-adaptive case. Due to spurious increase of the zonal flow shear with time, all non-adaptive cases show an eventual collapse of the heat flux. Under the adaptive scheme, not only is quasi-steady state achieved with non-collapsing heat fluxes, these fluxes have also been achieved using marker numbers as low as $1/4$ of that required with the non-adaptive one, allowing for longer unquenched turbulence resulting in higher ion temperature deviation from its initial state. The scheme is further shown to be effective in reducing noise accumulation in the physically undamped zonal flow via the measure of $\omega_{E\times B}$. Such noise accumulation is shown to be the result of marker sampling. SNR values of adaptive simulations remain high for long integration times. In contrast, the eventual drop of SNR for the non-adaptive scheme is only postponed by increasing $N_p$.

We have then investigated further to determine if conservative noise control alone would suffice to produce a similar noise reduction advantage. Via a systematic separation of adaptation and noise control, it is shown that the scheme with an adaptive control variate $f_0=f_M(t)$, coupled with a noise control operator $S_n$ which relaxes $f$ to the same $f_0$, gives the best results. There, the former is proved to be more important than the latter in noise control. Moreover, it is shown that further increase in the strength of $S_n$ not only un-physically damps zonal flows at early times, but also is only able to delay the latter's eventual indefinite rise due to noise accumulation. The adaptation of $f_0$ is therefore shown to be necessary, even with an adaptation rate as low as $\alpha_E=0.12\gamma_{max}$.

To mimick prolonged fluxes and high fluctuation levels not afforded by slab-ITG, the $\langle\phi\rangle$ term of the adiabatic electron response in the quasi-neutrality equation is attenuated. Such a measure is done to better simulate edge-plasma conditions, despite testing the adaptive scheme in sheared-slab geometry. This work has shown that for slightly attenuated $\langle\phi\rangle$, the adaptive scheme exhibited improved noise control as before, for relative fluctuation in $T_i$ as high as $20\%$. Such fluctuations are shown to be non-zonal. Therefore, it hints to a more sophisticated $f_0$, which extends beyond a f.s.a.~function that could prove to be useful for better noise control. Despite that, the adaptive scheme still gives SNR values orders of magnitude higher than that of the non-adaptive scheme, further increasing the credibility of simulated results. To further test the merits of the adaptive scheme, its implementation in a code which better simulates the plasma edge in toroidal geometry is required.

\amend
The obvious generalization of this adaptive scheme to include a time-dependent background density profile is left for future work. This will involve a similar relaxation equation to that of Eq.~(\ref{eq:adp_relax}), with its own relaxation rate parameter $\alpha_n$. However, recalculation of the left-hand side of the quasi-neutrality equation, Eq.~(\ref{eq:qn}), should also be performed at periodic time intervals to account for the time-dependence of the background density. Such an adaptive scheme involving a control variate with both time-dependent background density and temperature profiles could be useful even in the core, e.g. in simulating kinetic ballooning, tearing and internal kink modes. In the presence of fast ions, a time-dependent background density could be useful when simulating Alfv\'{e}n or energetic particle modes.
\color{black}

\begin{acknowledgements}
Helpful discussions with Peter Donnel, Mohsen Sadr and Giovanni Di Giannatale are gratefully acknowledged. This work is part of the EUROfusion `Theory, Simulation, Validation and Verification' (TSVV) Task, and has been carried out within the framework of the EUROfusion Consortium, funded by the European Union via the Euratom Research and Training Programme (Grant Agreement No 101052200 -- EUROfusion). Views and opinions expressed are however those of the author(s) only and do not necessarily reflect those of the European Union or the European Commission. Neither the European Union nor the European Commission can be held responsible for them. This work is also supported by a grant from the Swiss National Supercomputing Centre (CSCS) under project ID s1067, and was partly supported by the Swiss National Science Foundation. 
\end{acknowledgements}

\section*{Data Availability Statement}

The data that support the findings of this study are available from the corresponding author upon reasonable request.

\appendix

\section{Quadrature point convergence for r.h.s.~of quasi-neutrality equation} \label{sec:app1}

The aim of this section is to determine the number of quadrature points sufficient to integrate the r.h.s.~of Eq.~(\ref{eq:qn_df}), formally written as

\begin{eqnarray}
\delta n_{gd} &=& \frac{1}{2\pi}\int\dint{^3R}\dint{^3v}\dint{\alpha}\delta[\vec{R}+\vec{\rho}_L-\vec{r}]\delta f(\vec{R},\vp,\mu). \label{eq:rhs}
\end{eqnarray}

To proceed, we introduce a simple but non-trivial form of $\delta f$, so that Eq.~(\ref{eq:rhs}) can be solved analytically, namely,

\begin{eqnarray}
\delta f &=& \frac{\cos(\xi x)}{(2\pi T_i/m)^{3/2}}\exp\left[-\frac{m\vp^2/2+\mu B}{T_i}\right] \nonumber \\
&=& \frac{\cos(\xi x)}{(2\pi)^{3/2}v_{th}^3}\exp\left[-\frac{\vp^2+v_\perp^2}{2v_{th}^2}\right], \label{eq:df_anal}
\end{eqnarray}

with $v_{th} = \sqrt{T_i/m_i}$ the local thermal velocity, $\mu=mv_\perp^2/2B$ the magnetic moment, and $\xi$ a constant. To simplify further, we shall assume constant $T_i$ and $B$. Furthermore, given that $B_\parallel^\star=B[1+m_iB_y'(x)B_z\vp/(eB^3)]$ and all velocity variables $\vp$ and $v_\perp$ are contained in $\delta f$ and that $\delta f$ is an even function of $\vp$, the velocity integration effectively becomes

\begin{eqnarray}
\delta f\,\dint{^3v} &=& \delta f\,\frac{2\pi B_\parallel^\star}{m_i}\dint{\vp}\dint{\mu} \nonumber \\
&=& \delta f\,\frac{2\pi B}{m_i}\dint{\vp}\dint{\mu} \nonumber \\
&=& \delta f\,2\pi v_\perp\dint{v_\perp}\dint{\vp}.
\end{eqnarray} 

Inserting Eq.~(\ref{eq:df_anal}) into Eq.~(\ref{eq:rhs}) gives

\begin{eqnarray*}
\delta n_{gd}(\vec{r}) &=& \text{Re}\{I(x)\},
\end{eqnarray*}

with

\begin{eqnarray*}
& & I(x) \\
&=& \int\dint{\alpha}\int_{-\infty}^{\infty}\frac{\dint{\vp}e^{-\frac{\vp^2}{2v_{th}^2}}}{\sqrt{2\pi v_{th}^2}}\times\int_0^\infty\frac{\dint{v_\perp}v_\perp e^{-\frac{v_\perp^2}{2v_{th}^2}}}{2\pi v_{th}^2}\exp[\imag\xi(x+\rho_L\sin\alpha)] \\
&=& e^{\imag\xi x}\int\dint{\alpha}\frac{1}{2\pi}\int_0^\infty\frac{\dint{v_\perp}}{v_{th}}\frac{v_\perp}{v_{th}}e^{-\frac{v_\perp^2}{2v_{th}^2}}\exp\left(\imag\xi\rho_{th}\frac{v_\perp}{v_{th}}\sin\alpha\right) \\
&=& e^{\imag\xi x}\int_0^\infty\frac{\dint{v_\perp}}{v_{th}} J_0\left(\xi\rho_{th}\frac{v_\perp}{v_{th}}\right)\frac{v_\perp}{v_{th}}e^{-\frac{v_\perp^2}{2v_{th}}} \\
&=& \exp\left[\imag\xi x-\frac{(\xi\rho_{th})^2}{2}\right],
\end{eqnarray*}

where $\rho_{th}=v_{th}/\Omega$ is the thermal Larmor radius and $J_0$ is the zero-th order Bessel function of the first kind.

The evaluation of the field $\phi$, which is represented by a B-spline expansion, involves the contraction of the r.h.s.~of Eq.~(\ref{eq:qn_df}) with a B-spline element of order $p$,  $\Lambda^p(\vec{r}-\vec{r}_{ijk})=\Lambda^p(x-x_i)\Lambda^p(y-y_j)\Lambda^p(z-z_k)$. Here, the zero-th order B-spline is defined as

\begin{eqnarray*}
\Lambda^0(x) &=& \begin{cases}
1 & |x|<\Delta x/2 \\
0 & \text{else}
\end{cases}
\end{eqnarray*}

so that $\int\dint{x}\,\Lambda^0(x) = \Delta x$, and the higher order elements by the recurrence relation

\begin{eqnarray*}
\Lambda^p(x) &=& \frac{1}{\Delta x}\Lambda^{p-1}(x)*\Lambda^0(x) \hspace{1cm}p\ge1,
\end{eqnarray*}

\begin{figure*}
	\centering
	\begin{subfigure}[c]{0.33\textwidth}
	    \caption{}
		\centering
		\includegraphics[width=1.0\linewidth]{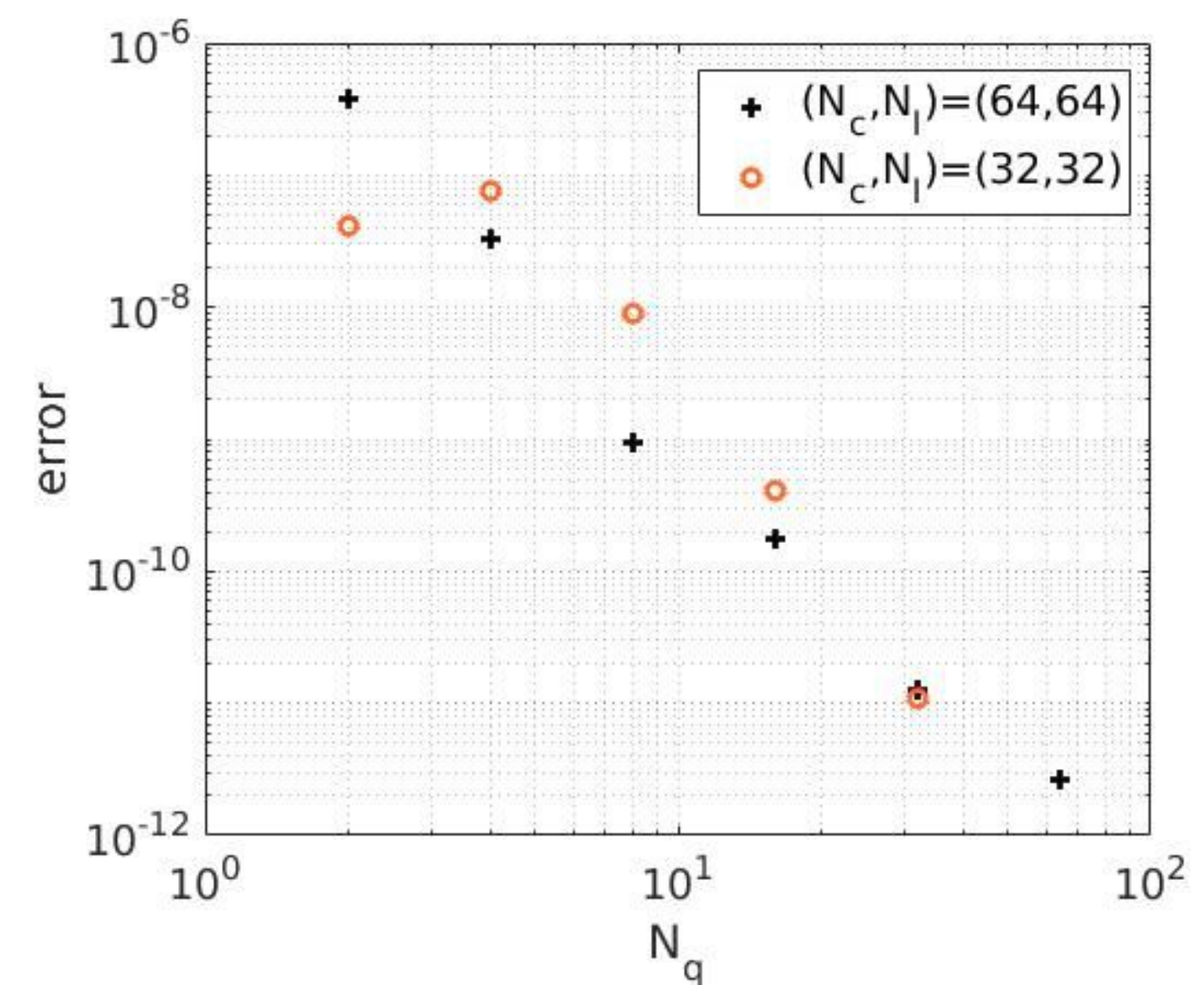}
	\end{subfigure}
	\begin{subfigure}[c]{0.33\textwidth}
	    \caption{}
		\centering
		\includegraphics[width=1.0\linewidth]{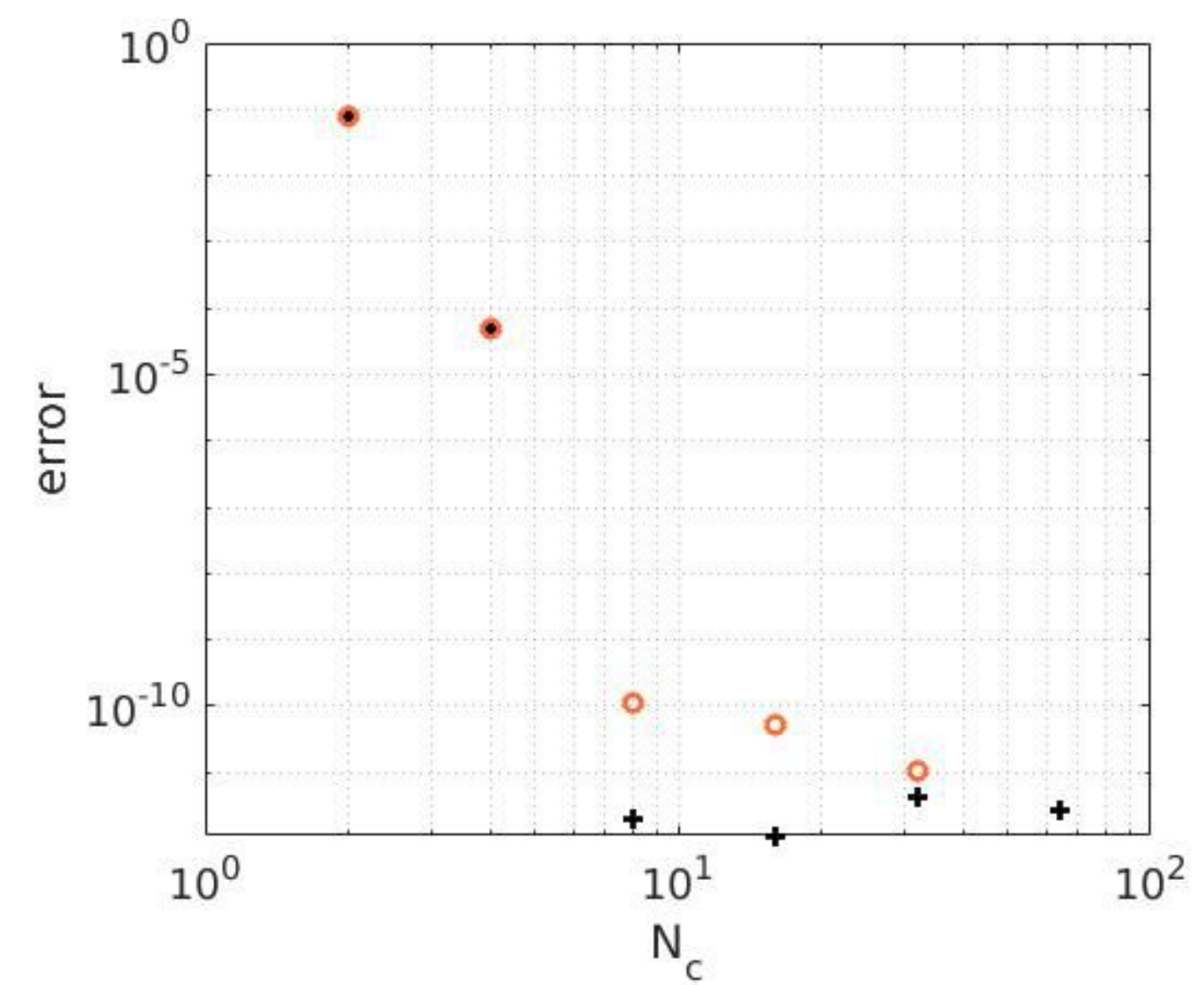}
	\end{subfigure}
	\begin{subfigure}[c]{0.33\textwidth}
	    \caption{}
		\centering
		\includegraphics[width=1.0\linewidth]{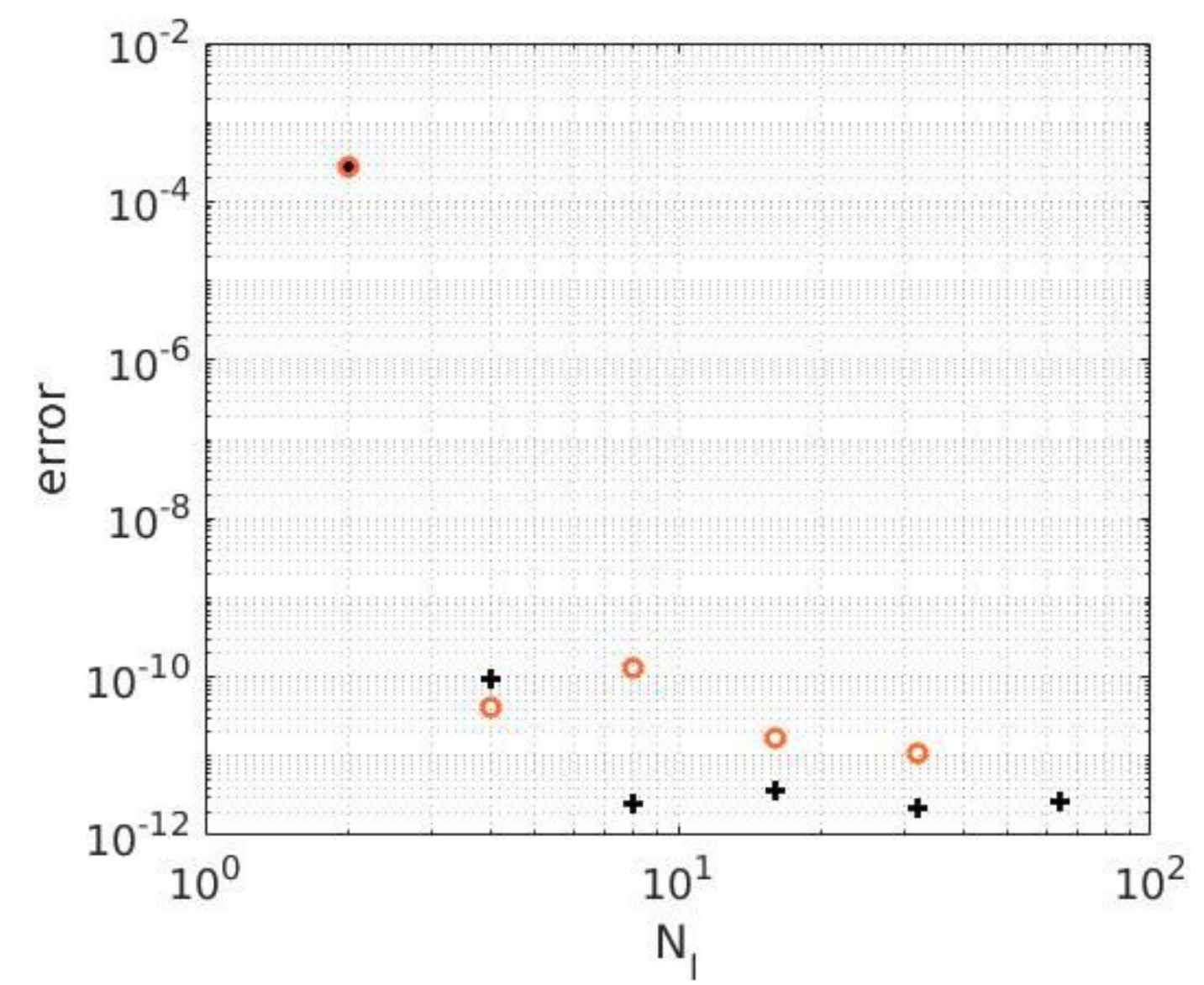}
	\end{subfigure}
	\caption{Convergence analysis for the (a) $x$ integration with Gauss-Legrendre, (b) $\alpha$ integration with Gauss-Chebyshev, and (c) $\mu$ integration with Gauss-Laguerre, quadrature points respectively, of the $f_0(t,\vec{R},\vp,\mu)-f_0(0,\vec{R},\vp,\mu)$ term of the r.h.s.~of Eq.~(\ref{eq:qn_df}). $(N_q,N_c,N_l)$ quadrature points are used for the integration of the dimensions $x$, $\alpha$ and $\mu$ respectively. For each of the above cases, the quadrature point number for a dimension is set to vary, while the other two are fixed at either $32$ or $64$.}
	\label{fig:rhsconv}
\end{figure*}

where $\Delta x$ is the grid size of equidistant points along dimension $x$, and $*$ stands for convolution. Therefore, we conduct a convergence analysis on the real part of the expression:

\begin{eqnarray}
& & \int\dint{^3r}\Lambda^p(\vec{r}-\vec{r}_{ijk})\delta n_{gd}(x) \nonumber \\
&=& \Delta y\Delta z\exp\left[\imag\xi x_i-\frac{(\xi\rho_{th})^2}{2}\right]\times\int\dint{x}\Lambda^p(x-x_i)e^{\imag\xi(x-x_i)} \nonumber \\
&=& \Delta x\Delta y\Delta z\exp\left[\imag\xi x_i-\frac{(\xi\rho_{th})^2}{2}\right]\left[\frac{2}{\xi\Delta x}\sin\left(\frac{\xi\Delta x}{2}\right)\right]^p. \label{eq:rhs_anal}
\end{eqnarray}

In practice, the $f_0(\vec{R},\vp,\mu,t)-f_0(\vec{R},\vp,\mu,0)$ term of the r.h.s. of Eq.~(\ref{eq:qn_df}) is calculated using Gaussian quadratures. Specifically, due to the limits of integration for each of the variables, $N_q$ Legendre points for the $x$ integral, $N_c$ Chebyshev points for the $\alpha$ integral, and $N_l$ Laguerre points for the $\mu$ integral have been used. The convergence study for $\xi L_x=32\pi$, which exceeds the typical wavelength of the integrand for this work, is show in Fig.~\ref{fig:rhsconv}.

\section*{References}

\bibliography{common_references}

\end{document}